\documentclass[]{aastex631}

\usepackage{amsmath}
\usepackage{mathtools}

\newcommand{\lalpha}{Lyman-$\alpha$}
\newcommand{\halpha}{H-$\alpha$}
\newcommand{\htwo}{H$_2$}

\newcommand{\hi}{H {\sc i}}

\submitjournal{Icarus}

\shorttitle{Photoelectrons in exoplanet atmospheres}
\shortauthors{A. Garc\'ia Mu\~noz}
\graphicspath{{./}{figures/}}

\begin{document}

\title{An efficient Monte Carlo model for the slowing down of photoelectrons. \\
Application to {\halpha} in exoplanet atmospheres.
}

\author[0000-0003-1756-4825]{A. Garc\'ia Mu\~noz}
\affiliation{
Universit\'e Paris-Saclay, Universit\'e Paris Cit\'e, CEA, CNRS, AIM, 91191, Gif-sur-Yvette, France}

\begin{abstract}

Photoelectrons, the fast electrons produced in the photoionization of planetary atmospheres, drive
transformations in the atmospheric gas that are often inhibited by energy considerations for thermal electrons. 
The transformations include excitation and ionization of atoms and molecules, which affect the detectability of these gases and constrain the fraction of incident stellar radiation that transforms into heat. 
To gain insight into these important questions, we build a Monte Carlo model that solves the slowing down of photoelectrons in a gas with arbitrary amounts of H and He atoms and thermal electrons. 
Our novel multi-score scheme differs from similar tools in that it efficiently handles rare collisional channels, 
as in the case of low-abundance excited atoms that undergo superelastic and inelastic collisions. 
The model is validated and its performance demonstrated. 
Further, we investigate whether photoelectrons might affect the population of the excited hydrogen H(2) detected at some exoplanet atmospheres by transmission spectroscopy in the {\halpha} line. 
For the ultra-hot Jupiter HAT-P-32b, 
we find that photoelectron-driven excitation of H(2) is inefficient at the pressures probed by the line core but becomes significant (yet sub-dominant) deeper in the atmosphere where the line wings form. The contribution of photoelectrons to the destruction of H(2) either by collisional deexcitation or ionization is entirely negligible, a conclusion likely to hold for exoplanet atmospheres at large. 
Importantly, photoelectrons dominate the gas ionization at the altitudes probed by the {\halpha} line, a fact that will likely affect, even if indirectly, the population of H(2) and other tracers such as metastable helium.  Future modeling of these excited levels should incorporate photoelectron-driven ionization.

\end{abstract}

\vspace{-1.5cm}


\section*{Highlights}

\begin{itemize}
    \item A novel Monte Carlo method to simulate the slowing down of photoelectrons in an atmosphere is proposed. 
    \item The model performs efficiently with both major and rare collisional channels, and treats superelastic collisions.
    \item The effect of photoelectrons on the upper atmosphere of an ultra-hot Jupiter is assessed. 
    \item Photoelectrons may control the ionization of the atmosphere where diagnostic lines such as {\halpha} are formed.\\
\end{itemize}

\section{Introduction} \label{sec:intro}

The bulk of the extreme ultraviolet photons (XUV; wavelengths $\lambda${$\lesssim$}912 {\AA} that may ionize the H and O atoms) incident on a planet becomes deposited in its upper atmosphere.
The most energetic photons eject one or more electrons from the heavy particles (atoms and molecules) in the gas. 
The nascent photoelectrons carry away as kinetic energy the excess of the photon energy over the 
ionization threshold of the heavy particles. 
For host stars that emit strongly in the XUV, the energy imparted to the photoelectrons may amount to hundreds of electron-Volts (eV). 
This is notably more than the mean kinetic energy of the thermal electrons ($\sim$0.65 eV at a temperature of 5,000 K) or the excitation and ionization thresholds of common heavy particles in an atmosphere. 
The photoelectrons interact through collisions with the heavy particles and thermal electrons, losing progressively their energy until they ultimately thermalize.
The collisions with the heavy particles result in 
a cascade of photoelectrons with a broad distribution of energies and with the
heavy particles adopting different excited and ionized forms.
 The elastic collisions with the thermal electrons, and to a much lesser extent with the heavy particles, contribute to the heating of the gas.\\ 

The slowing down of photoelectrons and non-thermal electrons in general has been investigated numerous times in astrophysical
\citep[][]{glassgoldlanger1973,lietal2012,ferlandetal2016}
and non-astrophysical problems \citep[][]{carron2007}.
The slowing down is  controlled by the composition and fractional ionization of the gas. 
The composition defines the energy thresholds for the heavy particles, and the relative abundance of thermal electrons dictates the partitioning between the inelastic collisions with heavy particles and the elastic collisions with thermal electrons. The absolute gas density, which plays a minor role in establishing the relative importance between the available collisional channels, is however fundamental at determining whether the heavy particles return the potential energy first stored as excitation and ionization energy back to the gas through collisions or radiate it away. Photoelectrons have received significant attention in the study of the solar system planets \citep{mendilloetal2002,campbellbrunger2016} because they
affect the energy budget of the atmospheres and  because the interaction of the photoelectrons with the atmospheric gas drives dayglow and auroral emissions that are useful to remotely probe the atmospheres.\\ 

Photoelectrons have so far received little attention in the study of exoplanet atmospheres. There are however good reasons to explore their effects,
the importance of which will depend on the planet's gas composition and exposure to XUV radiation, and on the specifics of the host star's spectrum. 
The available calculations for hydrogen-dominated atmospheres 
 suggest that the fraction of stellar XUV energy that goes into heating the atmosphere may be as low as 20{\%} if the gas remains neutral in the form of {\htwo} \citep{cecchipestellinietal2006,cecchipestellinietal2009,shematovichetal2014} 
but approaches 100{\%} if the gas is moderately ionized \citep{shaikhislamovetal2020}. 
Even if some of the early estimates omit thermal electrons in the slowing down process and do not track the chemical conversions that follow after {\htwo} ionization, which should in both cases modify upwards the heating efficiency, they nevertheless show that the effect of photoelectrons may be significant.
For the strongly irradiated exoplanets that are often targeted for characterization, 
variations of this order in the heating efficiency translate into similar uncertainties in the estimated lifetimes of their atmospheres \citep{garciamunozetal2020,garciamunozetal2021,shaikhislamovetal2020b}. 
Minimizing such uncertainties is highly desirable to gain confidence in our understanding of planetary evolution. 
Photoelectrons produced by X-ray photons, 
often defined by wavelengths shortwards of 100 {\AA}
and deposited at moderately high densities, 
can also modify the molecular chemistry in the deeper atmosphere \citep{loccietal2022}.
\\ 

There are additional reasons to consider the role of photoelectrons in exoplanet atmospheres. Excited hydrogen has been detected by means of transmission spectroscopy in the {\halpha} line at a few exoplanets \citep[][]{cauleyetal2017,yanhenning2018,czeslaetal2022}, 
and a number of models have been proposed for their interpretation \citep[][]{huangetal2017,garciamunozschneider2019,fossatietal2021,yanetal2022}. The impact of photoelectrons on the abundance of excited hydrogen remains however unexplored. 
A main recent finding in exoplanet science is that sub-Neptune-sized planets are abundant \citep{batalha2014}, and formation models predict that their atmospheric compositions may be very diverse
\citep[][]{venturinietal2020}. Elucidating the heating efficiency of atmospheres rich in atoms
such as oxygen or carbon, and the contribution of photoelectrons to it, will set valuable constraints on the long-term stability of  high-metallicity atmospheres. This problem has also received virtually no attention to date.\\ 

With the above in mind, we built a Monte Carlo (MC) scheme to simulate the slowing down of photoelectrons in a gas of atoms and thermal electrons. The scheme is quite general, but we focus for now on gases of H and He atoms in possibly arbitrary amounts, and leave for the future the addition of other atoms and molecules. 
The model overcomes some of the deficiencies of MC schemes in their treatment of rare collisional channels and naturally deals with inelastic and superelastic collisions of heavy particles in excited levels. 
The flexibility and efficiency of the model suggest that it is a viable option for online implementation in photochemical-hydrodynamical models where the gas composition and ionization fraction may vary 
in space and time. The paper is structured as follows. Sections {\S}\ref{sec:methods}--\ref{sec:validationetal} describe the scheme, its validation and main features; {\S}\ref{sec:implementation}
illustrates how the model might be incorporated into photochemical-hydrodynamical simulations; {\S}\ref{sec:halpha} addresses the effect of photoelectrons on the abundance of excited hydrogen and ionization in the upper atmosphere of an ultra-hot Jupiter. Last, {\S}\ref{sec:summary} summarizes the main findings and notes some possibilities for future work.

\section{Methods} \label{sec:methods}

Solutions to the problem of the slowing down of photoelectrons in a gas have been found by a variety of approaches, including the continuous slowing down approximation \citep{rees1963}, the time-dependent Boltzmann equation \citep{slinkeretal1988}, the
Spencer-Fano equation \citep{voit1991,kozmafransson1992} and the discrete energy bin method \citep{dalgarnolejeune1971}. 
Our focus here is on MC schemes \citep[][]{berger1963,habinggoldsmith1971,ciceroneetal1973,heapsetal1975,shull1979,bhardwajmichael1999,
valdesferrara2008,furlanettostoever2010}, which are flexible and intuitive to develop.
MC schemes work by numerically simulating experiments in which 
photoelectrons of a given initial energy are injected into the gas. The scheme
tracks how the initial and any newly-created photoelectrons exchange energy in collisions with the heavy particles and thermal electrons while becoming thermalized. 
Essential to any MC scheme is the idea that the slowing down of a photoelectron can be simulated by sequences
of individual collisions, and that at each collision the electron loses or gains a discrete amount of energy. 
Useful quantities for the macroscopic modeling of the gas 
(for example, excitation and ionization rates, or the fraction of injected energy that goes into heating) are obtained as ensemble averages over many simulations. 
\\

MC schemes are often deemed poor options when speediness is critical and there exist rare (low probability) collisional channels whose contributions must be quantified \citep{kozmafransson1992,furlanettostoever2010}. 
A main motivation for our work is to devise a MC scheme that remains efficient when rare collisional channels exist, partly as a way to test the effect of photoelectrons on the fate of gases in excited levels. 
We assume that the photoelectrons do not travel far before thermalizing, and that it is justified to omit transport effects. 
The validity of this local deposition approximation must be assessed on a case-by-case basis, and ideally the necessary information should be available as part of the model solution.
Most MC schemes used to study the slowing down of photoelectrons share some basic features. We review them here, and use those ideas to introduce our novel multi-score concept.

\subsection{MC simulation of the photoelectron cascade} \label{subsec:cascadesimulation}

The collisional channels available to a photoelectron of energy $E$($e'$) in a gas of atoms and thermal electrons are:
\begin{eqnarray}
 e' + {e_{\rm{th}}} & \xrightarrow{\sigma_{ee};\; E(e'')-E(e')=-f_{ee}E(e')<0} & e'' + {e_{\rm{th}}} \label{elastic_ee_equation} \\
 e' + X_{\rm{L}} & \xrightarrow{\sigma_{X_L}^{mt};\;E(e'')-E(e')=-2m_e/M_X E(e')<0 } & e'' + X_{\rm{L}} \label{elastic_eh_equation} \\
 e' + X_{\rm{L}} & \xrightarrow{\sigma_{X_L\rightarrow X_U}^{exc};\;E(e'')-E(e')=-E(X_U)+E(X_L)<0 } & e'' + X_{\rm{U}} \label{excitation_eh_equation} \\ 
 e' + X_{\rm{U}} & \xrightarrow{\sigma_{X_U\rightarrow X_L}^{deexc};\;E(e'')-E(e')=+E(X_U)-E(X_L)>0 } & e'' + X_{\rm{L}} \label{deexcitation_eh_equation} \\ 
 e' + X_{\rm{L}} & \xrightarrow{\sigma_{X_L\rightarrow X^+}^{ion};\;E(e'')+E(e''')-E(e')=-E(X^+)+E(X_L)=IP(X_L)<0 } & e'' + e''' + X^{+}. \label{ionization_eh_equation}
 \end{eqnarray}
Here, $e_{\rm{th}}$ stands for thermal electrons, and $X$ for the heavy particle. $X_{\rm{L}}$ and $X_{\rm{U}}$ are different excitation levels, with $E$($X_{\rm{L}}$){$<$}$E$($X_{\rm{U}}$). 
$X^+$ is the heavy particle in the ionization stage immediately above $X_{\rm{L}}$, and IP($X_{\rm{L}}$) stands for the ionization potential of $X_{\rm{L}}$. 
Channels as those of Eqs. \ref{elastic_eh_equation}--\ref{ionization_eh_equation} should be included for all the heavy particles in the gas, and for all the relevant combinations of lower and upper levels. Our implementation of the MC scheme handles such generalities, but for clarity we focus the present description on the above simplified scheme.
The outcome of an ionization collision is two electrons $e''$ and $e'''$. We adopt the convention that $e''$ is the slowest of the two and $E$($e''$){$\le$}($E$($e'$){$-$}IP($X_{\rm{L}}$))/2, and that $e'''$ is the fastest, which by energy conservation entails 
$E$($e'''$)=$E$($e'$){$-$}IP($X_{\rm{L}}$){$-$}$E$($e''$). 
Some information on cross sections and the change in energy experienced by the photoelectron at a collision is given 
above the arrows that connect the initial and final participants in each collisional channel.
For example, $f_{\rm{ee}}$ is the fractional energy lost by the photoelectrons in collisions with thermal electrons, and $m_e$/$M_X$ is the  electron-to-heavy particle mass ratio ($\sim$5$\times$10$^{-4}$ for $X$=H, and less for  other heavy particles). 
The electrons lose energy in elastic (Eqs. \ref{elastic_ee_equation}, \ref{elastic_eh_equation}) and inelastic  
(Eqs. \ref{excitation_eh_equation}, \ref{ionization_eh_equation}) collisions, and gain it in superelastic collisions (Eq. \ref{deexcitation_eh_equation}). 
The list of collisional channels omits the reverse of Eq. \ref{ionization_eh_equation} because being a three-body process it is of minor importance at low densities \citep{valdesferrara2008}.\\

In a MC scheme, the collisional channels are characterized by probabilities $p_j$($E'$) ({$\equiv$}$p_j$($E$($e'$); $j$=1--5 
based on Eqs. \ref{elastic_ee_equation}-\ref{ionization_eh_equation}) 
equal to the ratio of the channel collisional frequency $\nu_j$ and the total frequency 
$\nu_{\rm{T}}$=$\sum_j \nu_j$, namely $p_j$($E'$)=$\nu_j/\nu_{\rm{T}}$. 
The collisional frequency is given by the number density of the relevant collision target times its cross section times the photoelectron velocity, $\nu_j$=$n_j${$\sigma_j$}{$v_{\rm{e}}$}. For example, for elastic collisions of 
photoelectrons of energy $E'$ and thermal electrons, $n_j$=$n_{\rm{e}}$ is the number density of the thermal electrons, 
{$\sigma_j$}=$\sigma_{\rm{ee}}$($E'$) is the electron-electron cross section, 
and $v_{\rm{e}}$=$\sqrt{2E'/m_e}$. 
$p_j$($E'$) dictates the average fraction of photoelectrons of energy $E'$ that will scatter through 
channel $j$ or equivalently the fraction of times that any such photoelectron will pursue channel $j$ if this thought-out experiment is performed many times. 
For ionization collisions, $p_j$($E'$) bears the extra meaning of being 
the average number of newly-created photoelectrons relative to the number of incident photoelectrons. 
Each collision occurs with a net change in the energy of the participating photoelectrons  $\Delta${$E_j$}=$E(e'')$+($E(e''')$){$-$}$E(e')$ that depends on the specific channel $j$.\\

\begin{figure}[h]
    \centering
    \includegraphics[width=12.cm]{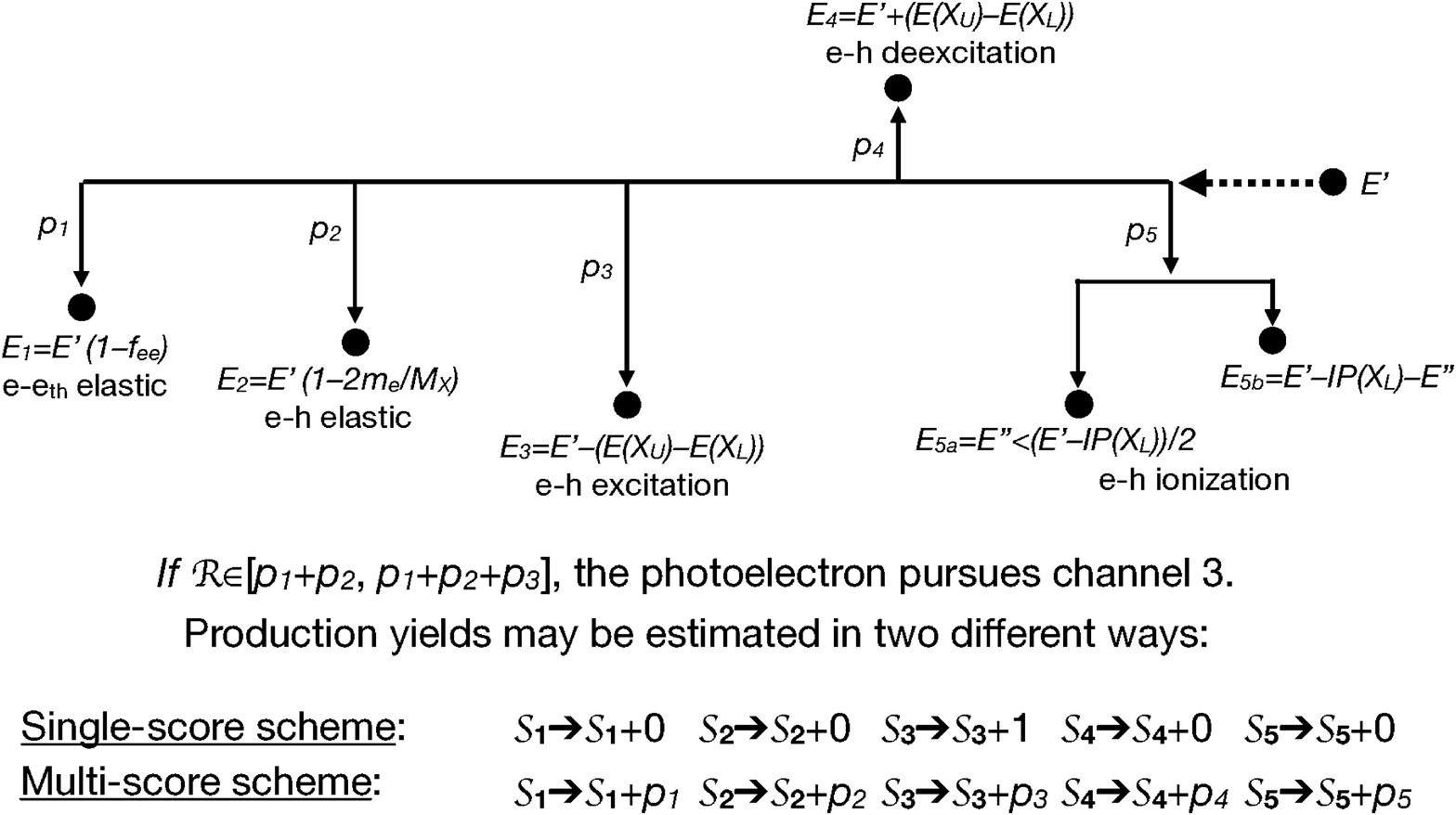}
    \caption{Flowchart of the MC scheme for the collisional channels of Eqs. 
    \ref{elastic_ee_equation}-\ref{ionization_eh_equation}.
    Following a collision a photoelectron of energy $E'$ will pursue one of the available channels that is decided on the basis of a randon number $\mathcal{R}$. If for example $\mathcal{R}${$\in$}[$p_1$+$p_2$, $p_1$+$p_2$+$p_3$], the photoelectron will pursue collisional channel 3 for heavy particle excitation, and its energy will be degraded from $E'${$\rightarrow${$E_3$}}. In the classical approach for estimating production yields, only the yield $\mathcal{S}_3$ is updated. In the multi-score scheme, all the production yields are updated thereby using more of the available information.
    } 
    \label{sketchMC_fig}
\end{figure}

The simulation with a MC scheme starts by injecting a photoelectron of energy $E_0$, continues by tracking 
in energy the cascade that develops with the injected photoelectron plus others that are newly created, 
and terminates when each of the photoelectrons falls below a prescribed cutoff energy $E_{\rm{cutoff}}$
(see below for a reasoned definition of this quantity).
Photoelectrons with energies {$>$}{$E_{\rm{cutoff}}$} are considered active in the simulation and are tracked further, but are removed from the simulation 
once their energies fall below {$E_{\rm{cutoff}}$}. 
The immediate fate of an active photoelectron of arbitrary energy $E'${$>$}{$E_{\rm{cutoff}}$} is decided based on the energy-dependent probabilities $p_j$, as sketched in Fig. \ref{sketchMC_fig}. 
To this end a random number $\mathcal{R}$ is generated and compared to the cumulative probability 
$P_j$=$\sum_{k=0}^j p_k$ (taking $p_0${$\equiv$}0, $P_j$ ranges between 0 and 1).
The photoelectron is then forced to pursue the collisional channel $j$ that satisfies the condition 
$\mathcal{R}${$\in$}[$P_{j-1}$, $P_j$].
If the photoelectron pursues an elastic, excitation or deexcitation channel, its energy is updated by a well-defined $\Delta${$E$} ($<$0 for elastic and inelastic collisions, $>$0 for superelastic collisions), and the post-collision photoelectron will go again through a similar sequence of decisions.
If however the photoelectron pursues an ionization channel, an extra photoelectron is created and added to the cascade. 
In this case,
the energy of the slowest of the two photoelectrons is determined from randomly sampling 
a probability distribution (Appendix \ref{appendix:xsections}), while the energy of the fastest one is derived from 
energy conservation. 
Once the energies of the two post-collision photoelectrons are updated, they are 
tracked further through the standard sequence of decisions in Fig. \ref{sketchMC_fig}. 
In practice it is advantageous to track first the slowest of the post-collision photoelectrons following ionization as this tends to keep to a minimum the list of active photoelectrons and the memory requirements.
\\

The adequate definition of $E_{\rm{cutoff}}$ may depend on the problem to be solved. 
If for example the priority is to determine the amounts of energy that go into each of the inelastic collisional channels,
then it is practical to define the cutoff energy by an arbitrary value below the smallest inelastic threshold. Indeed, once the photoelectron energy falls below $E_{\rm{cutoff}}$, and provided that only elastic or inelastic channels really matter, the photoelectron energy can only go into heating the gas.
Superlastic collisions occur without an energy threshold, and some extra attention must be paid to  $E_{\rm{cutoff}}$ 
if it is thought that an early termination of the simulation may bias the results. 
For our calculations, we have generally adopted $E_{\rm{cutoff}}$=2{$k_B$}$T_e$ ({$k_B$} is Boltzmann's constant), which terminates the simulation once the photoelectron reaches an energy slightly above that of the thermal electrons at temperature $T_e$ but well below the energy thresholds for excitation in most atoms of atmospheric interest. This choice ensures that the effect of the photoelectrons on deexcitation channels is well taken into account, but that 
the contributions of the thermal electrons and the photoelectrons remain separate. In any case, 
it is always possible to confirm the impact of this choice by a sensitivity analysis.
An additional question that arises is whether it is correct to remove from the simulation a photoelectron once 
its energy falls below $E_{\rm{cutoff}}$. Indeed, such an electron might conceivably gain energy through superelastic collisions and overcome again that barrier. 
The possibility is real, but the probability for superelastic collisions is $\ll$1 in most applications, and reintroducing the photoelectron in the simulation makes no practical difference.

\subsection{\label{subsec:yields}Estimating the production and energy yields}

We define the production yield for a collisional channel as the total number of times that the channel is pursued
relative to the number of photoelectrons that are injected at the top of the cascade. 
The production yields depend on the injection energy $E_0$. 
For their calculation, classical MC schemes utilize scores $\mathcal{S}_j$, one per channel, that keep track of the number of times that the channel is pursued within each simulation. 
If for example it is determined that following a collision the simulated photoelectron will pursue a channel $j$, then the corresponding score is increased by one and  $\mathcal{S}_j${$\rightarrow$}{$\mathcal{S}_j$}+1. 
Hereafter we refer to such schemes by the term of single-score schemes.
After a simulation is terminated, 
the production yields are simply estimated from the ratio between the scores and the number of injected photoelectrons that have been simulated. Production rates $W_j$ (in events cm$^{-3}$s$^{-1}$) may be obtained by multiplying the yields by the rate of injection of photoelectrons at the top of the cascade ${W}_0$ (in $e^-$cm$^{-3}$s$^{-1}$).
Typically, a calculation will simulate multiple cascades to obtain reliable statistics.\\

It is apparent why such an approach becomes inefficient at estimating the production yields for rare collisional channels. If the probability $p_j${$\ll$}1, the scheme needs a number of simulations $\gg$1/$p_j$ to pursue the rare channel a reasonable number of times and obtain reliable statistics.
In the past, this difficulty has been circumvented by removing the rare channel from the list of available channels and calculating its production yield a posteriori with the photoelectron statistics set by the dominant channels.  This idea is at the core of the yield spectra \citep{greenetal1977,bhardwajmichael1999} and works well when the rare collisional channel has a weak impact on the overall photoelectron statistics.
This approach however becomes inaccurate when the probability is not sufficiently small, and becomes impractical when it is not easy to anticipate which channels will be rare or dominant. 
Ideally, a MC scheme should treat all the available collisional channels in the same way and with the same expected accuracy.
\\

We propose a way to overcome the difficulty of single-score schemes to handle rare collisional channels that does not require further calculations. 
The idea follows from the sketch of Fig. \ref{sketchMC_fig}. If the collision could be simulated numerous times, it is clear that 
a better estimate of the average score for each channel would be $\mathcal{S}_j${$\rightarrow$}{$\mathcal{S}_j$}{+}{$p_j$}.
Thus, our proposed multi-score scheme follows the usual sequence of collisions of single-score schemes, but at each collision it updates the scores of \textit{all} channels according to their probabilities rather than simply updating the score of a single channel. 
As quantified in {\S}\ref{sec:validationetal},
by using all the information on probabilities available at each collision, the multi-score scheme 
provides more rapidly converging estimates of the production yields for rare channels than the single-score scheme.
\\

We also define energy yields as the fractional energy that goes into each collisional channel 
relative to the energy injected at the top of the cascade. The energy yields provide insight into how the slowing down of the photoelectrons might affect the heating of the gas. In particular, the total 
energy yield from all the elastic channels sets a minimum of the injected energy that goes into heating. 
Following the reasoning of the production yields, in the single-score scheme the energy yields
are estimated through scores $\mathcal{S'}_j${$\rightarrow$}{$\mathcal{S'}_j$}+$\Delta${$E_j$}, 
where $j$ is the channel pursued by the photoelectron and 
$\Delta${$E_j$} is the corresponding energy variation. 
Proceeding one photoelectron at a time, 
the single-score scheme is designed to conserve energy, a property that may be utilized for testing the consistency of the solution. In the multi-score scheme one may define the
scores $\mathcal{S'}_j${$\rightarrow$}{$\mathcal{S'}_j$}+{$p_j$}$\Delta${$E_j$} for all channels $j$
that must be updated at each collision even though in the scheme the photoelectron will pursue only one of the channels. 
We note that the latter approach does not strictly conserve energy because in the simulated cascade 
the photoelectron pursues a single channel $j$. 
As seen in {\S}\ref{sec:validationetal}, this inconsistency has no consequence as both treatments converge rapidly to the correct energy yields.

\subsection{\label{relaxdef_sec} Estimating the relaxation time and pathlength}

The time that it takes for the injected photoelectrons to 
slow down from $E_0$ to $E_{\rm{cutoff}}$, and the pathlength they travel in the process, is useful information to assess if the gas with which the photoelectrons interact can be assumed approximately uniform.
To that end, our implemented scheme tracks the time-of-arrival for each photoelectron in the cascade as they fall below $E_{\rm{cutoff}}$. 
It then identifies the longest time-of-arrival for each simulated cascade, and calculates the relaxation time $\tau_{\rm{relax}}$ as the average of the longest times-of-arrival from all the simulated cascades. 
The time-of-arrival for a photoelectron is calculated as the total time accumulated in between collisions. For photoelectrons that are newly created in the cascade, the time-of-arrival takes into account the history of the parent photoelectron. 
If for example the photoelectron pursues a collisional channel $j$ the time in between collisions is simply  $\Delta${$t_{\rm{step}}$}=1/$\nu_j$.
Following the same reasoning, the MC scheme calculates a relaxation pathlength 
$L_{\rm{relax}}$ as an estimate of the pathlength travelled by the photoelectrons while slowing down from 
$E_0$ to $E_{\rm{cutoff}}$. In this case, the pathlength travelled by the photoelectron in between collisions is  
$\Delta${$L_{\rm{step}}$}=$v_{\rm{e}}${$\Delta$}{$t_{\rm{step}}$}, and $L_{\rm{relax}}$ is calculated by accumulation and
averaging of the partial $\Delta${$L_{\rm{step}}$}. As calculated here, $L_{\rm{relax}}$ estimates the total pathlength travelled by a photoelectron regardless of the direction in which it moves. Because the photoelectrons will move at times away from the original injection location and other times towards it, 
$L_{\rm{relax}}$ is likely to overestimate the deposition range that is useful in some applications. We take $L_{\rm{relax}}$ as a first order estimate of that deposition range.
Both $\tau_{\rm{relax}}$ and $L_{\rm{relax}}$ depend on the total number density of the gas and not only on the relative abundances.

\subsection{Estimating the energy distribution function and flux of photoelectrons}

The photoelectrons in the cascade are characterized by an 
energy distribution function $dn_e$($E$)/$dE$ [e$^-$cm$^{-3}$eV$^{-1}$] and a (direction-integrated) flux
$\phi_e$($E$) [e$^-$cm$^{-2}$s$^{-1}$eV$^{-1}$], which are related through the identity  $\phi_e$($E$)$\equiv${$v_e$($E$)}$dn_e/dE$($E$).
Knowing these properties of the photoelectron cascade
is useful in itself. It also enables to calculate the production rate for any channel $j$ through:
\begin{equation}
    W_j=n_j \int \sigma_j(E) v_e(E) \frac{d n_e}{dE}(E) dE
    \approx n_j \sum_{i} \sigma_j(E_i) v_e(E_i) \frac{d n_e}{dE}(E_i) \Delta_g E_i,
    \label{Wj_eedf_eq}
\end{equation}
the discrete form of which assumes an energy grid of bin sizes $\Delta_g${$E_i$} centered at $E_i$.
\\

To estimate the photoelectron flux, 
the MC scheme uses additional scores, one per energy bin, that record the number of collisions occurring within the bin during the simulation. Every time a collision occurs, the score is updated
$N_{\rm{colls}}$($E_i$){$\rightarrow$}$N_{\rm{colls}}$($E_i$)+1. 
Averaging over all the simulations, the energy distribution function is finally obtained from:
\begin{equation}
W_0 N_{\rm{colls}}(E_i) = \nu_T(E_i) \frac{dn_e}{dE}(E_i) \Delta_g E_i. 
\label{Wj_eedf2_eq}
\end{equation}
This expression establishes the steady-state balance between the rate at which collisions occur within the energy bin centered at $E_i$ (left hand side), and how the electron density within the bin reacts to being lost with a total frequency 
$\nu_T(E_i)$ (right hand side). 
$W_0$ scales the number of collision events per energy bin to the rate at which photoelectrons are injected into the gas.
In our implementation the photoelectron flux, rather than the energy distribution function, is calculated 
from Eq. \ref{Wj_eedf2_eq} plus 
$\phi_e$($E_i$)$\equiv${$v_e$($E_i$)}$dn_e/dE$($E_i$) using the energy grid at which the cross sections are prescribed. 
As it should be, application to Eq. \ref{Wj_eedf_eq}
of the photoelectron flux $\phi_e$($E_i$) inferred from Eq. \ref{Wj_eedf2_eq} results in the production yields obtained in {\S}\ref{subsec:yields}. 
\\

\section{\label{sec:validationetal} Model validation and performance} 

\subsection{Comparison with earlier work} 

We set out to validate our multi-score scheme with calculations for H and He gases, for which there is a significant literature
\citep[][]{spitzerscott1969,habinggoldsmith1971,bergeroncollin-souffrin1973,shull1979,shullvansteenberg1985,xumccray1991,dalgarnoetal1999,valdesferrara2008}. 
Past inter-model comparisons \citep[][]{furlanettostoever2010} show that the agreement is overall satisfactory and that when minor discrepancies exist they are mainly attributable to differences in the adopted cross sections rather than to the methods for simulating the slowing down of the photoelectrons. 
We have collected the up-to-date set of cross sections described in Appendix \ref{appendix:xsections}, and focus here on reproducing the past calculations.  
Particularly relevant are the calculations by \citet{xumccray1991} and \citet{dalgarnoetal1999}, who obtained excitation and ionization yields without tracking the ensuing radiative decays and recombinations.
Appendix \ref{validation_appendix} summarizes our extensive comparison against them, which covers conditions for a pure H gas, a pure He gas, and a H:He mixture in a 1:0.1 proportion (in number density). The fractional ionization in the calculations ranges from virtually zero to near full ionization.
Typically, each calculation includes a number of simulated photoelectrons of at least 10$^4$ 
that insures the accuracy of most quantities to the sub-percent level. 
\\

The agreement between our calculations and those by \citet{xumccray1991} and \citet{dalgarnoetal1999} is very satisfactory. 
For example we find that for $E_0$=1 keV 
the number of ions formed per  unit energy in neutral gases of H and He is 0.0268 and 0.0217 eV$^{-1}$, respectively. 
These quantities compare well with the 1/36.1=0.0277 and 1/48.5=0.0206 eV$^{-1}$ quoted by \citet{dalgarnoetal1999}. Also for a neutral gas and $E_0$=1 keV, our heating efficiency
for neutral gases of H and He atoms is 0.119 and 0.171, respectively. 
\citet{dalgarnoetal1999} report for them 0.11 and 0.16.
We also find that the trends with energy and fractional ionization are quantitatively and qualitatively consistent between models. Indeed, the structure in the excitation yield and heating efficiency that occurs in a gas of H atoms for low fractional ionization (Fig. \ref{xumccray1991_figures45} in Appendix \ref{validation_appendix}) mimics the structure seen in the calculations by \citet{shull1979} and \citet{xumccray1991}.  \citet{dalgarnoetal1999} smoothed out the resonances that occur in the cross sections at low energies, which prevents a more in-detail comparison between their findings and ours  at low energies and low fractional ionizations.
The discrepancies between our calculations and those by \citet{xumccray1991} and \citet{dalgarnoetal1999} are in line with what others have reported \citep{valdesferrara2008,furlanettostoever2010}. 
We consider validated our multi-score scheme.

\subsection{\label{convergencerate_subsec}Convergence} 

We explore the performance (convergence, accuracy, computational cost) of the multi-score scheme  and compare it to the performance of the single-score one.
We take for the exploration a gas of H atoms in the ground level plus a trace amount of the metastable H(2$s$), [H(2$s$)]/[H(1)]$=$10$^{-7}$. 
We identify the dominant channels:
\begin{eqnarray*}
    \rm{\textbf{a}}: e' + \mbox{H}(1) & \rightarrow & e'' + \mbox{H}(2p)\\
    \rm{\textbf{b}}: e' + \mbox{H}(1) & \rightarrow & e'' + e''' + \mbox{H}^+
\end{eqnarray*}
and the rare channels: 
\begin{eqnarray*}
    \rm{\textbf{c}}: e' + \mbox{H}{(2s)} & \rightarrow & e'' + \mbox{H}(1)\\
    \rm{\textbf{d}}: e' + \mbox{H}(2s) & \rightarrow & e'' + e''' + \mbox{H}^+.
\end{eqnarray*}
The latter channels involve superelastic and ionization collisions of H(2$s$), respectively. 
We explore the impact of fractional ionization by adopting two different values (10$^{-5}$ and 10$^{-2}$) for the relative abundance of thermal electrons, 
 $x_e$=[$e_{\rm{th}}$]/[H(1)]. 
Table \ref{convrate_table} summarizes the calculations obtained with both schemes for a range of energies $E_0$ from 30 eV to 1 keV, 
and a number of simulations $N_{\rm{simul}}$ from 10 to 10$^7$. The columns show the production yields $\Phi_{\bf{a}}$, $\Phi_{\bf{b}}$, $\Phi_{\bf{c}}$ and $\Phi_{\bf{d}}$, and the heating efficiency $\eta_h$.
\\

Table \ref{convrate_table} offers a few conclusions: 
\textit{i)} Both schemes converge to the same production yields and heating efficiencies when a sufficiently large number of simulations are carried out; 
\textit{ii)} Both schemes perform similarly in the estimation of the production yields for the dominant channels and the heating efficiency. This is not unexpected as those quantities are largely dictated by the dominant channels and both schemes track the photoelectrons in the cascade in identical ways; 
\textit{iii)} The multi-score scheme estimates the production yields for the rare channels to within 1--2 significant digits with about 100 simulations and sometimes with fewer. The single-score scheme however requires {$N_{\rm{simul}}$}{$\sim$}10/yield to estimate the specific yield to within 1--2 significant digits, and when 
{$N_{\rm{simul}}$}{$\lesssim$}1/yield the single-score scheme typically outputs a zero yield.
In other words, estimating the production yield for a rare channel to a similar accuracy involves orders-of-magnitude more simulations with the single-score scheme than with the multi-score scheme;  
\textit{iv)} In the multi-score scheme the convergence with {$N_{\rm{simul}}$} to the exact production yields or heating efficiency does not depend strongly on the fractional ionization; 
\textit{v)} In both schemes the convergence with {$N_{\rm{simul}}$} is notably faster at high energies. The reason for this is probably that energetic photoelectrons undergo many collisions and create many photoelectrons, and that these outcomes 
have a randomizing effect on the calculations that reduces their statistical errors.

\subsection{\label{yields_subsec}Production yields for channels that involve excited levels} 

\begin{figure}[h]
\centering
\includegraphics[width=8.9cm]{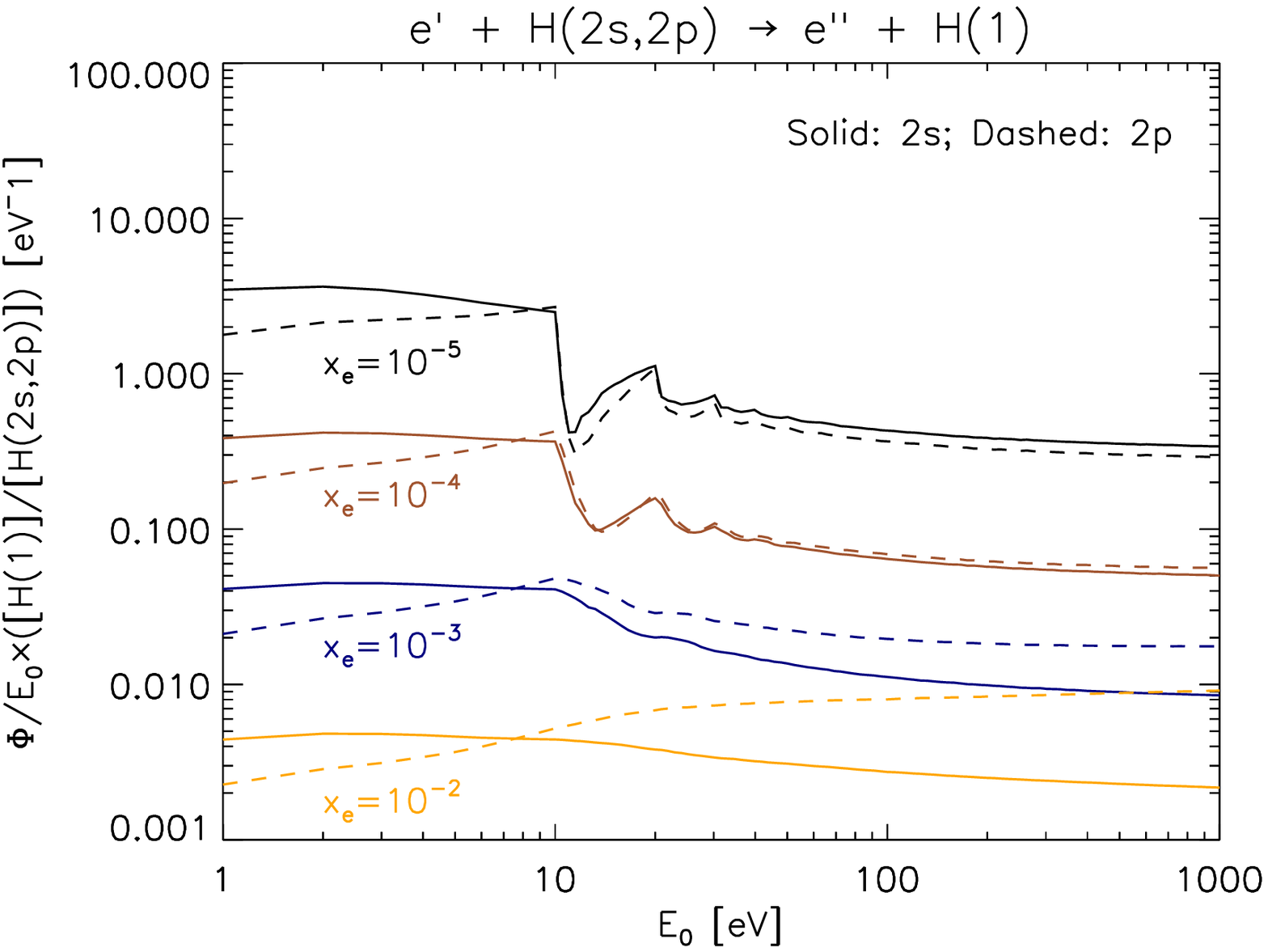} \\ 
\vspace{-0.0cm}
\includegraphics[width=8.9cm]{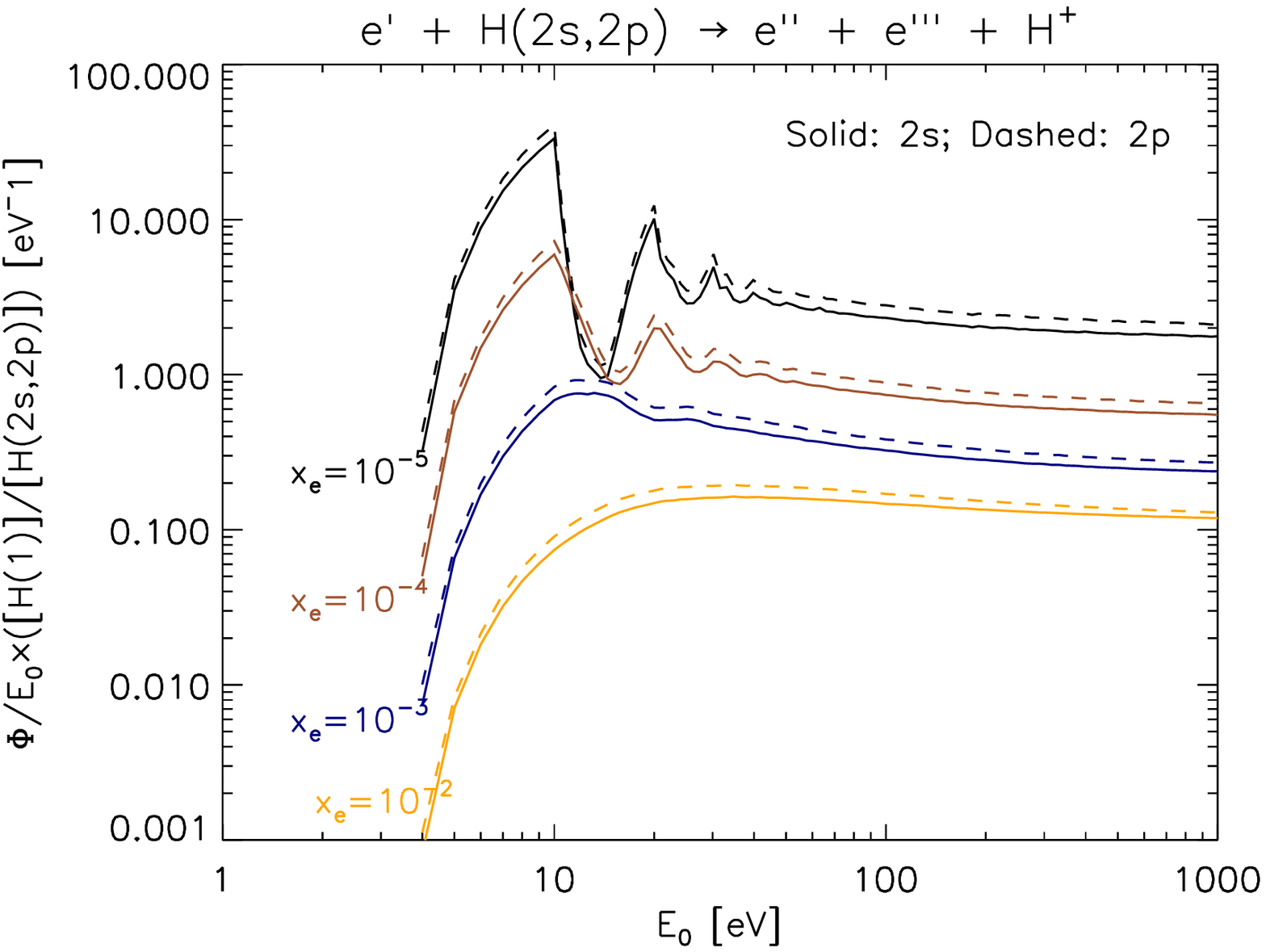} \\ 

    \caption{Doubly-normalized yields $\Psi_j$($E_0$)/$E_0$ for collisional deexcitation and ionization of H(2$s$) (solid lines) and H(2$p$) (dashed lines) as a function of the fractional ionization $x_e$ and injection energy $E_0$. The calculations are for a gas of hydrogen atoms. As an example, for $x_e$=10$^{-5}$ the number of electrons created from H(2$s$) when the gas is targeted by photoelectrons of energy $E_0$=1 keV amounts to $\sim$2$\times$1,000$\times$[H(2$s$)]/[H(1)].
    } 
    \label{normyields_fig}
\end{figure}

Production yields for the excitation and ionization of gases of H and He atoms in their ground levels have been published before \citep[][]{shull1979,shullvansteenberg1985,xumccray1991,dalgarnoetal1999,valdesferrara2008,furlanettostoever2010}.
Appendix \ref{validation_appendix} summarizes our calculations of some of those properties. 
We have not found published production yields for channels involving H(2$s$) or H(2$p$) as collision targets of the photoelectrons. 
Such calculations are needed to understand how photoelectrons might affect the population of these levels in atmospheres. 
Figure \ref{normyields_fig} extends the calculations in Table \ref{convrate_table} by including both H(2$s$) and H(2$p$) as collision targets.
The results are cast into doubly-normalized yields (one for each excited level), 
$\Psi_j$/$E_0$=$\Phi_j$($E_0$)/$E_0${$\times$}[H(1)]/[H(2$s$,2$p$)]. The doubly-normalized yields 
are nearly independent of the abundance ratios as long as [H(2$s$,2$p$)]/[H(1)]{$\ll$1}
(for a  discussion on this see {\S}\ref{small_abundance_sec}), and tend to asymptotic values at large energies $E_0$.
We are particularly interested
in the deexcitation channel \textbf{c} and the ionization channel \textbf{d} ({\S}\ref{convergencerate_subsec}), plus the similar channels for H(2$p$). 
We omit any channels for excitation of H(2$s$) or H(2$p$) into levels H($n${$\ge$}3), even though their cross sections may be large (Appendix \ref{appendix:xsections}; Fig. \ref{xs_fig2}). We justify this by noting that 
for moderate H(1) densities that cause the emission line connecting with the ground level to be optically thick, 
the H($n${$\ge$3}) will radiate back into H(2$s$,2$p$) and thus the net losses of H(2$s$) and H(2$p$) are practically zero.
\\

Inspection of Fig. \ref{normyields_fig} offers a few conclusions. 
As a destruction mechanism for H(2$s$) and H(2$p$), collisional ionization is more efficient than deexcitation (or quenching) at most injection energies and fractional ionizations. The exception to this occurs for $E_0${$<$}a few eV, when the injected photoelectron cannot overcome the 3.4 eV needed for ionization. 
Interestingly, comparing the ionization yields in Fig. \ref{normyields_fig}
against the ionization yields from the ground level (Appendix \ref{validation_appendix}; Figs. \ref{dalgarnoetal1999_fig6}, \ref{dalgarnoetal1999_h-em}) 
shows that ionization of the excited levels is more efficient 
(after correction by their relative abundances). 
For example, at low fractionizal ionizations, a 1 keV photoelectron 
(Appendix \ref{validation_appendix}; Fig. \ref{dalgarnoetal1999_fig6}) produces $\sim$27 new photoelectrons, whereas the equivalent metric for H(2$s$) or H(2$p$) (Fig. \ref{normyields_fig}, bottom) is $\sim$2,000.
There are three reasons for the enhanced efficiency in the ionization of the excited levels: firstly, the ionization threshold for
H(2$s$, 2$p$) is only 3.4 eV whereas that for the ground level is 13.6 eV; 
secondly, the ionization cross sections are typically a factor of a few larger for the excited levels 
(Appendix \ref{appendix:xsections}); 
thirdly, the photoelectron fluxes peak at the small energies that may ionize the excited levels but not the ground level
(see the discussion in {\S}\ref{flux_sec}, in particular Fig. \ref{fluxe_panel}).

\subsection{\label{meanfreecolumn_subsec} Relaxation properties} 

\begin{figure}[h]
\centering
\includegraphics[width=8.9cm]{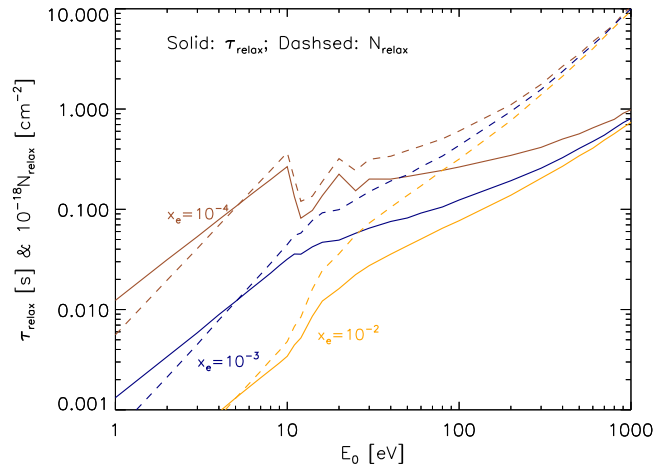}
\caption{Relaxation time and column in a gas of H atoms. The calculations adopted a reference [H(1)]=10$^{10}$ cm$^{-3}$. At other densities, 
the relaxation time can be estimated by multiplying the values in the graph by 10$^{10}$/[H(1)], with 
[H(1)] in cm$^{-3}$.
    } 
    \label{relax_fig}
\end{figure}

We have calculated the relaxation time $\tau_{\rm{relax}}$ and pathlength $L_{\rm{relax}}$
as defined in {\S}\ref{relaxdef_sec} for a gas of H atoms over a range of fractional ionizations. 
Both quantities scale linearly with the inverse of density, and it was adopted
[H(1)]=10$^{10}$ cm$^{-3}$. 
The solid lines in Fig. \ref{relax_fig} show $\tau_{\rm{relax}}$ for $E_{\rm{cutoff}}$=2$k_B${$T_e$}{$\sim$}0.17 eV ($T_e$=1,000 K).  
It is found that $\tau_{\rm{relax}}${$<$}1 s at the prescribed density and $E_0${$<$}1 keV. 
As expected, the photoelectrons relax more rapidly as the relative abundance of thermal electrons increases and electron-electron collisions become dominant. 
In the same figure, the dashed lines show the column of gas travelled by the photoelectrons $N_{\rm{relax}}$=[H(1)]$L_{\rm{relax}}$. 
$N_{\rm{relax}}${$<$}10$^{19}$ cm$^{-2}$ at 1 keV, and notably less than that at lower energies.
$N_{\rm{relax}}$ is a more meaningful metric than $L_{\rm{relax}}$ in non-uniform gases. 
In a real atmosphere, one should compare the
$N_{\rm{relax}}$ calculated over the relevant range of injection energies to the column of gas overhead to assess the validity of the local deposition approximation in the slowing down of the photoelectrons. 
\\

\subsection{\label{flux_sec} Photoelectron flux and energy distribution function}

The slowing down of monoenergetic photoelectrons is primarily dictated by the energy $E_0$ and the gas composition
(in particular, by the relative abundances of thermal electrons and heavy particles). 
Elucidating how the photoelectron flux $\phi_e$($E$) is affected by these factors provides valuable insight that can subsequently be generalized to planetary atmospheres, for which the nascent photoelectrons will have a broad range of energies
that reflect the broad range of wavelengths of the incident stellar photons.
This insight also provides a basis for interpretation of the production yields under different conditions.
\\

\begin{figure}[h]
    \centering
    \includegraphics[width=8.9cm]{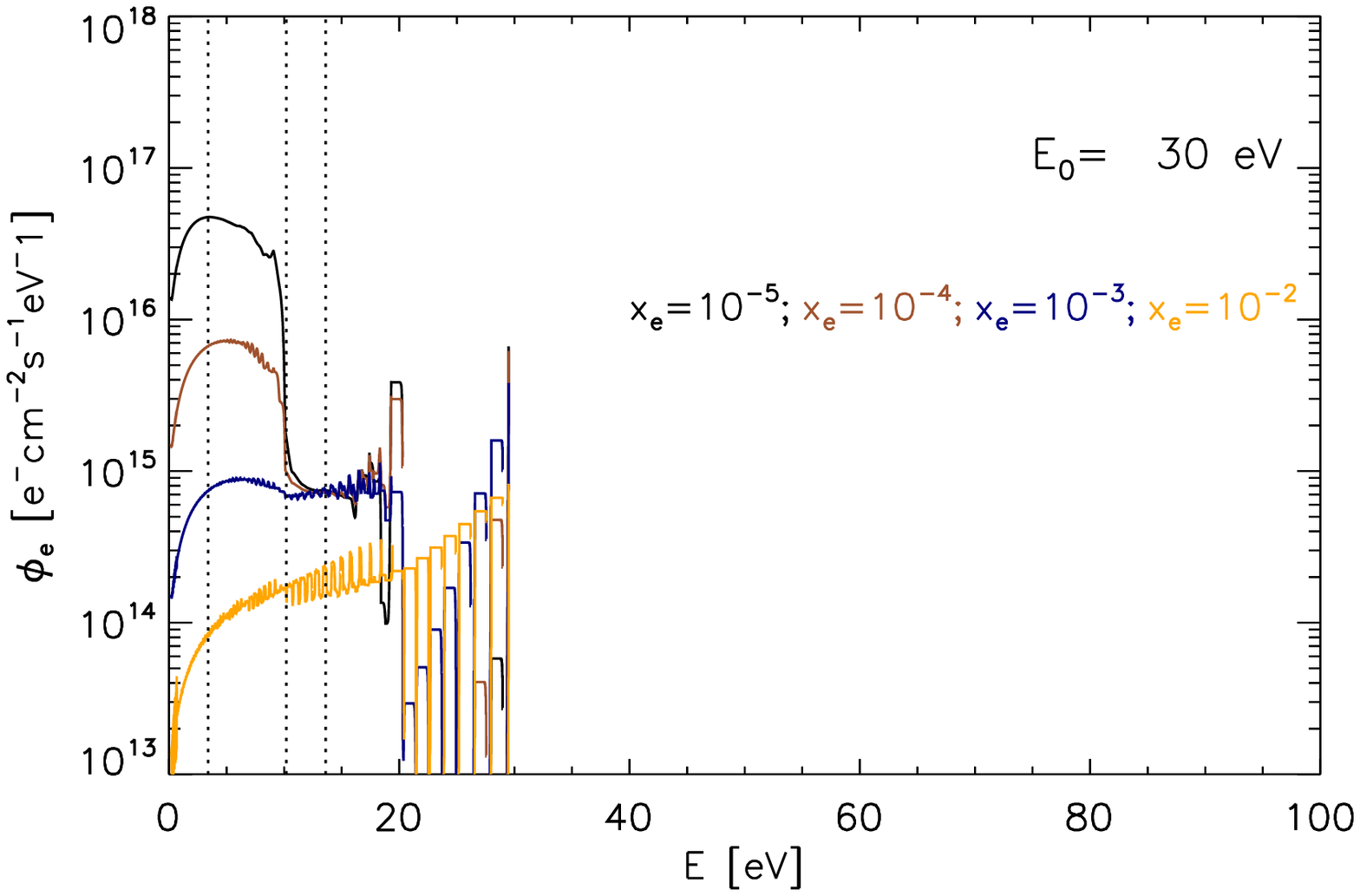} \\
    \vspace{-0.2cm}
    \includegraphics[width=8.9cm]{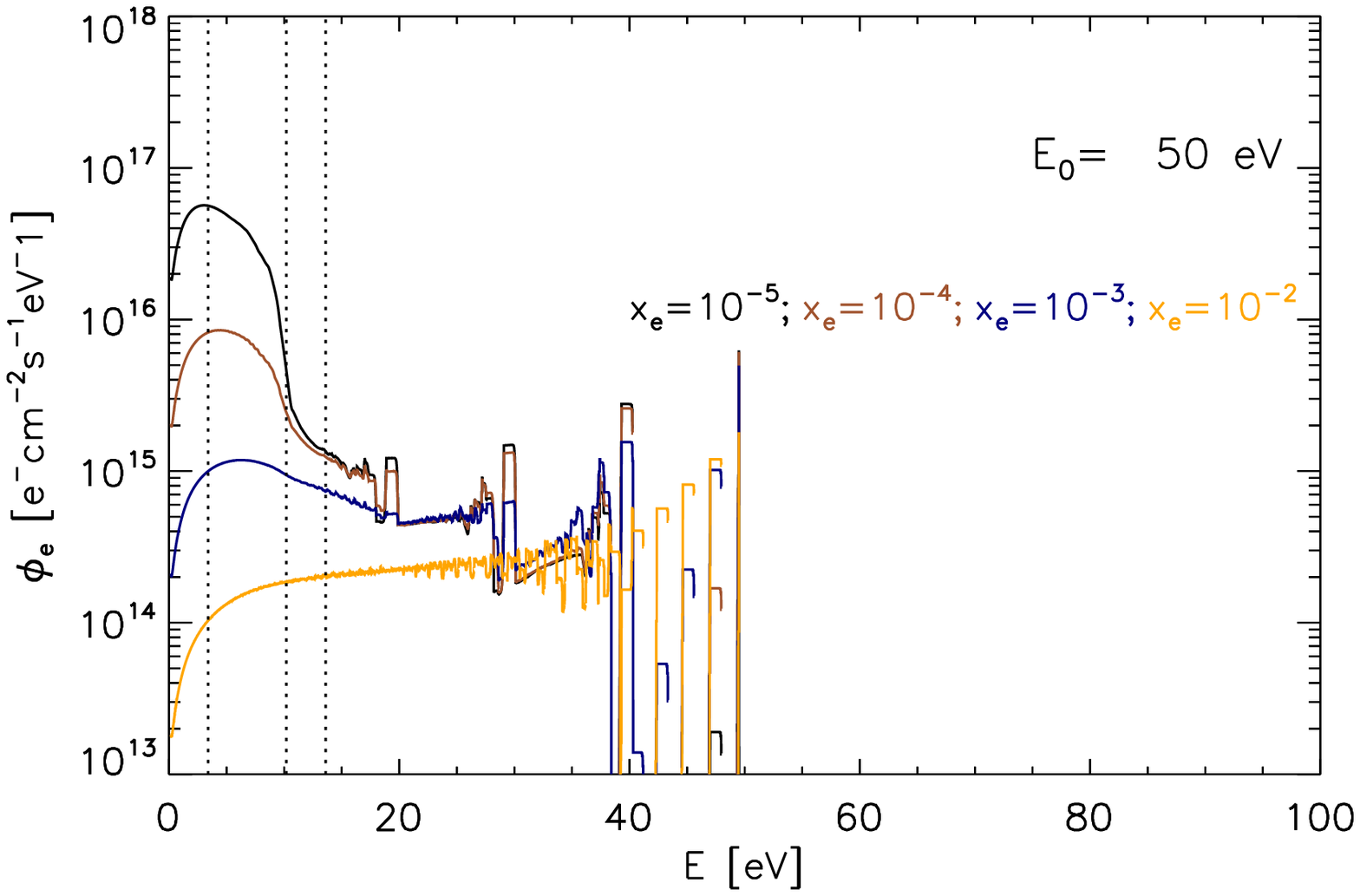} \\
    \vspace{-0.2cm}    
    \includegraphics[width=8.9cm]{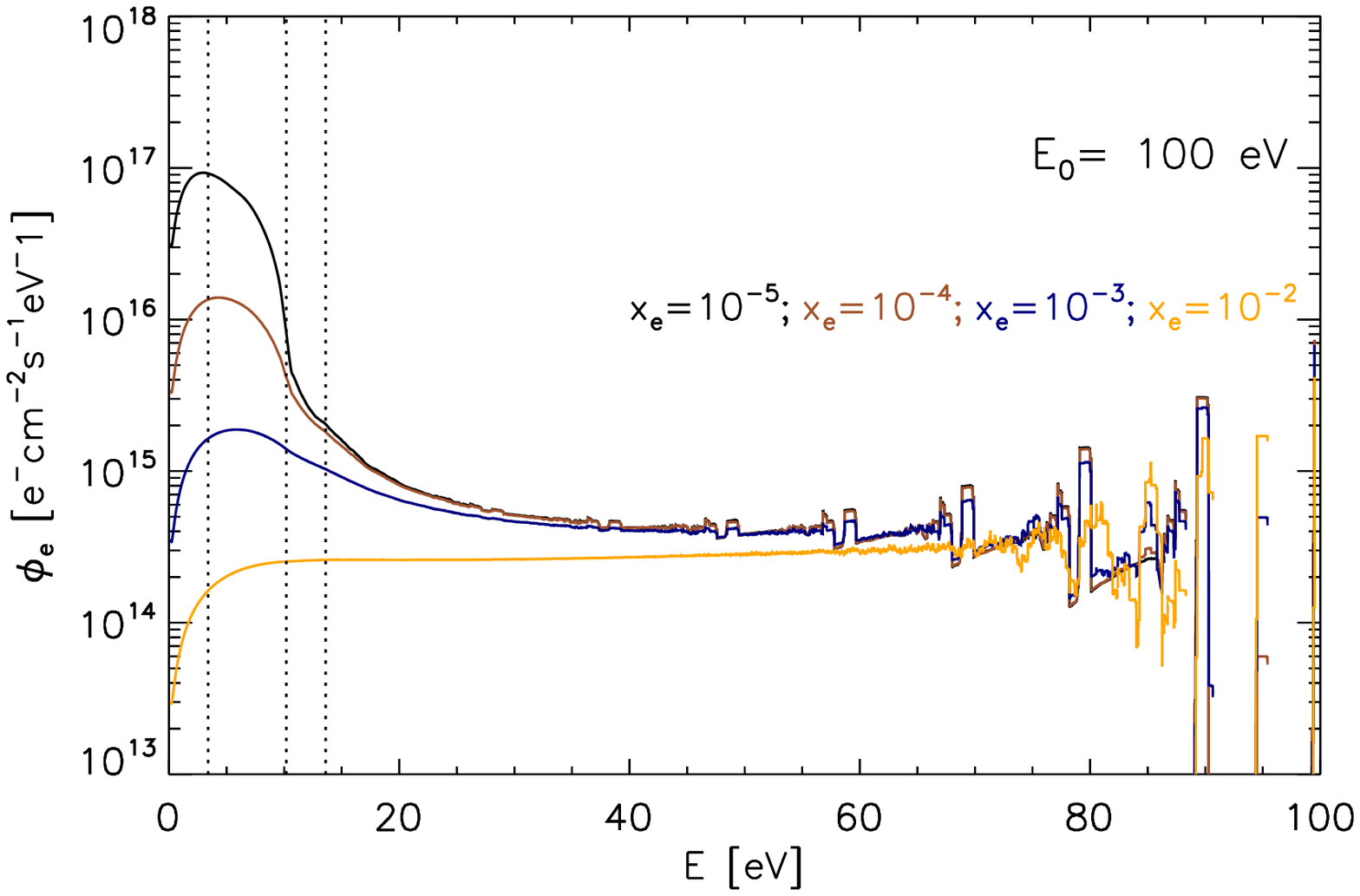} \\
    \vspace{-0.2cm}
    \includegraphics[width=8.9cm]{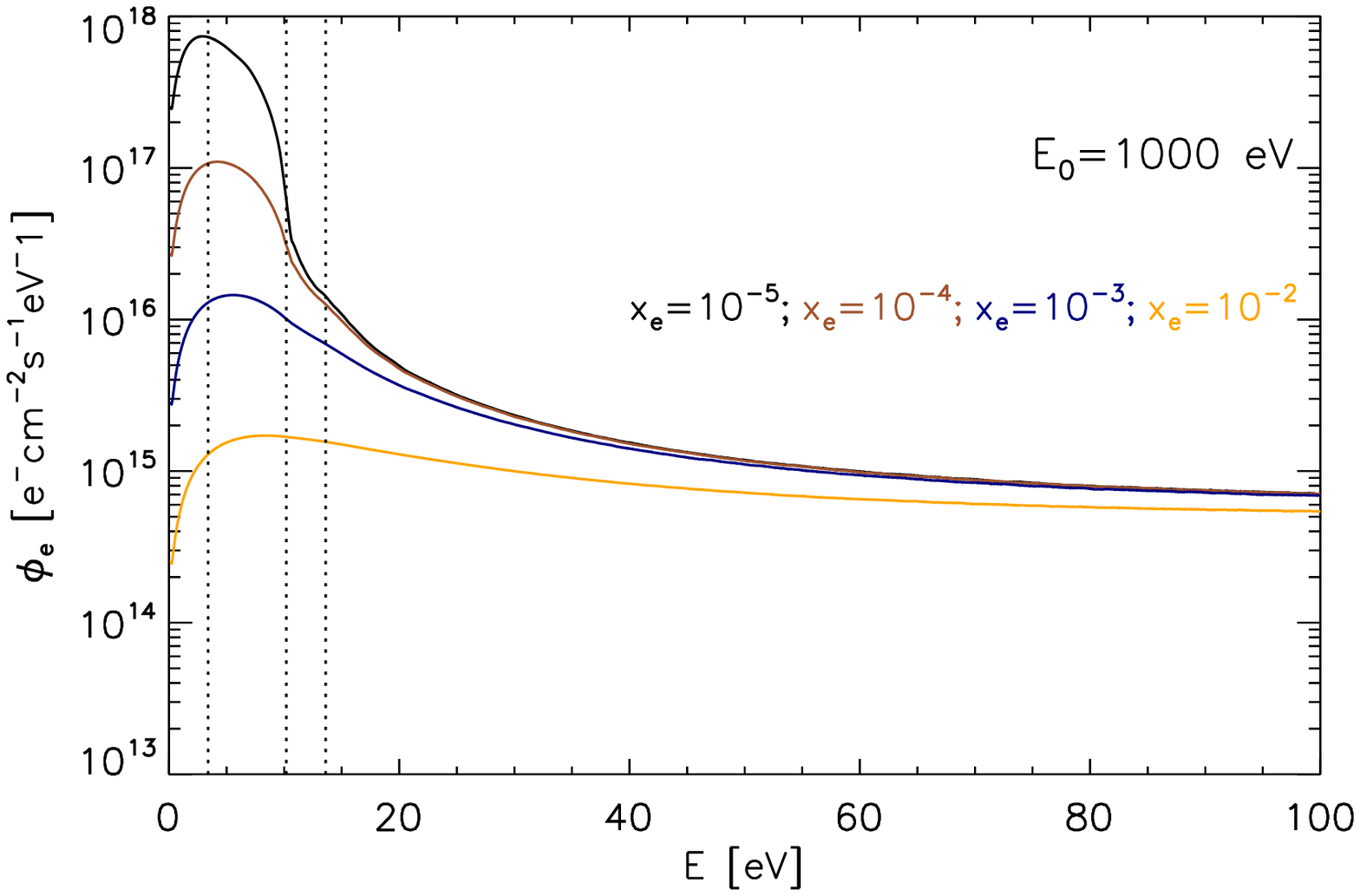} \\
    \caption{Photoelectron fluxes calculated for a gas of H atoms with the prescribed $x_e$ and $E_0$. 
    The dashed vertical lines mark the 
    3.4, 10.2 and 13.6 eV thresholds. 
    All plots have been smoothed with a boxcar average of 1 eV width to eliminate some of the structure that appears especially at high energies. The calculations were done with $W_0$=1 $e^-$cm$^{-3}$s$^{-1}$ and [H(1)]=1 cm$^{-3}$.
    } 
    \label{fluxe_panel}
\end{figure}

We have calculated $\phi_e$($E$; $E_0$, $x_e$) for a gas of H atoms, 
$x_e$=[$e_{\rm{th}}$]/[H(1)] between 10$^{-5}$ and 10$^{-2}$, and $E_0$ from 30 eV to 1 keV. 
Figure \ref{fluxe_panel} summarizes the calculations.
For guidance, the energies of 10.2 and 13.6 eV for first excitation and ionization of H(1), respectively, and 3.4 eV, for ionization of H(2) are marked. 
For any combination of $x_e$ and $E_0$, 
$\phi_e$($E$) shows some structure at sufficiently high energy. 
The structure is real and portrays the discrete nature of the slowing down at high energies. 
The structure smooths out at lower energies, after subsequent ionization collisions produce multiple photoelectrons with a broad range of energies.
\\

For a fixed $x_e$, 
higher fluxes occur for higher $E_0$, as the result of the larger number of newly created electrons in the cascade. 
For a fixed $E_0$, $\phi_e$($E$) may exhibit three differentiated behaviors. Firstly, at {$E$}$>$10.2 eV, which marks the threshold for the first inelastic channel in H(1), $\phi_e$($E$) is relatively flat with energy (unless $E${$\rightarrow$}{$E_0$} and some structure appears, as noted above). Secondly, 
crossing the threshold $E${$<$}10.2 eV, the lack of inelastic channels to slow down the photoelectrons may result in the build-up of a photoelectron population. This population is particularly abundant for the lower fractional ionizations. The reason is that 
as $x_e$ increases, Coulomb collisions quickly remove the low-energy photoelectrons from the cascade and their abundance drops. 
The threshold for ionization of H(2$s$) and H(2$p$) is at 3.4 eV. Because the cross sections for these channels are large between 3.4 and 10.2 eV 
(Appendix \ref{appendix:xsections}; Fig. \ref{xs_fig2}), the ionization rates of both levels by photoelectron collisions are particularly sensitive to $x_e$.
Thirdly, at $E$ less than a few eV Coulomb collisions dominate, efficiently slowing down the photoelectrons and causing a drop in $\phi_e$($E$) towards the thermal energies. 
This drop dominates over the rise in the deexcitation cross sections of H(2$s$) and H(2$p$), which partly explains why superelastic collisions remain of minor importance for the loss of these levels (Fig. \ref{normyields_fig}).\\

\section{Implementation in photochemical-hydrodynamical models} \label{sec:implementation}

Only some of the outputs from the MC model for the slowing down of photoelectrons may be required to set up a photochemical-hydrodynamical model calculation. 
The most likely quantities to be required are the production yields and heating efficiency. This section demonstrates how to implement them in a photochemical-hydrodynamical model for different formulations of mass and energy conservation.

\subsection{\label{subsec:production_rates} Production rates}

As defined here, the dimensionless production yield $\Phi_j$ quantifies the number of excited or ionized heavy particles produced through the collisional channel $j$ relative to the number of injected photoelectrons of a given energy $E_0$. In term of rates, if $W_0$($E_0$) [$e^-$cm$^{-3}$s$^{-1}$] is the number of injected photoelectrons per time and volume units then:
\begin{equation*}
    W_j(E_0)=W_0(E_0) \Phi_j(E_0) 
\end{equation*}
is the rate, same units as $W_0$($E_0$), of excited or ionized heavy particles produced in the cascade.
\\

In the atmosphere, $W_0$($E_0$) is dictated by the local photoionization rates and nascent photoelectron energies. 
Photoionization of a heavy particle $X$ by photons within a (narrow) interval of wavelengths
[$\lambda_{a}$, $\lambda_{b}$] ($\Delta${$\lambda_{ab}$}=$\lambda_{b}${$-$}$\lambda_{a}$, $\lambda_{ab}$={($\lambda_a$+$\lambda_b$)/2})
produces photoelectrons at the rate:
\begin{equation*}
W_{0,X}(E_0)\approx [X] \sigma^{\rm{pi}}_{X} (\lambda_{ab}) \mathcal{F}_{\lambda} (\lambda_{ab}) \frac{\lambda_{ab}}{hc} \Delta \lambda_{ab} 
\end{equation*}
of energies between $E_{0,b}$=$hc/\lambda_b${$-$}IP($X$) and $E_{0,a}$=$hc/\lambda_a${$-$}IP($X$). 
 $\sigma_X^{pi}$ stands for the photoionization cross section, and $\mathcal{F}_{\lambda}$ 
for the local irradiance (in energy units; ideally including diffuse and nondiffuse components). 
Photoionization typically occurs over wavelengths from X-rays to the near UV (NUV), and each interval [$\lambda_a$, $\lambda_b$] maps onto an energy interval [$E_{0,b}$, $E_{0,a}$]. 
Further, photoionization of different heavy particles $X$, $X'$, $X''$, etc., by photons within a given wavelength range produces photoelectrons of different energies at different rates. 
In summary, the production rate calculation 
must consider the summation over all heavy particles that undergo photoionization in the atmosphere and over all the energies with which the nascent photoelectrons are formed:
\begin{equation}
W_j=\sum_{k} W_{0,X}(E_{0,k}) \Phi_j(E_{0,k})
+ \sum_{k} W_{0,X'}(E'_{0,k}) \Phi_j(E'_{0,k})
+ \sum_{k} W_{0,X''}(E''_{0,k}) \Phi_j(E''_{0,k})
+ ...
\label{wj_eq}
\end{equation}
This expression and others introduced later use a summation over energy or wavelength bins because that is the way they are usually coded in models. If needed, the summations can be straightforwardly turned into integrals.
\\

\subsubsection{Application to an atmosphere of H atoms}

For demonstration, we specify the above to the situation in which photoionization of H atoms is the source of photoelectrons 
and the cascade develops through collisions with other H atoms and thermal electrons. 
This description is exact for an idealized atmosphere that contains only H atoms. 
If the atmosphere contains additional heavy particles, however, a proper calculation should track the photoelectrons produced through photoionization of each heavy particle. Indeed, in the case of an atmosphere composed of H and He atoms in solar proportions,  
because the photoionization cross sections of He are much larger than those of H shortwards of 504 {\AA}, He photoionization may dominate the production of photoelectrons at {\hi} columns $\gtrsim$10$^{18}$ cm$^{-2}$ for which the stellar radiation at longer wavelengths has been sufficiently attenuated.\\ 

For an atmosphere of H atoms, the net production rate of electrons that results from local photoionization of H(1) plus photoelectron-driven collisional ionization is:
$$
\frac{\delta [{\rm{H}}(1)]}{\delta t}=
{-}\frac{\delta [{\rm{e_{th}}}]}{\delta t}=
{-}\frac{\delta [{\rm{H}}^+]}{\delta t}=
{-}{[\rm{H}(1)]} \sum_k {\sigma^{\rm{pi}}_{\rm{H(1)}}} (\lambda_{k}) 
\mathcal{F}_{\lambda} (\lambda_{k}) 
{\Delta \lambda_k}
{\frac{\lambda_k}{hc}} (1+\Phi_{\rm{H(1)}\rightarrow \rm{H}^+}(E_{0,k})),
$$
where $E_{0,k}$=$hc$/$\lambda_k${$-$}13.6 eV. 
The ratio of electrons produced in the photoelectron cascade and from photoionization in a neutral H gas  
amounts to $\Phi_{\rm{H(1)}\rightarrow \rm{H}^+}${$\sim$}2.5 and 27 (Table \ref{convrate_table}) for energies of 100 and 1000 eV that are produced by photons of wavelengths $\sim$124 and 12 {\AA}, respectively.
\\

Tracking the population of excited levels H($n${$\ge$}2) may not be required in each photochemical-hydrodynamical model. 
This is true in particular when the interest lies only in determining the impact of such levels on the energy budget. 
In that case, and provided that it can be assumed that the excited levels will ultimately radiate away their excitation energy, the model may only require the heating efficiency. 
Indeed, the heating efficiency introduced above includes the relevant information about the energy that is not radiated away and 
it can be calculated without explicitly solving for the excited level abundances. 
Calculating the abundances may however be required when they are useful for remotely sensing the atmosphere
(as in {\S}\ref{sec:halpha}) or when they do not radiate away all their energy. 
The latter occurs for example at some ultra-hot Jupiters  \citep{garciamunozschneider2019}, for which NUV photoionization of H(2)
is a main source of H$^+$ at pressure levels deep enough that most of the XUV stellar radiation has been deposited above.
\\

If tracking the population of excited levels is required, it can be done as described below for H(2$s$). 
The production  rate for this sublevel from excitation collisions of photoelectrons with H(1) is:
$$
\frac{\delta [{\rm{H}}(2s)] }{\delta t}=
{[\rm{H}(1)]} \sum_k {\sigma^{\rm{pi}}_{\rm{H(1)}}} (\lambda_{k}) 
\mathcal{F}_{\lambda} (\lambda_{k}) 
{\Delta \lambda_k}
{\frac{\lambda_k}{hc}} 
\Phi_{\rm{H(1)}\rightarrow \rm{H}(2s)}(E_{0,k}). 
$$
One could conceivably add to this term a contribution from excitation
into H(2$p$) and subsequent thermalization between sublevels, but it is easier to deal with that separately. 
One could also add a contribution from excitation into H($n${$\ge$}3), some of which will radiate into H(2$s$), 
but again it is easier to deal with that on a case-by-case basis.
\\

The loss of H(2$s$) from ionization collisions with photoelectrons is given by:
\begin{equation}
\frac{\delta [{\rm{H}}(2s)] }{\delta t}=-
[{\rm{H}}(1)]  \sum_k  {\sigma^{\rm{pi}}_{\rm{H(1)}}} (\lambda_{k}) 
\mathcal{F}_{\lambda} (\lambda_{k}) 
{\Delta \lambda_k}
{\frac{\lambda_k}{hc}} 
\Phi_{\rm{H}(2s) \rightarrow
\rm{H}^+}(E_{0,k}),
\label{h2s_lossionization_eq}
\end{equation}
and from superelastic collisions: 
\begin{equation}
\frac{\delta [{\rm{H}}(2s)] }{\delta t}=-
[{\rm{H}}(1)]  \sum_k  {\sigma^{\rm{pi}}_{\rm{H(1)}}} (\lambda_{k}) 
\mathcal{F}_{\lambda} (\lambda_{k}) 
{\Delta \lambda_k}
{\frac{\lambda_k}{hc}} 
\Phi_{\rm{H}(2s) \rightarrow
\rm{H(1)}}(E_{0,k}).
\label{h2s_lossquenching_eq}
\end{equation}
Following the above reasoning, the loss of 
H(2$s$) in collisions with photoelectrons may also proceed through the production of H(2$p$) and 
H($n${$\ge$}3), and additional loss terms should be considered. We omit them here because such contributions are often negligible when compared to collisions with the thermal population of electrons or protons. It is also easier to deal with the specifics of such contributions on a case-by-case basis. 
\\

\subsubsection{\label{small_abundance_sec}The small abundance approximation}

Continuing with the above example, the MC model determines the production yields 
$\Phi_{\rm{H}(2s) \rightarrow \rm{H}^+}(E_{0})$ and $\Phi_{\rm{H}(2s) \rightarrow
\rm{H(1)}}(E_{0})$ by prescribing a number density for H(2$s$) (see Table \ref{convrate_table}).
Because all the cross sections that enter the formulation are independent of the absolute density (except the electron-electron cross section, although this dependence is very weak), these yields depend approximately on the ratio of number densities [H(2$s$)]/[H(1)] rather than on [H(2$s$)] or [H(1)] separately. 
Further, if [H(2$s$)]/[H(1)]{$\ll$1} the production yields scale linearly with this ratio, and we can define the abundance-normalized yields $\Psi$($E_0$) by
$\Phi_{\rm{H}(2s) \rightarrow \rm{H}^+}(E_{0})${$\approx$}$\Psi_{\rm{H}(2s) \rightarrow \rm{H}^+}(E_{0})$[H(2$s$)]/[H(1)] and
$\Phi_{\rm{H}(2s) \rightarrow \rm{H(1)}}(E_{0})${$\approx$}$\Psi_{\rm{H}(2s) \rightarrow \rm{H(1)}}(E_{0})$[H(2$s$)]/[H(1)].
This approximation works well as long as the H(2$s$) number density is too small to affect the cascade of photoelectrons, 
the fate of which is dominated by the more abundant components of the gas such as H(1) and the thermal electrons. 
In this approximation, Eqs. \ref{h2s_lossionization_eq}-\ref{h2s_lossquenching_eq} can be rewritten as: 
\begin{equation}
\frac{\delta [{\rm{H}}(2s)] }{\delta t}=-
[{\rm{H}}(2s)]  \sum_k  {\sigma^{\rm{pi}}_{\rm{H(1)}}} (\lambda_{k}) 
\mathcal{F}_{\lambda} (\lambda_{k}) 
{\Delta \lambda_k}
{\frac{\lambda_k}{hc}} 
\Psi_{\rm{H}(2s) \rightarrow
\rm{H}^+}(E_{0,k}),
\label{h2s_lossionization_minor_eq}
\end{equation}
and
\begin{equation}
\frac{\delta [{\rm{H}}(2s)] }{\delta t}=-
[{\rm{H}}(2s)]  \sum_k  {\sigma^{\rm{pi}}_{\rm{H(1)}}} (\lambda_{k}) 
\mathcal{F}_{\lambda} (\lambda_{k}) 
{\Delta \lambda_k}
{\frac{\lambda_k}{hc}} 
\Psi_{\rm{H}(2s) \rightarrow
\rm{H(1)}}(E_{0,k}), 
\label{h2s_lossquenching_minor_eq}
\end{equation}
which are independent of [H(1)] even if H(1) photoionization is the original source of photoelectrons. 
We have tested with some of the calculations for $\Phi_{\bf{c}}$ and $\Phi_{\bf{d}}$ shown in Table \ref{convrate_table}, and confirmed that the linear scaling between the yields $\Phi$ and $\Psi$ remains accurate to within a few {\%} for [H(2$s$)]/[H(1)] ratios as high as 10$^{-3}$.
The same simplifications will hold valid for other heavy particles in the gas 
as long as they are present in small abundances.

\subsection{Energy balance}

All the transformations that occur from collisional and radiative (de-)excitations and chemical conversions have a bearing on the energy balance. There are two main ways to incorporate such transformations into the energy conservation equation of the gas.\\ 

The first
`photon-based' treatment relies on tracking the photon interactions and adding to the `right hand side (RHS)' of the energy equation the integral of the radiance gradient over solid angle and wavelength $\Gamma$=$\int 
\textbf{s} \cdot \nabla I_{\lambda}(\textbf{s})  d\Omega(\textbf{s}) d\lambda$, 
where \textbf{s} is a direction vector. In this treatment, the total energy of the gas on the `left hand side (LHS)' of the energy equation must include the excitation and ionization energies of the particles.
A main advantage of this treatment is that the solution of the radiative transfer problem directly provides the net heating/cooling rate $\Gamma$. The contribution of the photoelectrons to the energy budget goes into the rates that (de)populate the different heavy particles that participate in the radiative transfer problem. 
\\

The second approach relies on tracking the electron interactions. In this `electron-based' treatment, the `RHS' of the energy equation contains the net energy balance in all the transformations after subtracting the changes in the excitation and ionization energies from $\Gamma$, and the total energy on the `LHS' contains only the contribution from the translational modes to the internal energy of the gas.\\

Both treatments should be equivalent if they are properly formulated. 
Although the `photon-based' treatment is probably easier to set up in media that transition from optically thin to thick conditions, the `electron-based' treatment is more popular in models of planetary atmospheres. 
For that reason, in what follows we elaborate on the contribution of photoelectrons to the energy equation in the `electron-based' treatment.\\

Recalling the expression for the production rate of photoelectrons from photoionization of heavy particles $X$, the energy that is carried away by them is simply 
$W_{0,X}$($E_0$) times $E_0$ ({$\equiv$}{$hc$}/$\lambda${$-$}IP($X$)). 
A fraction of this energy $\eta_h$ is transferred to the gas in elastic collisions and is certain to heat it up. The remaining 1{$-$}{$\eta_h$} fraction goes into the excitation and ionization of the heavy particles. 
The fate of the corresponding excitation and ionization energies, and whether it is returned as heat back to the gas, will depend on the competition between radiative channels and collisional deexcitation (or chemical conversion). A critical factor in that competition is the density of the background gas. Clearly, the amount of excitation and ionization energy that is returned into heat will be larger at higher densities. Properly quantifying that transfer of energy can be complicated and is not attempted here. 
If however the densities are relatively low and none of the energy stored as excitation or ionization is converted into translational energy, the rate at which heat is deposited in the gas
$\mathcal{Q}$ [erg cm$^{-3}$s$^{-1}$] is:
\begin{eqnarray*}
\begin{aligned}
\mathcal{Q}= & \sum_{k} W_{0,\rm{X}}(E_{0,k}) \left(\frac{hc}{\lambda_k} - \rm{IP}(X) \right)  \eta_h(E_{0,k}) + \\
& \sum_{k} W_{0,\rm{X}'}(E'_{0,k}) \left(\frac{hc}{\lambda_k} - \rm{IP}(X') \right)  \eta_h(E'_{0,k}) + \\ 
& \sum_{k} W_{0,\rm{X}''}(E''_{0,k}) \left(\frac{hc}{\lambda_k} - \rm{IP}(X'') \right)  \eta_h(E''_{0,k})
+ ...
\end{aligned}
\end{eqnarray*}
The conditions under which this expression is valid are often met for the low densities found in upper atmospheres. 
\\

\underline{Application to an atmosphere of H atoms}. The heating rate takes on a especially simple form: 
\begin{equation*}
\mathcal{Q}=[{\rm{H}}(1)]
\sum_k  {\sigma^{\rm{pi}}_{\rm{H(1)}}} (\lambda_{k}) 
\mathcal{F}_{\lambda} (\lambda_{k}) {\Delta \lambda_k} \left(1-\frac{\lambda_k}{\lambda_0} \right) \eta_h (E_{0,k}).
\end{equation*}
The parenthesis is related to the energy left after removal of the H(1) ionization potential from the incident photon energy. 
That is indeed the excess energy imparted to the nascent photoelectrons. A fraction of it determined by $\eta_h (E_{0})$ and 
that depends on the photoelectron energy 
goes into heating, while the rest goes into excitation and ionization. 
For low densities of the background gas, it can reasonably be assumed that only the $\eta_h (E_{0})$ heats the gas.\\

\section{\label{sec:halpha} An application: The H-$\alpha$ line in the atmosphere of HAT-P-32b}

We investigate the role of photoelectrons in the formation of the {\halpha} line identified in the transmission spectrum of HAT-P-32b \citep{czeslaetal2022}. 
This ultra-hot Jupiter orbits at 0.034 AU an active F-type of XUV output $\sim$100$\times$ solar. 
The fraction of this output that occurs at X-ray wavelengths is surely variable, but \citet{czeslaetal2022} estimate that it is $\sim$20{\%}. Given the large amount of XUV photons received by the planet, HAT-P-32b represents an interesting target to explore the interaction of photoelectrons with its atmosphere.
The transmission spectrum reveals the transition $\mbox{H(2)}$+$h\nu$ (6565 {\AA}){$ \rightarrow$}H(3) occurring at high altitude and thus that the planet is surrounded by an extended envelope of excited hydrogen. 
\citet{czeslaetal2022} presented some modeling of the hydrogen atmosphere of HAT-P-32b using the NLTE-hydrodynamical model by \citet{garciamunozschneider2019}. 
The question we ask ourselves now is whether including the effect of photoelectrons might modify the H(2) populations presented in \citet{czeslaetal2022}. 
To answer it, we review the processes for H(2) loss and production in the published model, and analyze how they might be affected by photoelectrons. 
We leave for future work the self-consistent implementation of photoelectrons in the NLTE-hydrodynamical model, 
which will be updated in additional ways with respect to the version used in \citet{czeslaetal2022}, 
and keep the current exercise as a first assessment.
The original model treats the sublevels H(2$s$) and H(2$p$) separately. 
Such a complexity is unnecessary here, and we work instead with properties averaged over the H(2) population.
For example, the total loss rate for H(2)$\rightarrow$H(1)+$h\nu$ (1216 {\AA}) 
is calculated as the summation of the loss rates from the allowed transition 
H(2$p$){$\rightarrow$}H(1)+$h\nu$ and the two-photon emission H(2$s$)$\rightarrow$H(1)+$h\nu$, and the corresponding lifetime is calculated as the ratio of the total loss rate by the H(2) number density.\\

We calculate the rates for photoelectron production integrated over all energies $W_{0,T}$  and the energy spectra  $dW_0/dE_0$ describing the photoelectron production rate per unit energy.\footnote{$W_{0,T}$ [$e^-$cm$^{-3}$s$^{-1}$]=${\int_0^{\infty}}$ {$dW_0/dE_0$}{$dE_0$}. Conversion from
$dW_0/dE_0$ [$e^-$cm$^{-3}$s$^{-1}$eV$^{-1}$] to the bin-integrated $W_0$($E_{0,k}$) 
[$e^-$cm$^{-3}$s$^{-1}$] introduced in {\S}\ref{subsec:production_rates} is done in practice by multiplying the former by the energy bin size $\Delta${$E_{0,k}$}, which is a model-dependent quantity related to the wavelength grid at which the radiative transfer problem is solved.} 
Figure \ref{W0T_photoe_fig} (top) shows the atmospheric profiles of [H], [H$^+$], [H(2)] and temperature published in \citet{czeslaetal2022}, and Fig. \ref{W0T_photoe_fig} (bottom) the contribution of H(1) and H(2) photoionization to $W_{0,T}$ over the same range of pressures. 
Photoionization of H($n${$\ge$}3) makes a negligible contribution and is not discussed. 
The H(1) contribution clearly dominates over that by H(2), except in a narrow region near (2--7)$\times$10$^{-2}$ dyn cm$^{-2}$ where they are comparable. 
Figure \ref{dW0dE_fig} shows the $dW_0/dE_0$ spectra at a few selected altitudes. 
At the higher pressures, H(1) photoionization results in $dW_0/dE_0$ spectra that exhibit a distinct high-energy peak near $E_0${$\sim$}100 eV. 
The peak occurs because the star emits strongly at X-ray wavelengths, and given that the photoionization cross sections behave as {$\propto$}$\lambda^3$ the high-energy photons become deposited relatively deep in the atmosphere.
The high-energy peak disappears at lower pressures, and the $dW_0/dE_0$ spectrum adopts the soft shape expected when most of the photoelectrons are produced by photons with wavelengths slightly shorter than the Lyman continuum threshold at 912 {\AA}. 
The $dW_0/dE_0$ spectrum that arises from H(2) photoionization is also soft.  
The reason is that H(2) photoionization is driven mainly by photons near the Balmer continuum threshold at 3646 {\AA}, where both the star emits strongly and the H(2) photoionization cross sections are large, 
a combination that favors the production of low-energy photoelectrons. 
The $dW_0/dE_0$ spectra that arise from H(2) photoionization show a distinct peak near 6.8 eV due to photoionization by {\lalpha} radiation.
Most of these {\lalpha} photons are from internally-generated diffuse radiation rather than from direct stellar radiation.\\

\begin{figure}[h]
\centering
\includegraphics[width=9cm]{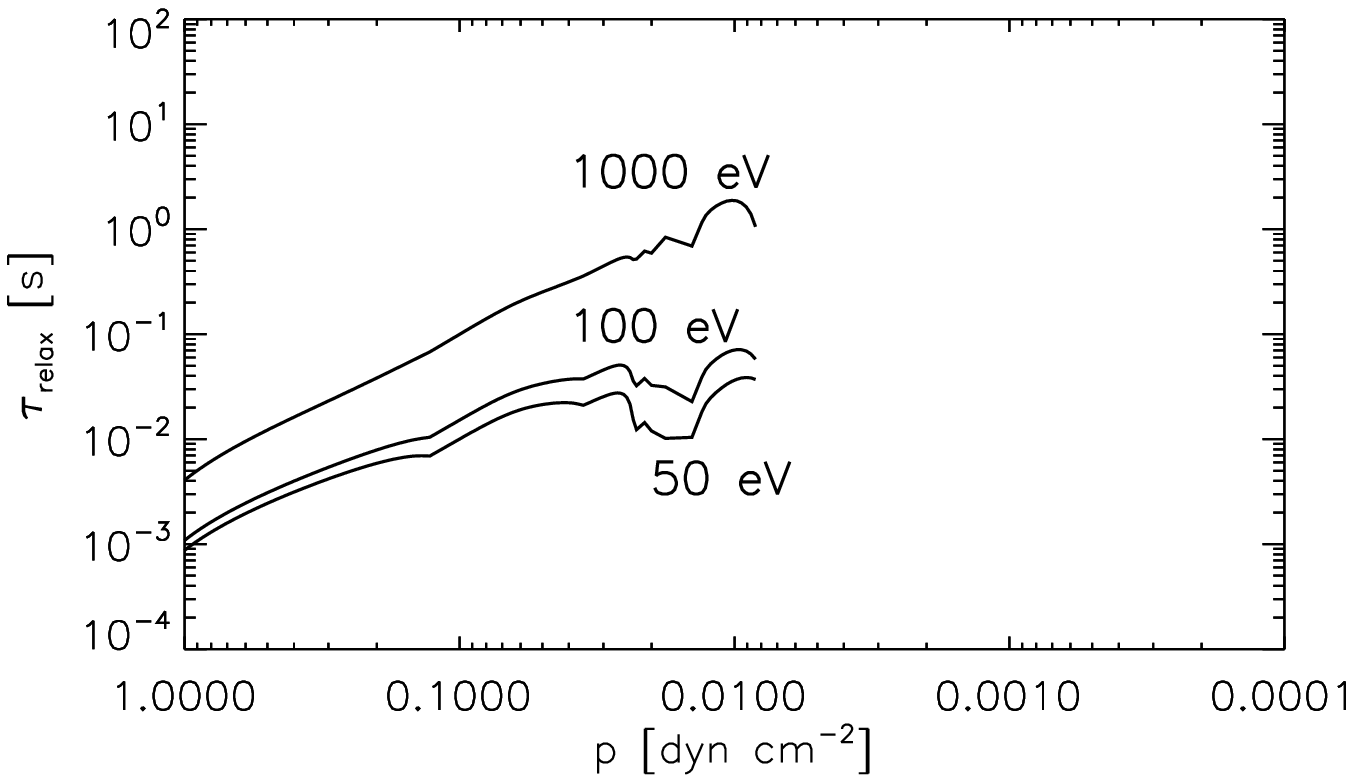} \\
\includegraphics[width=9cm]{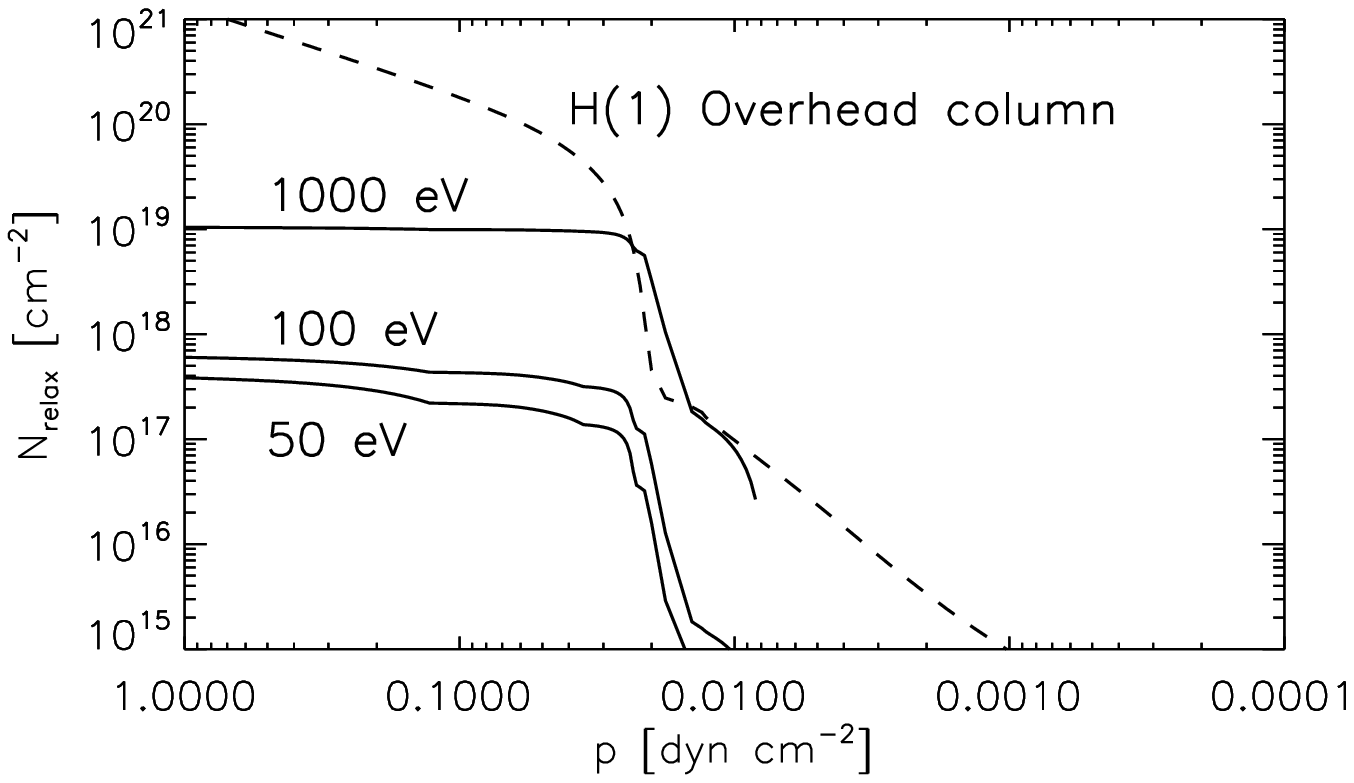} 
\caption{Top. Relaxation time in the atmosphere of HAT-P-32b for photoelectrons of energy 50 eV, 100 eV and 1 keV.
Bottom. Corresponding relaxation columns of H(1) (solid). Also shown, the overhead column of H(1) at each altitude (dashed).
} 
\label{relax_profiles_fig}
\end{figure}

Figure \ref{relax_profiles_fig} (top) shows the relaxation time calculated as described in {\S}\ref{relaxdef_sec} for photoelectrons of energies $E_0$=50, 100 and 10$^3$ eV. We have truncated the 
 profiles to altitudes for which $x_{\rm{e}}$=[$e_{\rm{th}}$]/[H(1)]{$<$}10$^2$ to emphasize the region where the {\halpha} line is formed. 
The calculations show that $\tau_{\rm{relax}}$ increases with altitude as the H(1) density drops. 
The photoelectrons relax in a fraction of a second, in particular those with $E_0${$<$}100 eV.
Similarly, Fig.  \ref{relax_profiles_fig} (bottom) shows the relaxation column for the same injection energies. For comparison, it is also shown the overhead column of H(1). The photoelectrons with energies $E_0${$\le$}100 eV relax at all altitudes over a H(1) column that is much less than the overhead column, which justifies the local deposition approximation. For photoelectrons with energies 1 keV, both columns are comparable at the peak of the H(2) density, and the local deposition approximation might be less justifiable there. We note however that the calculated $N_{\rm{relax}}$ represent  upper limits on the gas column required to thermalize the photoelectrons, and that at high altitudes the production of very energetic photoelectrons is minor and therefore will have no bearing on the overall state of the gas.
\\

\subsection{Loss of H(2)}

\begin{figure}[h]
\centering
\includegraphics[width=8.9cm]{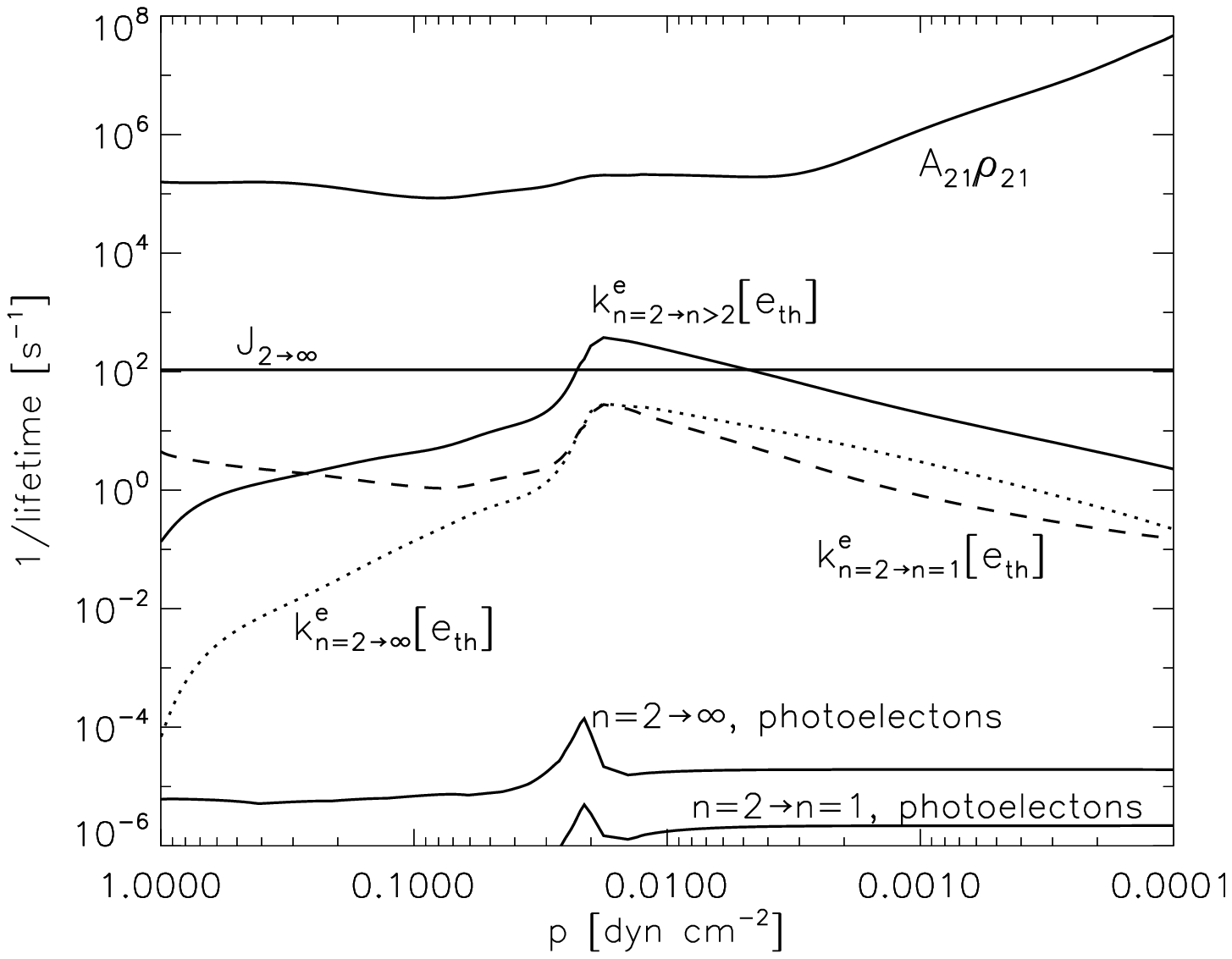} \\ 
\includegraphics[width=8.9cm]{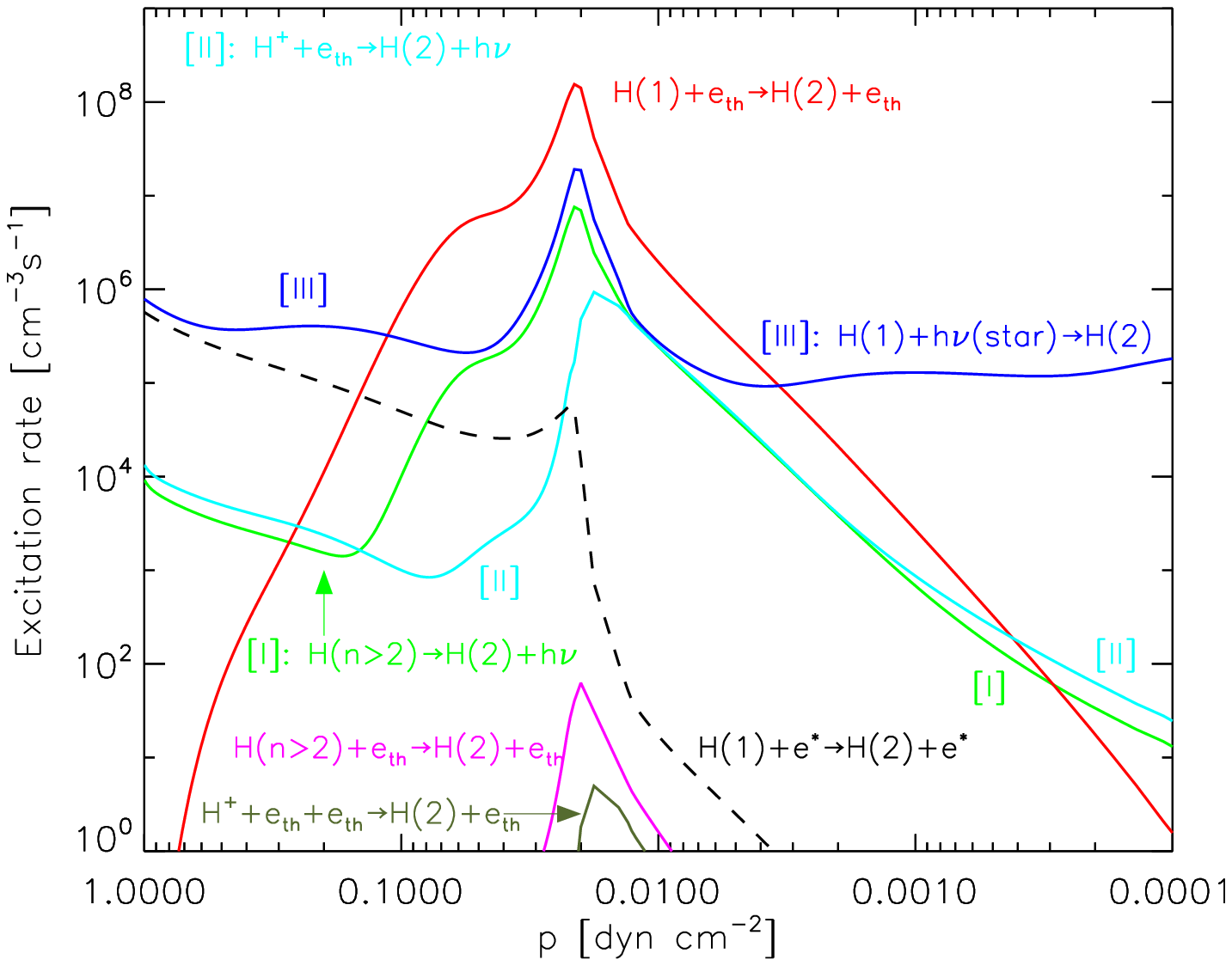} \\ 
\includegraphics[width=8.9cm]{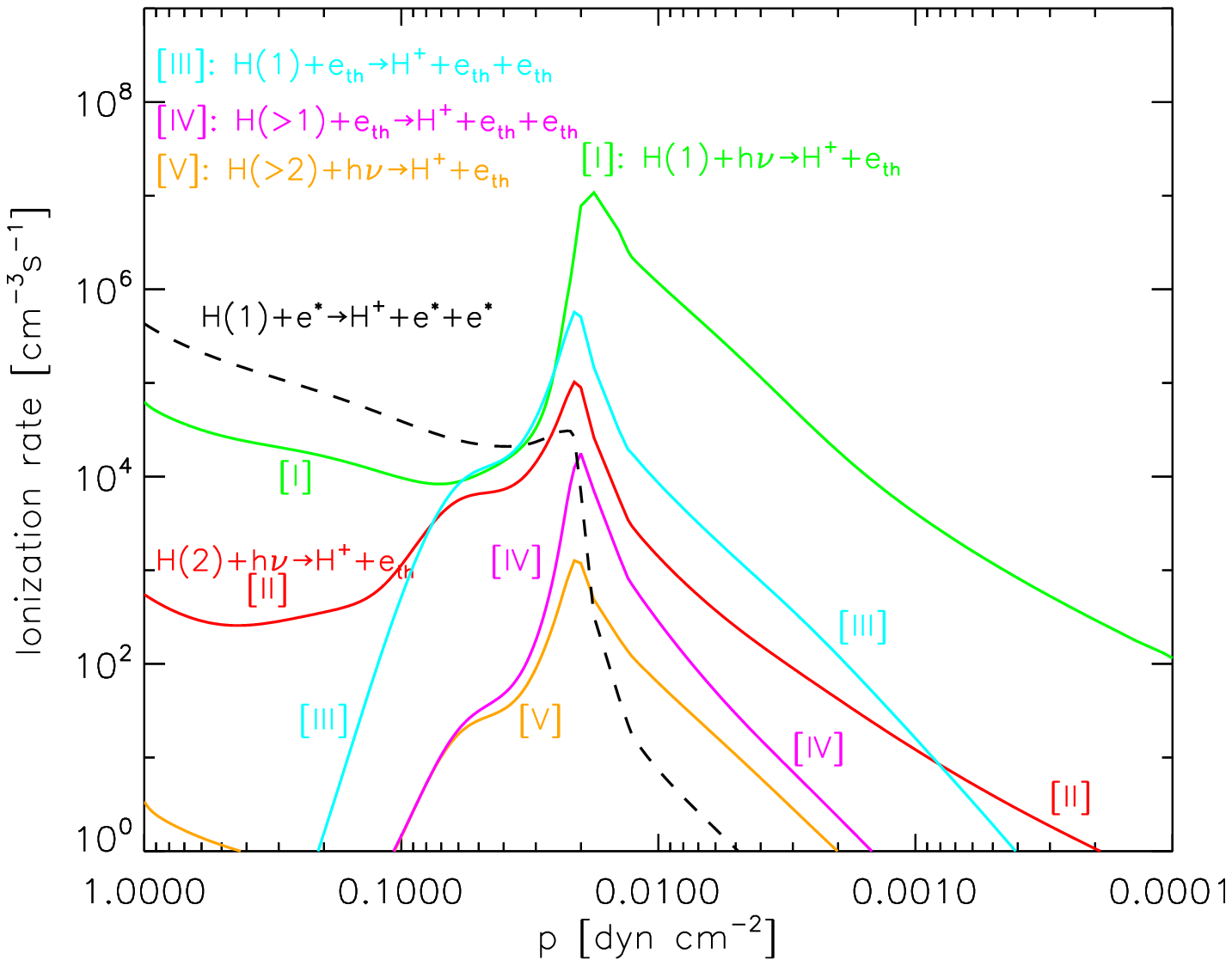} 
\caption{Top. Inverse of lifetimes for processes that result in H(2) losses. 
Middle. Rates for processes that result in the production of H(2). Bottom. Rates for processes that result in the production of H$^+$. In the middle-bottom panels, the labels use $e_{\rm{th}}$ and $e^*$ to differentiate between thermal electrons and photoelectrons. 
} 
\label{h02_loss_gain_lifetime_photoe_fig}
\end{figure}

Figure \ref{h02_loss_gain_lifetime_photoe_fig} (top) shows the H(2) lifetimes inferred from the NLTE-hydrodynamical model.
The loss mechanisms include: radiative decay to the ground level with emission of a {\lalpha} photon and photoionization; collisions with thermal electrons leading to ionization, deexcitation and excitation to 
H($n${$\ge$}3) (the latter is not a true loss process though, as the newly excited H($n${$\ge$}3) rapidly decay back to H(2)); collisions with photoelectrons leading to ionization and deexcitation.
In the figure, some of these terms are labelled by reference to their rate coefficients.  
\\

$A_{21}${$\rho_{21}$} is the radiative lifetime
in the {\lalpha} line after correction for self-absorption or opacity effects.\footnote{For a transition between bound levels 1 and 2 of statistical weights $g_1$ and $g_2$, $\rho_{21}$ is defined as {$\rho_{21}$}=$1 - \frac{g_2}{g_1} \frac{\lambda^5_{12}}{2 h c^2} \Phi^d_{12} \frac{[{\rm{H}}(1)]}{[{\rm{H}}(2)]}$, where $\Phi^d_{12}$ is the integral of the average diffuse intensity over the line \citep{garciamunozschneider2019}. 
The subtracting term results from self-absorption of diffuse photons, and effectively reduces the transition probability $A_{21}$ from its value in vacuum. 
It is convenient to include such a term in the losses of H(2) because its interpretation is quite natural that way. 
In contrast, it is convenient to keep the contribution of stellar photons 
$A_{21} \frac{g_2}{g_1} \frac{\lambda^5_{12}}{2 h c^2} \Phi^{\star}_{12} [{\rm{H}}(1)]$ to the excitation of H(1) into H(2) as a production mechanism for H(2).} 
It differs from the radiative lifetime in vacuum ($\rho_{21}$=1) in that if the line is optically thick the emitted {\lalpha} photon is promptly reabsorbed, and an emission-absorption cycle will occur many times before the photon finally escapes the line. 
As a consequence, the transition probability is effectively reduced by $\rho_{21}$. This factor becomes as small as 10$^{-4}$ at pressures $\ge$10$^{-2}$ dyn cm$^{-2}$ in our model of HAT-P-32b.
The phenomenon of line trapping is treated in \citet{czeslaetal2022}
by solving exactly the coupled problem of radiation-population,
and is often treated in other models by estimating $\rho_{21}$ with the help of escape factors  \citep[][]{elitzurasensioramos2006}. 
Although self-absorption reduces significantly the effective transition probability in the {\lalpha} line, 
radiative decay into the ground level remains by far the leading mechanism for depopulating H(2) at all altitudes in our model. 
The emission of H(2$s$) to the ground level is optically forbidden, but fast H(2$s$)--H(2$p$) interconversion in collisions with protons ensures that 
the H(2$s$) atom will eventually transition into the ground level passing first through H(2$p$).
\\

An effectively longer radiative lifetime will not affect the cooling potential of the {\lalpha} line because, based on Fig. \ref{h02_loss_gain_lifetime_photoe_fig}, the photons will ultimately escape before the excited level is quenched or ionized. 
An effectively longer radiative lifetime will however affect the H(2) population because for a prescribed production rate of this level $\mathcal{P}_2$, the H(2) density responds to the effective lifetime as $A_{21}${$\rho_{21}$}[H(2)]=$\mathcal{P}_2$. 
Self-absorption is thus critical in establishing the correct H(2) density in the model, which is key towards comparing the model predictions with observations. 
Importantly, the {\halpha} line in transmission spectroscopy provides simultaneous insight into the excitation mechanisms for H(2) and into the opacity of the medium where the line is formed.
\\

\begin{figure}[h]
    \centering
    \includegraphics[width=8.9cm]{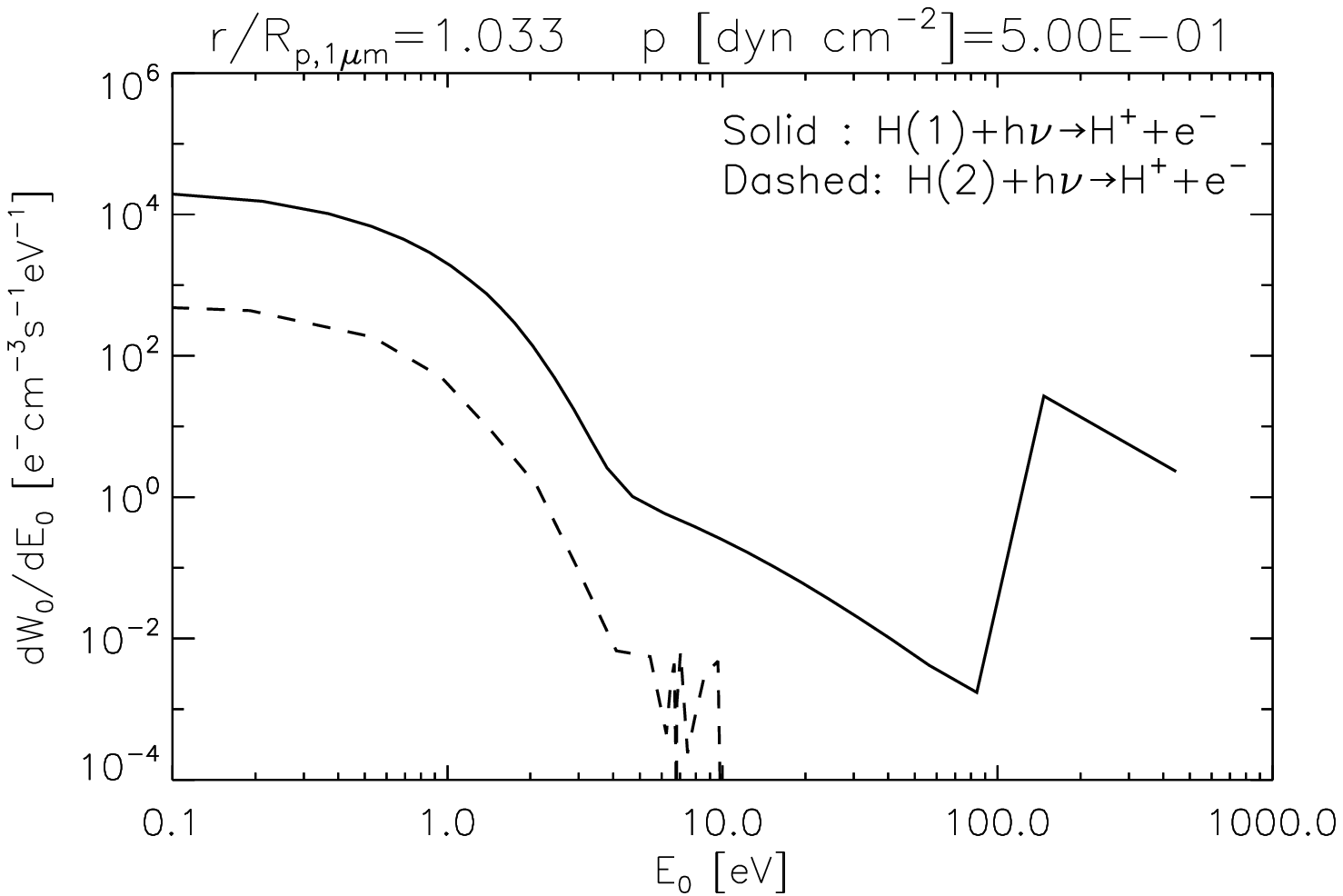}  \includegraphics[width=8.9cm]{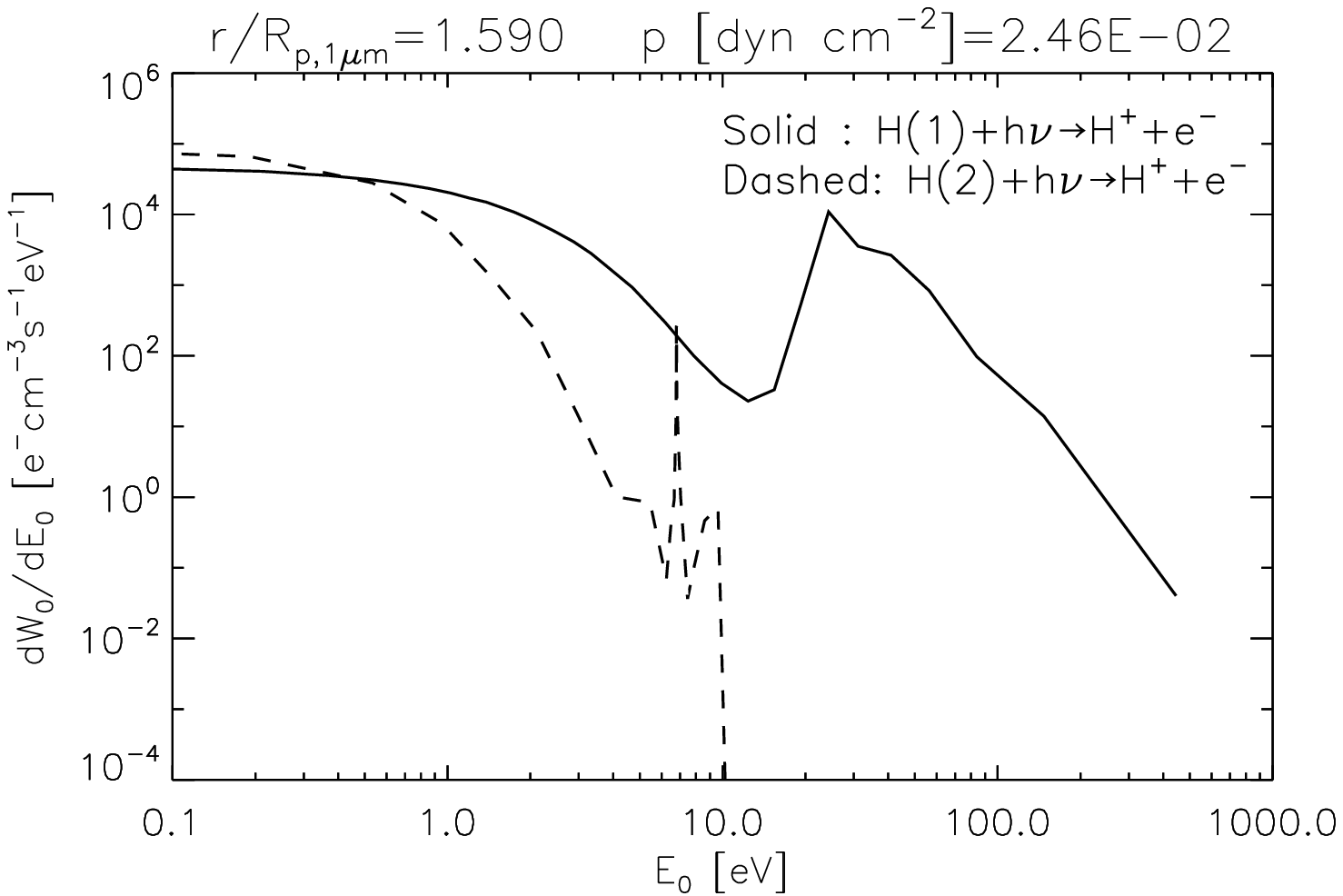} \\
    \includegraphics[width=8.9cm]{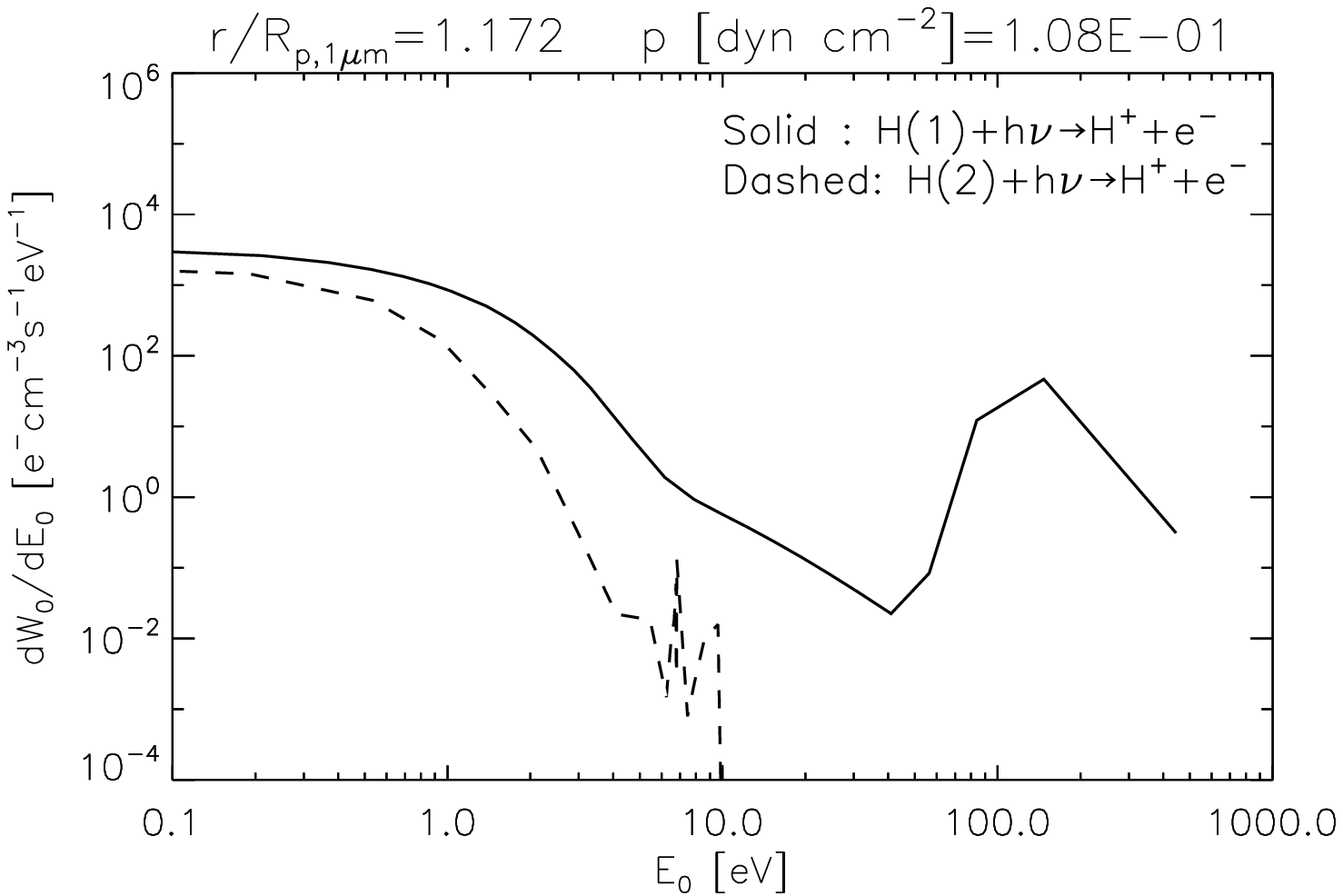}  \includegraphics[width=8.9cm]{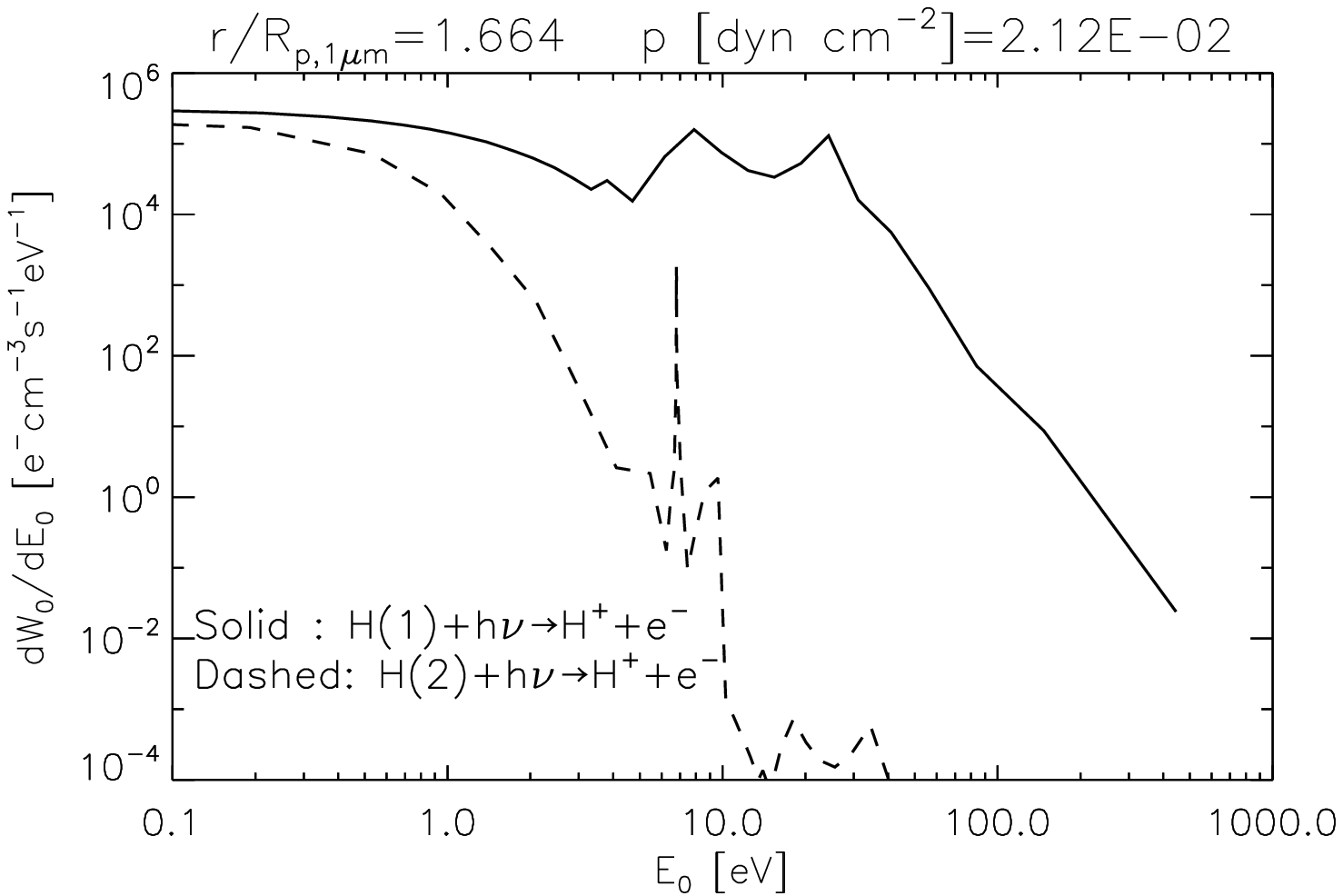} \\     \includegraphics[width=8.9cm]{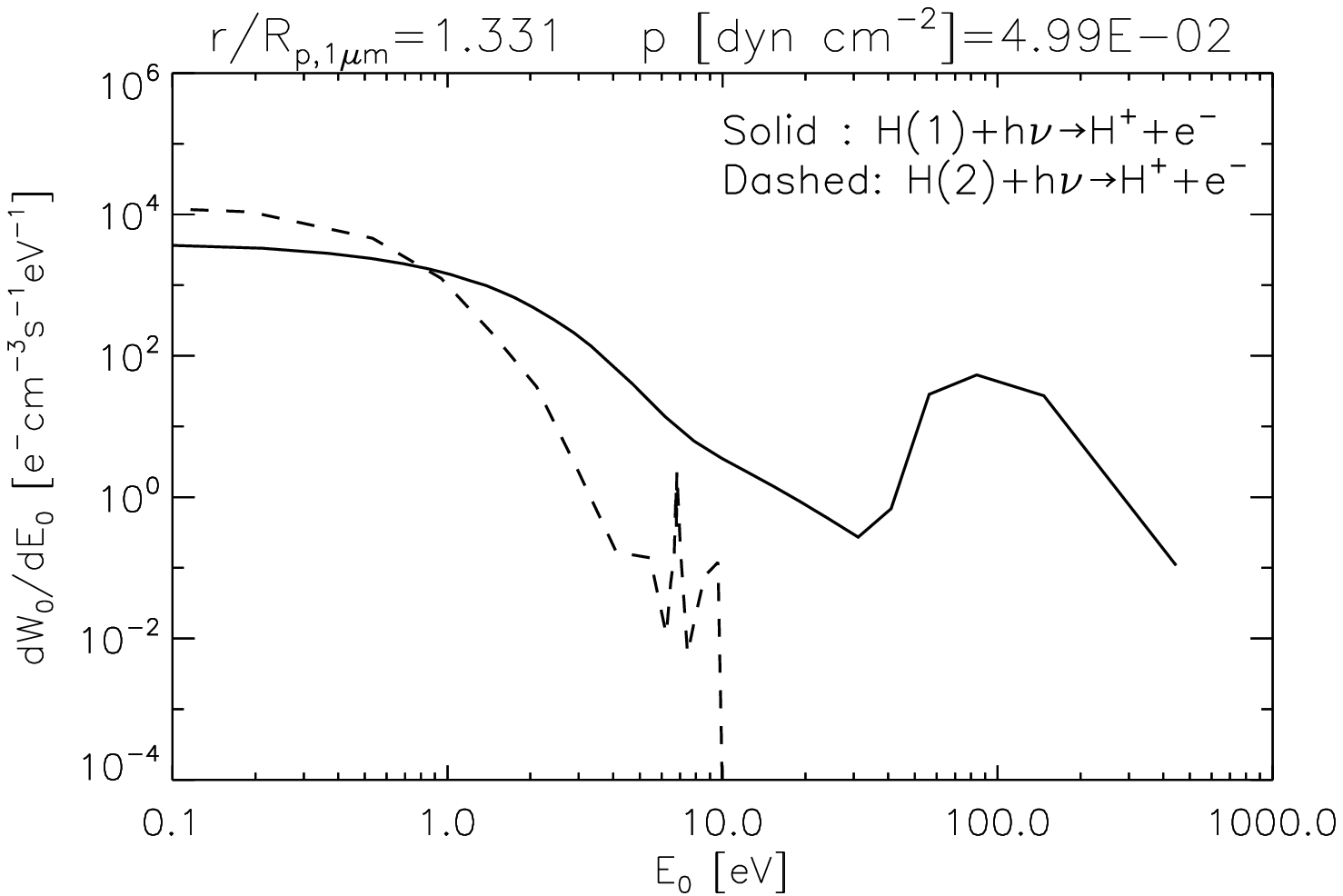}  \includegraphics[width=8.9cm]{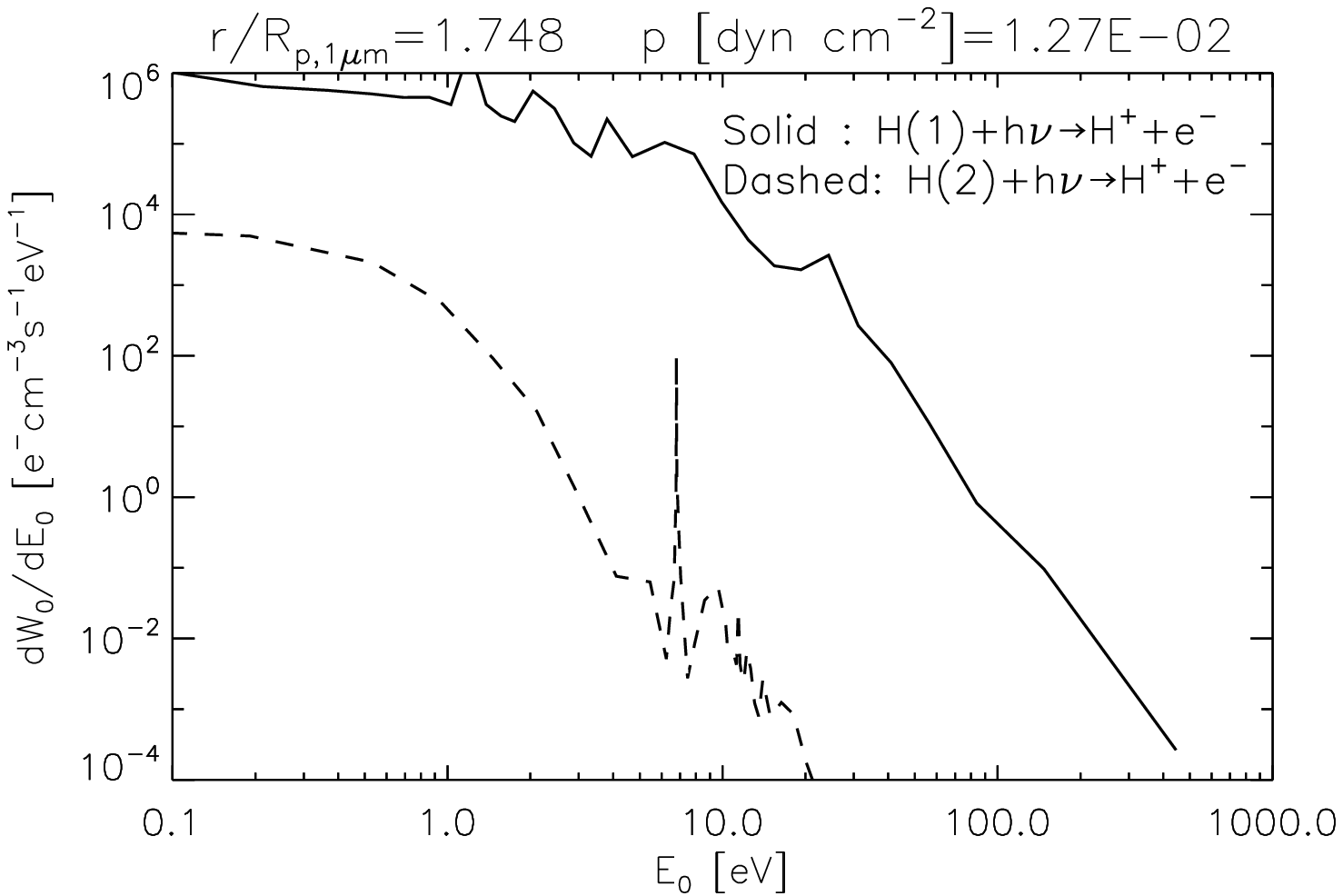} \\
    \caption{Energy spectra of the nascent photoelectrons produced through photoionization of H(1) and H(2) at different levels in the atmosphere. 
    } 
    \label{dW0dE_fig}
\end{figure}

The photoelectrons contribute negligibly to the H(2) loss either through superelastic or ionization collisions. We calculated their contributions with the aid of $W_{0,T}$ and $dW_0/dE_0$,  Figs. \ref{W0T_photoe_fig}-\ref{dW0dE_fig}, and the abundance-normalized yields of Fig. \ref{normyields_fig}.
The calculated lifetimes are shown in Fig. \ref{h02_loss_gain_lifetime_photoe_fig} (top). 
Deexcitation is less efficient than ionization,
even though superelastic collisions occur with no energy threshold, 
partly because the relevant cross sections are generally lower than the ionization cross sections over a broad range of energies. 
To understand why H(2) ionization in collisions with photoelectrons turns out to be inefficient, we 
recall Eqs. \ref{h2s_lossionization_minor_eq}-\ref{h2s_lossquenching_minor_eq}. 
For example, the relevant timescale for H(2$s$) is determined by the absolute value of ${\delta \ln{ [{\rm{H}}(2s)]}}$/{$\delta t$}, which according to Eq. \ref{h2s_lossionization_minor_eq} has the same form as the rate coefficient $J_{1\rightarrow\infty}$ for H(1) photoionization but weighted by
$\Psi_{\rm{H}(2s) \rightarrow \rm{H}^+}$
at each wavelength/energy bin in the summation. 
For reference, Fig. \ref{W0T_photoe_fig} shows $J_{1\rightarrow\infty}$ in our HAT-P-32b model. Its value is
$\sim$5$\times$10$^{-2}$ s$^{-1}$ at the top of the atmosphere, and drops rapidly as the stellar photons with wavelengths closer to 912 {\AA} (and that produce low-energy photoelectrons) become deposited in the atmosphere. 
$\Psi_{\rm{H}(2s) \rightarrow \rm{H}^+}$ is exactly zero for $E_0${$<$}3.4 eV and rather small at $E_0${$>$}3.4 eV for moderately high fractional ionizations, a combination of factors that cause that 
the lifetime for H(2) ionization in collisions with photoelectrons is much smaller than 
$J_{1\rightarrow\infty}$ in the uppermost atmosphere. 
$\Psi_{\rm{H}(2s) \rightarrow \rm{H}^+}$ becomes of order one only for
$E_0${$\gtrsim$}10 eV, which corresponds to photons of wavelengths
$\lesssim$525 {\AA} that contribute modestly to $J_{1\rightarrow\infty}$. 
The compensation between $\Psi_{\rm{H}(2s) \rightarrow \rm{H}^+}$ (larger for ionization by more energetic photons)
and the ionizing capacity of the radiation that reaches a specific layer of the atmosphere (which depends on the stellar spectrum but also on the photoionization cross section of the background gas)
explains both the absolute contribution of photoelectrons to H(2) ionization, and the 
fact that the characteristic timescale remains nearly independent of altitude.
\\

\subsection{Production of H(2) and H$^+$}

\begin{figure}[h]
\centering
\includegraphics[width=9cm]{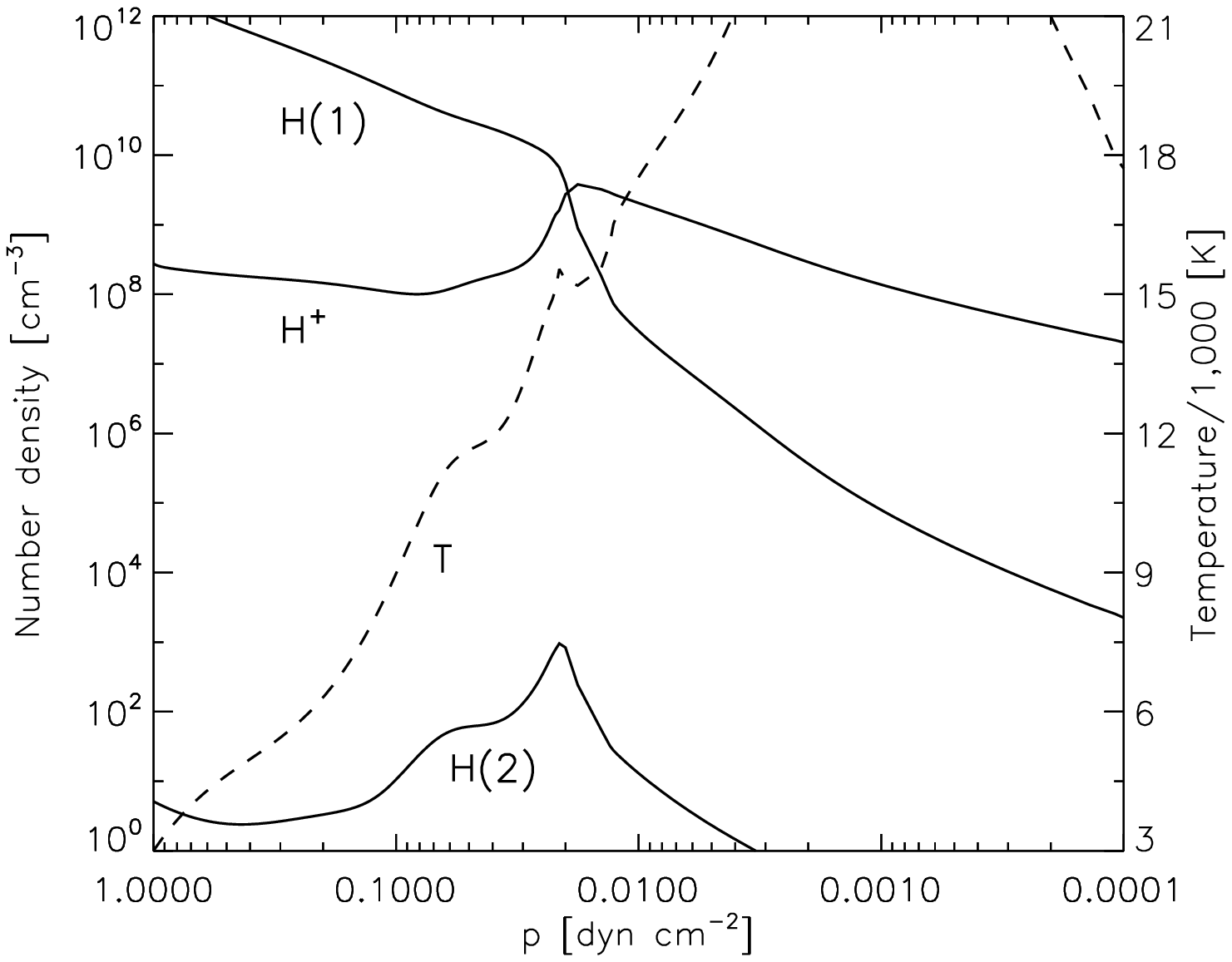} \\
\includegraphics[width=9cm]{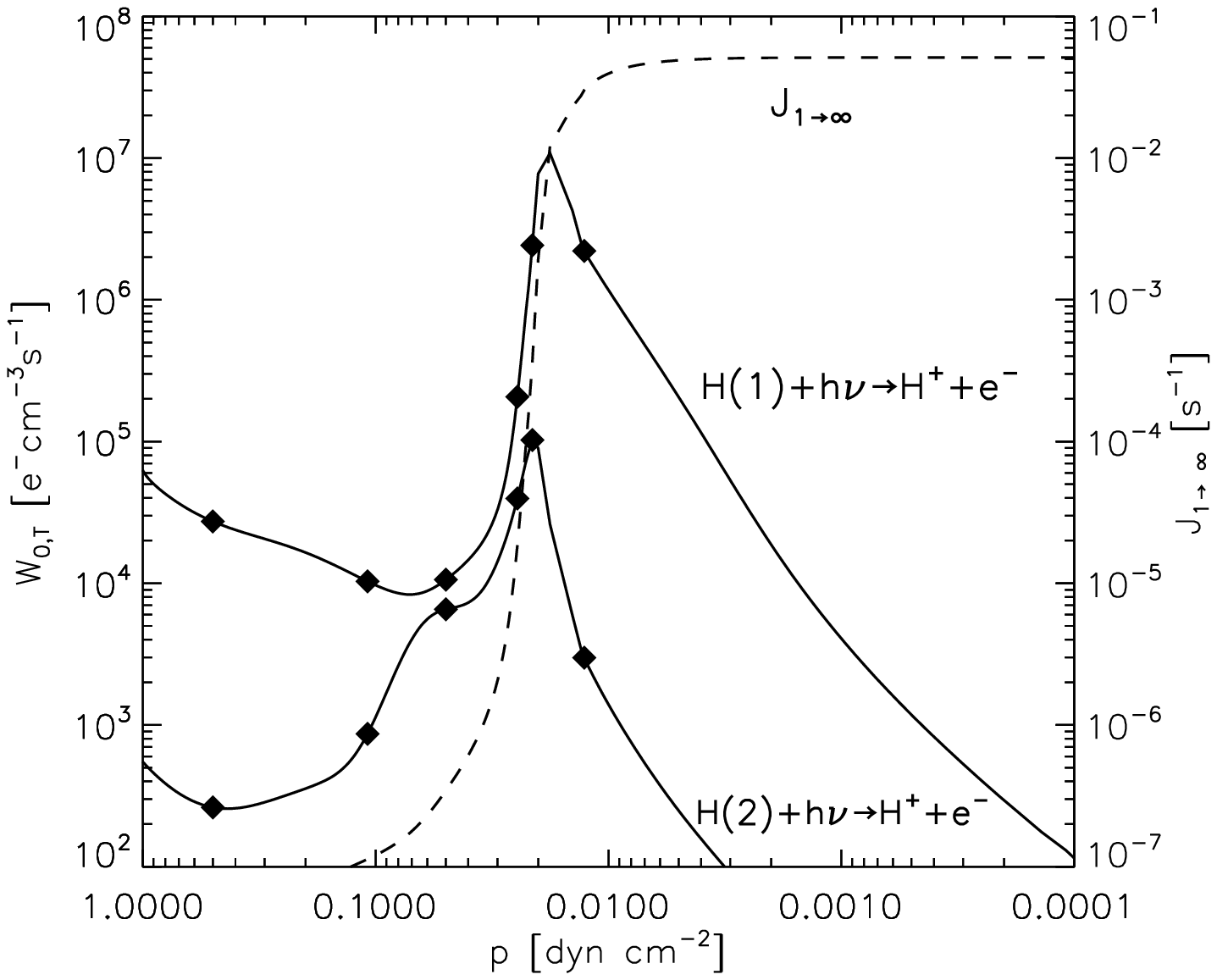} 
\caption{Top. Profiles of [H], [H$^+$], [H(2)] and temperature for HAT-P-32b
from \citet{czeslaetal2022}. Bottom. Direct photoionization rate (multiple ionizations by photoelectrons excluded) over the same range of pressures. Note that H(2) makes a subdominant yet non-negligible contribution to the direct photoionization rate at $\sim$(2--7)$\times$10$^{-2}$ dyn cm$^{-2}$.
The diamonds mark the pressure levels specified in the panel of Fig. \ref{dW0dE_fig}. In the same panel, also the rate coefficient $J_{1\rightarrow\infty}$ for H(1) photoionization.
} 
\label{W0T_photoe_fig}
\end{figure}

Figure \ref{h02_loss_gain_lifetime_photoe_fig} (middle) shows the rates at which H(2) is formed
in the atmosphere of HAT-P-32b. Collisional excitation by thermal electrons dominates the H(2) production in the region where the H(2) abundance peaks (see Fig. \ref{W0T_photoe_fig}) and the {\halpha} line forms. 
Photoexcitation of the ground level by direct stellar photons (photoexcitation by diffuse radiation is incorporated in $\rho_{21}$ and budgeted in the H(2) loss terms) dominates at low and high pressures, but this contribution is relatively minor and insufficient to explain the H(2) observed in the transmission spectra. 
The other mechanisms contributing to H(2) production are relatively inefficient. 
In particular, collisional excitation by photoelectrons becomes comparable to the dominant excitation mechanism only at the base of the model atmosphere. We note also that direct radiative recombination into H(2) and 
radiative decay from H($n${$\ge$3}) (either populated from radiative recombination or from collisional excitation by thermal electrons of H(1)) remain subdominant at the pressures where the {\halpha} line forms.
\\

Figure \ref{h02_loss_gain_lifetime_photoe_fig} (bottom) shows the production rates for H$^+$. H(1) photoionization plays a major role, with H(1) collisional ionization by thermal electrons and H(2) photoionization contributing comparably over narrow regions of the atmosphere. 
More interesting, collisional ionization by photoelectrons dominates the H$^+$ production at pressures $\gtrsim$3$\times$10$^{-2}$ dyn cm$^{-2}$, a region that overlaps with the peak in the H(2) abundance (Fig. \ref{W0T_photoe_fig}). 
It is apparent that collisional ionization by photoelectrons should be included in models of the extended atmospheres of exoplanets as their neglect may bias the local fractional ionization. 
In particular, increased ionization should reduce the heating efficiency locally, and thus the mass loss rate of the planet to some extent. Increased ionization will also affect the rate at which H(2) is excited and at which H$^+$ recombines, and both factors should affect the 
strength of the {\halpha} line. 
We defer to a dedicated investigation the quantification of these possibilities and the implicit feedbacks in the photochemical-hydrodynamical modeling.
\\

\subsection{Impact on other diagnostic lines}

On a more tentative basis, 
we note the possibility that a bias in the electron density due to the neglect of photoelectrons may have affected the He/H ratios inferred by inverse modeling of the transition He(2$^3S$)+$h\nu$ (10830{\AA}){$\rightarrow$}He(2$^3P$) detected in transmission spectroscopy at some exoplanets 
\citep{lamponetal2020,czeslaetal2022,fossatietal2022}. 
Radiative recombination, which probably controls the formation of the 
metastable He(2$^3S$), and singlet-triplet mixing through collisions with thermal electrons, which partly controls the destruction of He(2$^3S$), both depend on the electron densities. A biased electron density may affect the competition between these and the other collisional-radiative processes that dictate the He(2$^3S$) abundance. 
Unlike the H(2) level that emits rapidly to the ground level (even after accounting for self-absorption), the radiative lifetime of He(2$^3S$) is very long (transition probability of 1.26{$\times$}10$^{-4}$ s$^{-1}$; \citet{hodgmanetal2009}). 
This means that the He(2$^3S$) densities probed in transmission spectroscopy potentially reflect the conditions from deeper down in the atmosphere. 
A biased estimate of the electron density in those deeper regions may have implications on the predicted metastable helium that are worth investigating.
\\

\section{\label{sec:summary} Summary} 
 
Photoionization in a planetary atmosphere produces photoelectrons that alter the chemical and energy balance of the atmospheric gas in ways different to thermal electrons. Many of these effects remain unexplored in exoplanet atmospheres, but  there are reasons that suggest they are important.
 To list a few: most of the known exoplanets are strongly irradiated; they orbit stars with possibly high (and variable) levels of XUV emission; the smaller exoplanets may contain gases other than hydrogen/helium that react to photoelectrons in  specific ways because they have relatively low-energy excitation levels. \\
 
We have developed a MC model to investigate these possibilities 
and facilitate new investigations in other astrophysical objects. 
The model is efficient in its treatment of rare collisional channels, and takes into account collisions of the photoelectrons with heavy particles in their ground and excited levels. 
The idea behind the multi-score MC scheme proposed here may prove useful in other 
problems concerned with the slowing down of charged heavy particles, as in studies of aurorae and cosmic ray deposition. 
\\

We have taken a published model of the atmosphere of the ultra-hot Jupiter HAT-P-32b
to explore the effect of photoelectrons on the population of the excited level H(2) that is probed with transmission spectroscopy in the {\halpha} line. 
Focusing on the direct role of photoelectrons, 
we find that they contribute negligibly to the loss of H(2) through either ionization or deexcitation, 
but that they contribute in a non-negligible way 
to the excitation of H(2) in collisions with the ground level H(1)
in the region of formation of the {\halpha} line wings.
More importantly, we find that the photoelectrons will alter the ionization balance  in the region of the atmosphere
where the {\halpha} line is formed. Because the collisions of H(1) with thermal electrons remain the main channel to form H(2), it is anticipated that
photoelectrons will contribute through this indirect path to shaping the {\halpha} line. 
For similar reasons, it is anticipated that photoelectron-driven ionization will offer new insight into the (de)population of other atoms that 
are probed with transmission spectroscopy. 
In this respect, metastable helium remains an interesting possibility.
The modelling of HAT-P-32b confirms the importance of self-absorption or opacity effects at predicting the correct abundance of H(2) or any other excited level that connects radiatively with the ground level. 
A key advantage of our multi-score MC scheme with respect to other MC schemes for the investigation of these issues is that it reduces the computational cost associated with rare collisional channels, some of them being potentially significant. 
An efficient scheme facilitates the self-consistent implementation of photoelectrons in the 
simulation of exoplanet atmospheres with photochemical-hydrodynamical models, thereby enabling an accurate look into feedbacks and into situations when rare channels become dominant and vice versa.
\\

Ongoing work will examine the impact of photoelectrons on the energy budget of hydrogen atmospheres of close-in exoplanets \citep[][]{gilletetal2022}. Future work will assess the response to photoelectrons of hydrogen-dominated and non-hydrogen-dominated atmospheres, and the impact of photoelectrons on some atoms that are often targeted in transmission spectroscopy observations.
\\

The MC model developed here is publicly available through the website https://antoniogarciamunoz.wordpress.com/ or by e-mail request at antonio.garciamunoz@cea.fr or tonhingm@gmail.com. The package includes the FORTRAN sources, the implemented cross sections and a variety of input/output files to reproduce most of the calculations presented in this paper.
\\

\clearpage
\newpage

\begin{table}[t]
\scriptsize
\caption{Production yields $\Phi_{\bf{a}}$, $\Phi_{\bf{b}}$, $\Phi_{\bf{c}}$ and $\Phi_{\bf{d}}$ and heating efficiency $\eta_h$ 
for the study of \S\ref{convergencerate_subsec}. In each calculation photoelectrons of energy $E_0$ are injected into a gas of H(1), H(2$s$) and thermal electrons in proportions specified by [H(2$s$)]/[H(1)]=10$^{-7}$ and $x_e$ (see text). 
Each calculation is based on $N_{\rm{simul}}$ simulations and $E_{\rm{cutoff}}$=0.17 eV ($T_{\rm{e}}$=10$^3$ K).
For each $E_0$ and $N_{\rm{simul}}$ entry, the top and bottom rows refer to the multi- and single-score schemes,  respectively. 
}             
\label{convrate_table}      
\centering                          
\begin{tabular}{|c c | c c c c c | c c c c c |}        
\hline
      &                  & \multicolumn{5}{c|}{$x_e$=10$^{-5}$} & \multicolumn{5}{c|}{$x_e$=10$^{-2}$} \\
$E_0$ & $N_{\rm{simul}}$ & 
\textbf{a} & \textbf{b} & \textbf{c} & \textbf{d} & $\eta_h$ &  
\textbf{a} & \textbf{b} & \textbf{c} & \textbf{d} & $\eta_h$ \\
\hline\hline
   30 & 10$^1$               & 0.9942(+0) & 0.5402(+0) &   0.2730($-$5) &   0.2196($-$4) & 0.1708 & 0.2814(+0) & 0.2024(+0) &   0.8973($-$8) &   0.4165($-$6) & 0.7448 \\    
   30 & 10$^1$               & 0.8000(+0) & 0.6000(+0) &   0.0000(+0) &   0.0000(+0) & 0.3389 & 0.3000(+0) & 0.2000(+0) &   0.0000(+0) &   0.0000(+0) & 0.6463 \\    
\hline
   30 & 10$^2$               & 0.9156(+0) & 0.5367(+0) &   0.2166($-$5) &   0.1285($-$4) & 0.2236 & 0.2923(+0) & 0.1851(+0) &   0.1031($-$7) &   0.4859($-$6) & 0.7451 \\    
   30 & 10$^2$               & 0.8400(+0) & 0.6100(+0) &   0.0000(+0) &   0.0000(+0) & 0.2524 & 0.2900(+0) & 0.1600(+0) &   0.0000(+0) &   0.0000(+0) & 0.7454 \\    
\hline
   30 & 10$^3$               & 0.8619(+0) & 0.5151(+0) &   0.2155($-$5) &   0.1395($-$4) & 0.2668 & 0.3015(+0) & 0.1945(+0) &   0.1023($-$7) &   0.4884($-$6) & 0.7356 \\    
   30 & 10$^3$               & 0.8400(+0) & 0.5390(+0) &   0.0000(+0) &   0.0000(+0) & 0.2561 & 0.2950(+0) & 0.1950(+0) &   0.0000(+0) &   0.0000(+0) & 0.7409 \\    
\hline
   30 & 10$^4$               & 0.8716(+0) & 0.5156(+0) &   0.2179($-$5) &   0.1457($-$4) & 0.2595 & 0.2987(+0) & 0.1939(+0) &   0.1016($-$7) &   0.4825($-$6) & 0.7376 \\    
   30 & 10$^4$               & 0.8692(+0) & 0.5191(+0) &   0.0000(+0) &   0.0000(+0) & 0.2587 & 0.3026(+0) & 0.1947(+0) &   0.0000(+0) &   0.0000(+0) & 0.7359 \\    
\hline
   30 & 10$^5$               & 0.8718(+0) & 0.5186(+0) &   0.2173($-$5) &   0.1430($-$4) & 0.2579 & 0.2982(+0) & 0.1932(+0) &   0.1017($-$7) &   0.4825($-$6) & 0.7382 \\    
   30 & 10$^5$               & 0.8697(+0) & 0.5222(+0) &   0.1000($-$4) &   0.0000(+0) & 0.2573 & 0.3014(+0) & 0.1946(+0) &   0.0000(+0) &   0.0000(+0) & 0.7366 \\    
\hline
   30 & 10$^6$               & 0.8721(+0) & 0.5188(+0) &   0.2172($-$5) &   0.1434($-$4) & 0.2577 & 0.2989(+0) & 0.1936(+0) &   0.1019($-$7) &   0.4837($-$6) & 0.7376 \\    
   30 & 10$^6$               & 0.8723(+0) & 0.5193(+0) &   0.2000($-$5) &   0.1700($-$4) & 0.2575 & 0.2983(+0) & 0.1937(+0) &   0.0000(+0) &   0.0000(+0) & 0.7379 \\    
\hline
   30 & 10$^7$               & 0.8723E+00 & 0.5189(+0) &   0.2172($-$5) &   0.1435($-$4) & 0.2575 & 0.2988(+0) & 0.1935(+0) &   0.1019($-$7) &   0.4836($-$6) & 0.7376 \\    
   30 & 10$^7$               & 0.8723E+00 & 0.5188(+0) &   0.1000($-$5) &   0.1280($-$4) & 0.2574 & 0.2987(+0) & 0.1935(+0) &   0.0000(+0) &   0.7000($-$6) & 0.7378 \\    

\hline\hline

   50 & 10$^1$               & 0.1836(+1) & 0.1141(+1) &   0.2507($-$5) &   0.1011($-$4) & 0.0684 & 0.6515(+0) & 0.5308(+0) &   0.1528($-$7) &   0.7767($-$6) & 0.6370 \\    
   50 & 10$^1$               & 0.1800(+1) & 0.1200(+1) &   0.0000(+0) &   0.0000(+0) & 0.1681 & 0.6000(+0) & 0.6000(+0) &   0.0000(+0) &   0.0000(+0) & 0.6190 \\    
\hline
   50 & 10$^2$               & 0.1592(+1) & 0.1058(+1) &   0.2441($-$5) &   0.1303($-$4) & 0.1709 & 0.6856(+0) & 0.5686(+0) &   0.1528($-$7) &   0.7946($-$6) & 0.6159 \\    
   50 & 10$^2$               & 0.1480(+1) & 0.1140(+1) &   0.0000(+0) &   0.0000(+0) & 0.1662 & 0.7300(+0) & 0.5300(+0) &   0.0000(+0) &   0.0000(+0) & 0.6023 \\    
\hline
   50 & 10$^3$               & 0.1571(+1) & 0.1077(+1) &   0.2581($-$5) &   0.1388($-$4) & 0.1724 & 0.6874(+0) & 0.5695(+0) &   0.1546($-$7) &   0.8001($-$6) & 0.6149 \\    
   50 & 10$^3$               & 0.1553(+1) & 0.1086(+1) &   0.0000(+0) &   0.0000(+0) & 0.1781 & 0.6950(+0) & 0.5710(+0) &   0.0000(+0) &   0.0000(+0) & 0.6159 \\    
\hline
   50 & 10$^4$               & 0.1563(+1) & 0.1066(+1) &   0.2600($-$5) &   0.1437($-$4) & 0.1782 & 0.6853(+0) & 0.5683(+0) &   0.1537($-$7) &   0.7964($-$6) & 0.6160 \\    
   50 & 10$^4$               & 0.1568(+1) & 0.1056(+1) &   0.0000(+0) &   0.0000(+0) & 0.1802 & 0.6819(+0) & 0.5758(+0) &   0.0000(+0) &   0.0000(+0) & 0.6115 \\    
\hline
   50 & 10$^5$               & 0.1558(+1) & 0.1065(+1) &   0.2614($-$5) &   0.1431($-$4) & 0.1803 & 0.6892(+0) & 0.5714(+0) &   0.1543($-$7) &   0.8012($-$6) & 0.6138 \\    
   50 & 10$^5$               & 0.1556(+1) & 0.1068(+1) &   0.0000(+0) &   0.0000(+0) & 0.1809 & 0.6877(+0) & 0.5704(+0) &   0.0000(+0) &   0.0000(+0) & 0.6139 \\    
\hline
   50 & 10$^6$               & 0.1555(+1) & 0.1065(+1) &   0.2616($-$5) &   0.1436($-$4) & 0.1813 & 0.6882(+0) & 0.5703(+0) &   0.1544($-$7) &   0.8009($-$6) & 0.6145 \\    
   50 & 10$^6$               & 0.1553(+1) & 0.1066(+1) &   0.6000($-$5) &   0.1300($-$4) & 0.1811 & 0.6889(+0) & 0.5703(+0) &   0.0000(+0) &   0.0000(+0) & 0.6141 \\    
\hline
   50 & 10$^7$               & 0.1555(+1) & 0.1065(+1) & 0.2616($-$5) & 0.1435($-$4) & 0.1812 & 0.6885(+0) & 0.5707(+0) &   0.1544($-$7) &   0.8011($-$6) & 0.6143 \\    
   50 & 10$^7$               & 0.1555(+1) & 0.1066(+1) & 0.2900($-$5) & 0.1450($-$4) & 0.1811 & 0.6886(+0) & 0.5707(+0) &   0.0000(+0) &   0.8000($-$6) & 0.6143 \\    

\hline\hline

   100 & 10$^1$               & 0.2836(+1) & 0.2510(+1) &   0.5061($-$5) &   0.3311($-$4) & 0.1938 & 0.1820(+1) & 0.1733(+1) &   0.2748($-$7) &   0.1490($-$5) & 0.4717 \\    
   100 & 10$^1$               & 0.2800(+1) & 0.2600(+1) &   0.0000(+0) &   0.0000(+0) & 0.1768 & 0.1500(+1) & 0.1600(+1) &   0.0000(+0) &   0.0000(+0) & 0.5077 \\    
\hline
   100 & 10$^2$               & 0.3104(+1) & 0.2469(+1) &   0.4472($-$5) &   0.2564($-$4) & 0.1505 & 0.1650(+1) & 0.1537(+1) &   0.2726($-$7) &   0.1427($-$5) & 0.5247 \\    
   100 & 10$^2$               & 0.2990(+1) & 0.2610(+1) &   0.0000(+0) &   0.0000(+0) & 0.1566 & 0.1580(+1) & 0.1760(+1) &   0.0000(+0) &   0.0000(+0) & 0.4928 \\    
\hline
   100 & 10$^3$               & 0.3158(+1) & 0.2474(+1) &   0.4358($-$5) &   0.2299($-$4) & 0.1407 & 0.1725(+1) & 0.1611(+1) &   0.2745($-$7) &   0.1472($-$5) & 0.5025 \\    
   100 & 10$^3$               & 0.3084(+1) & 0.2504(+1) &   0.0000(+0) &   0.0000(+0) & 0.1499 & 0.1674(+1) & 0.1671(+1) &   0.0000(+0) &   0.0000(+0) & 0.4990 \\    
\hline
   100 & 10$^4$               & 0.3116(+1) & 0.2458(+1) &   0.4332($-$5) &   0.2310($-$4) & 0.1512 & 0.1732(+1) & 0.1622(+1) &   0.2738($-$7) &   0.1470($-$5) & 0.5002 \\    
   100 & 10$^4$               & 0.3129(+1) & 0.2481(+1) &   0.0000(+0) &   0.1000($-$3) & 0.1492 & 0.1728(+1) & 0.1647(+1) &   0.0000(+0) &   0.0000(+0) & 0.4966 \\    
\hline
   100 & 10$^5$               & 0.3124(+1) & 0.2469(+1) &   0.4306($-$5) &   0.2300($-$4) & 0.1484 & 0.1740(+1) & 0.1629(+1) &   0.2745($-$7) &   0.1477($-$5) & 0.4979 \\    
   100 & 10$^5$               & 0.3121(+1) & 0.2469(+1) &   0.0000(+0) &   0.4000($-$4) & 0.1484 & 0.1745(+1) & 0.1625(+1) &   0.0000(+0) &   0.1000($-$4) & 0.4980 \\    
\hline
   100 & 10$^6$               & 0.3125(+1) & 0.2470(+1) &   0.4302($-$5) &   0.2298($-$4) & 0.1481 & 0.1739(+1) & 0.1628(+1) &   0.2746($-$7) &   0.1477($-$5) & 0.4982 \\    
   100 & 10$^6$               & 0.3125(+1) & 0.2467(+1) &   0.5000($-$5) &   0.2000($-$4) & 0.1482 & 0.1736(+1) & 0.1629(+1) &   0.0000(+0) &   0.2000($-$5) & 0.4982 \\    
\hline
   100 & 10$^7$               & 0.3125(+1) & 0.2469(+1) & 0.4303($-$5) & 0.2299($-$4) & 0.1483 & 0.1739(+1) & 0.1628(+1) &   0.2746($-$7) &   0.1477($-$5) & 0.4981 \\    
   100 & 10$^7$               & 0.3126(+1) & 0.2469(+1) & 0.3900($-$5) & 0.2270($-$4) & 0.1483 & 0.1739(+1) & 0.1628(+1) &   0.0000(+0) &   0.1600($-$5) & 0.4981 \\    

\hline\hline

   1000 & 10$^1$               & 0.3409(+2) & 0.2722(+2) &   0.3195($-$4) &   0.1567($-$3) & 0.0998 & 0.2388(+2) & 0.2142(+2) &   0.2133($-$6) &   0.1191($-$4) & 0.3475 \\    
   1000 & 10$^1$               & 0.3360(+2) & 0.2640(+2) &   0.0000(+0) &   0.0000(+0) & 0.1090 & 0.2300(+2) & 0.2050(+2) &   0.0000(+0) &   0.0000(+0) & 0.3643 \\    
\hline
   1000 & 10$^2$               & 0.3297(+2) & 0.2671(+2) &   0.3530($-$4) &   0.1892($-$3) & 0.1243 & 0.2375(+2) & 0.2127(+2) &   0.2177($-$6) &   0.1195($-$4) & 0.3517 \\    
   1000 & 10$^2$               & 0.3269(+2) & 0.2714(+2) &   0.0000(+0) &   0.0000(+0) & 0.1217 & 0.2455(+2) & 0.2108(+2) &   0.0000(+0) &   0.0000(+0) & 0.3508 \\    
\hline
   1000 & 10$^3$               & 0.3337(+2) & 0.2678(+2) &   0.3423($-$4) &   0.1776($-$3) & 0.1159 & 0.2355(+2) & 0.2106(+2) &   0.2173($-$6) &   0.1191($-$4) & 0.3573 \\    
   1000 & 10$^3$               & 0.3338(+2) & 0.2677(+2) &   0.0000(+0) &   0.0000(+0) & 0.1177 & 0.2358(+2) & 0.2113(+2) &   0.0000(+0) &   0.0000(+0) & 0.3568 \\    
\hline
   1000 & 10$^4$               & 0.3326(+2) & 0.2676(+2) &   0.3417($-$4) &   0.1770($-$3) & 0.1181 & 0.2362(+2) & 0.2114(+2) &   0.2170($-$6) &   0.1188($-$4) & 0.3553 \\    
   1000 & 10$^4$               & 0.3323(+2) & 0.2682(+2) &   0.0000(+0) &   0.1000($-$3) & 0.1175 & 0.2364(+2) & 0.2114(+2) &   0.0000(+0) &   0.0000(+0) & 0.3551 \\    
\hline
   1000 & 10$^5$               & 0.3331(+2) & 0.2678(+2) &   0.3409($-$4) &   0.1763($-$3) & 0.1170 & 0.2362(+2) & 0.2114(+2) &   0.2170($-$6) &   0.1188($-$4) & 0.3554 \\    
   1000 & 10$^5$               & 0.3331(+2) & 0.2678(+2) &   0.7000($-$4) &   0.1400($-$3) & 0.1171 & 0.2361(+2) & 0.2114(+2) &   0.0000(+0) &   0.2000($-$4) & 0.3553 \\    
\hline
   1000 & 10$^6$               & 0.3330(+2) & 0.2678(+2) &   0.3410($-$4) &   0.1764($-$3) & 0.1172 & 0.2363(+2) & 0.2114(+2) &   0.2170($-$6) &   0.1189($-$4) & 0.3553 \\    
   1000 & 10$^6$               & 0.3330(+2) & 0.2678(+2) &   0.3200($-$4) &   0.1960($-$3) & 0.1172 & 0.2363(+2) & 0.2114(+2) &   0.0000(+0) &   0.1500($-$4) & 0.3553 \\    
\hline
   1000 & 10$^7$               & 0.3330(+2) & 0.2678(+2) &   0.3411($-$4) &   0.1764E-03 & 0.1172 & 0.2362(+2) & 0.2114(+2) &   0.2170($-$6) &   0.1189($-$4) & 0.3554 \\    
   1000 & 10$^7$               & 0.3330(+2) & 0.2678(+2) &   0.3050($-$4) &   0.1783E-03 & 0.1172 & 0.2362(+2) & 0.2114(+2) &   0.2000($-$6) &   0.1210($-$4) & 0.3554 \\    

\hline                        

\end{tabular}
\end{table}

\clearpage

\clearpage

\appendix
\section{\label{appendix:xsections}Cross sections}

This Appendix describes the cross sections implemented in the MC model for the collisions of the photoelectrons with the other particles in the gas.  The data available in compilations and the literature usually refer to elastic and inelastic collisions. Thus when needed we obtained the deexcitation cross sections for superelastic collisions from application of detailed balancing, 
{$g_{X_L}$}{($E+E^\ast$)}{$\sigma^{exc}_{X_L \rightarrow X_U}$}($E+E^\ast$)={$g_{X_U}$}{$E$}{$\sigma^{deexc}_{X_U \rightarrow X_L}$}($E$), where $E^{\ast}$=$E$($X_U$){$-$}$E$($X_L$) is the energy difference between the upper and lower levels and 
{$g_{X_L}$} and {$g_{X_U}$} their statistical weights. 
We omitted Coulomb collisions with charged heavy particles; their cross sections are comparable to those for electron-electron collisions, yet the energy exchanged in them is lower by about the electron-to-heavy particle mass ratio
 \citep[][pgs. 417-421]{zeldovichraizer2002}. 
\\ 

\begin{figure}[b]
\centering
\includegraphics[width=13cm]{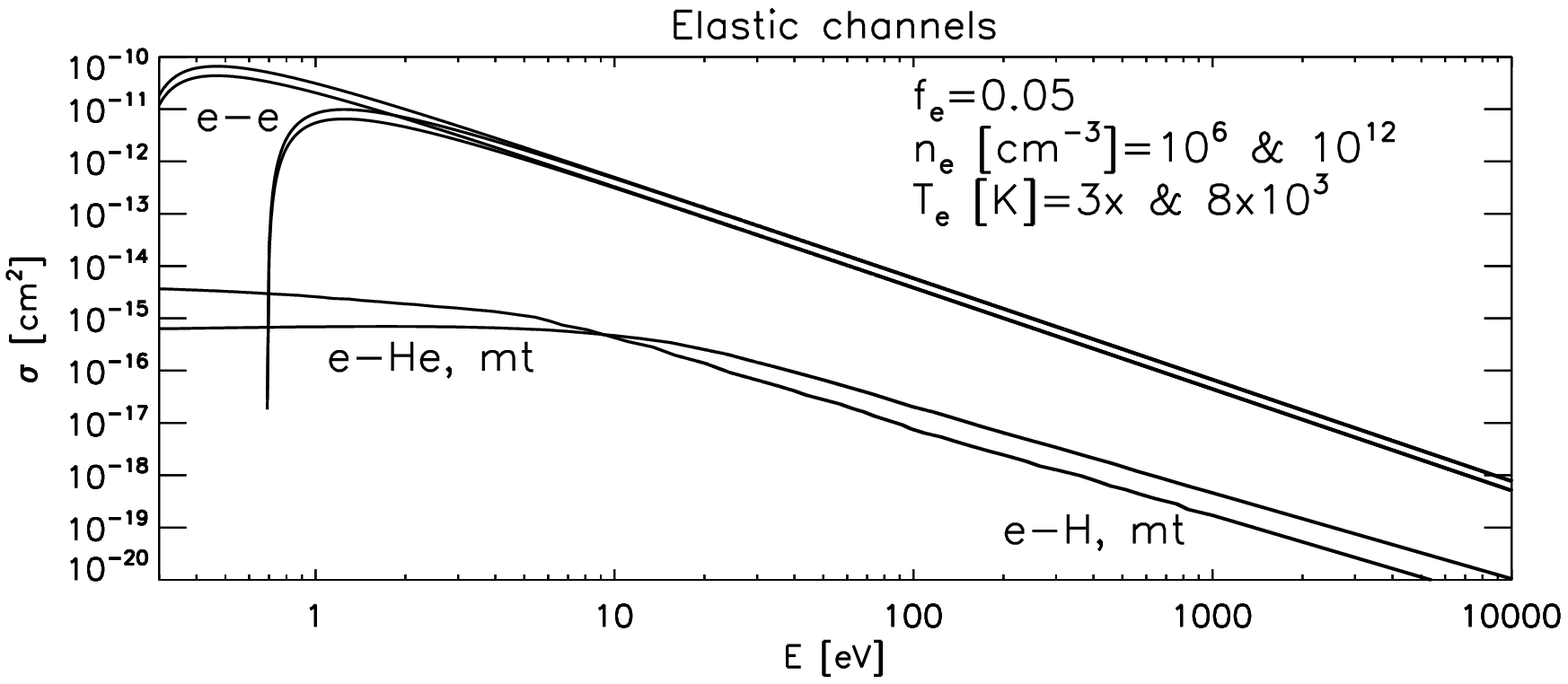} \\ 
\vspace{-0.2cm}
\includegraphics[width=13cm]{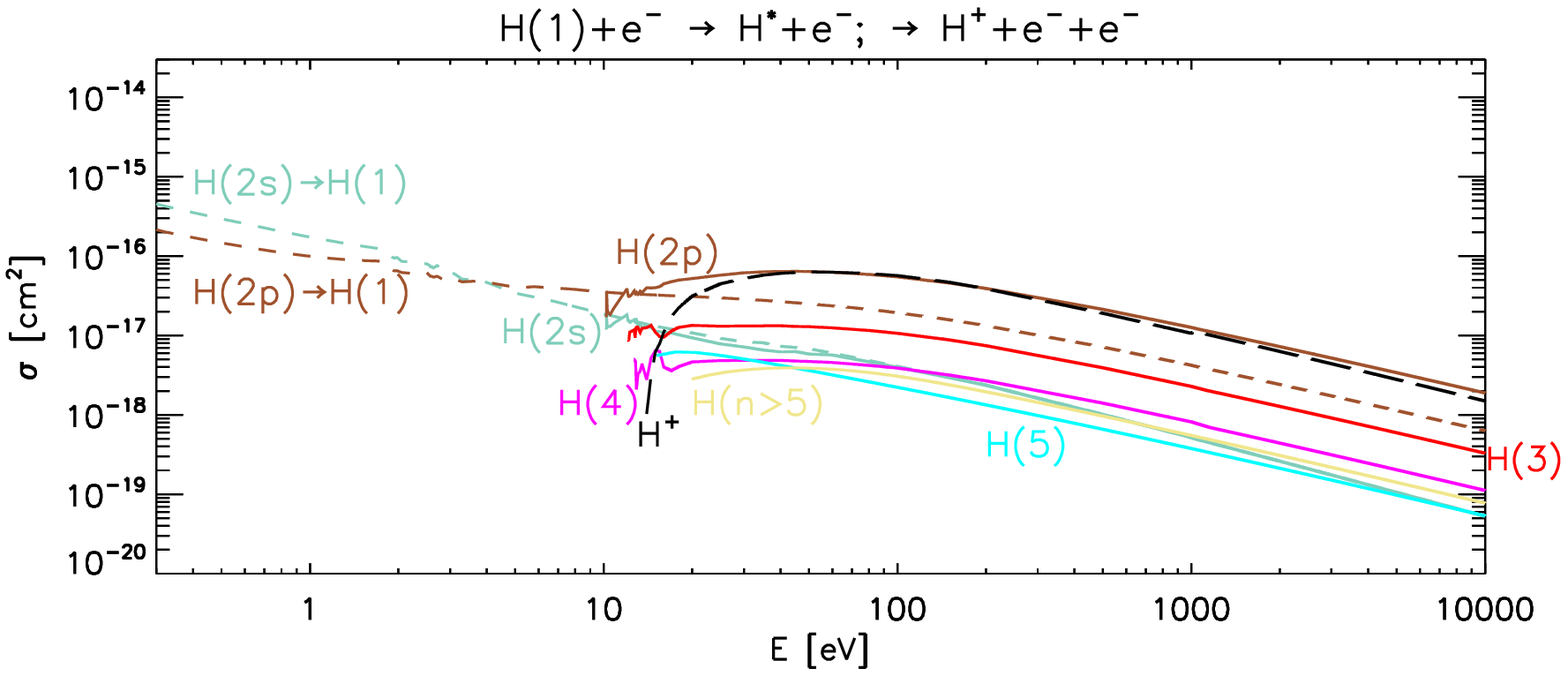} \\ 
\vspace{-0.2cm}
\includegraphics[width=13cm]{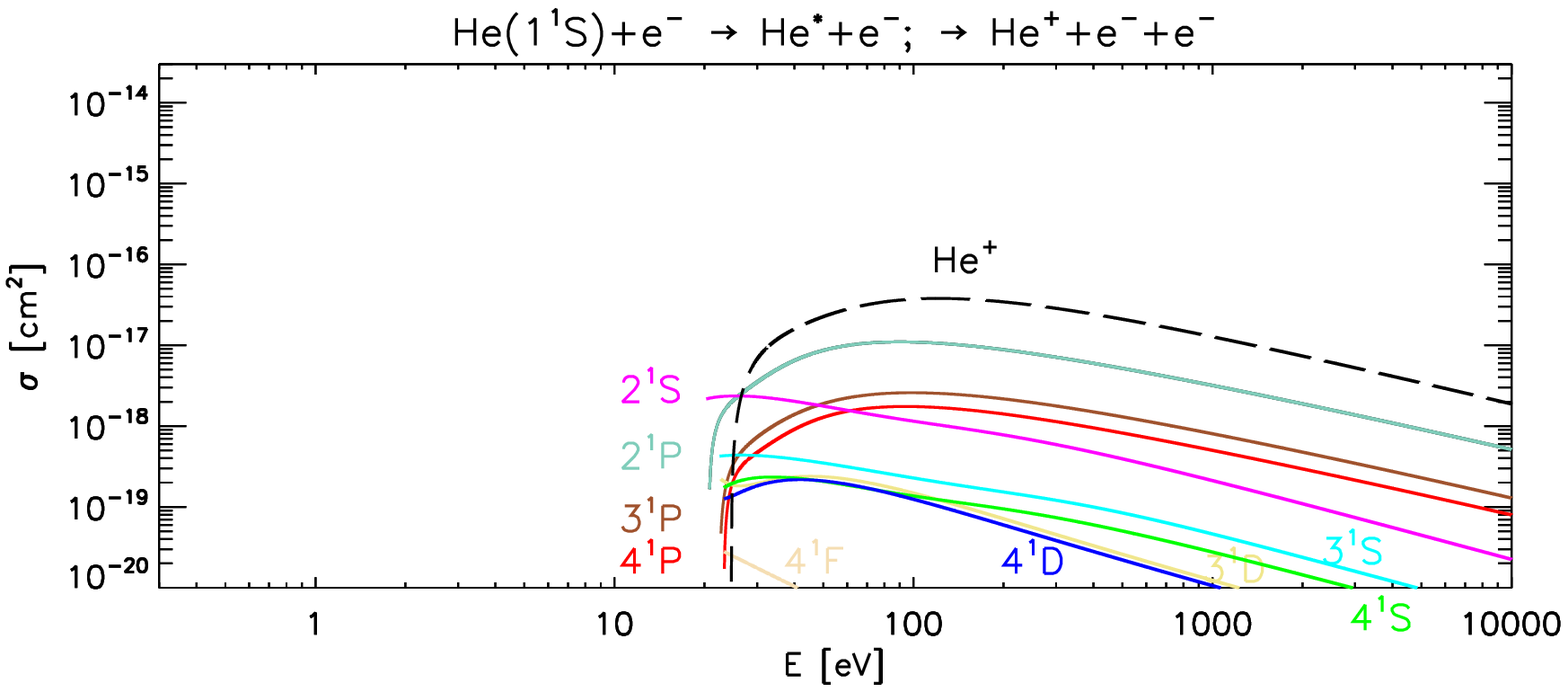} \\ 
\vspace{-0.2cm}
\includegraphics[width=13cm]{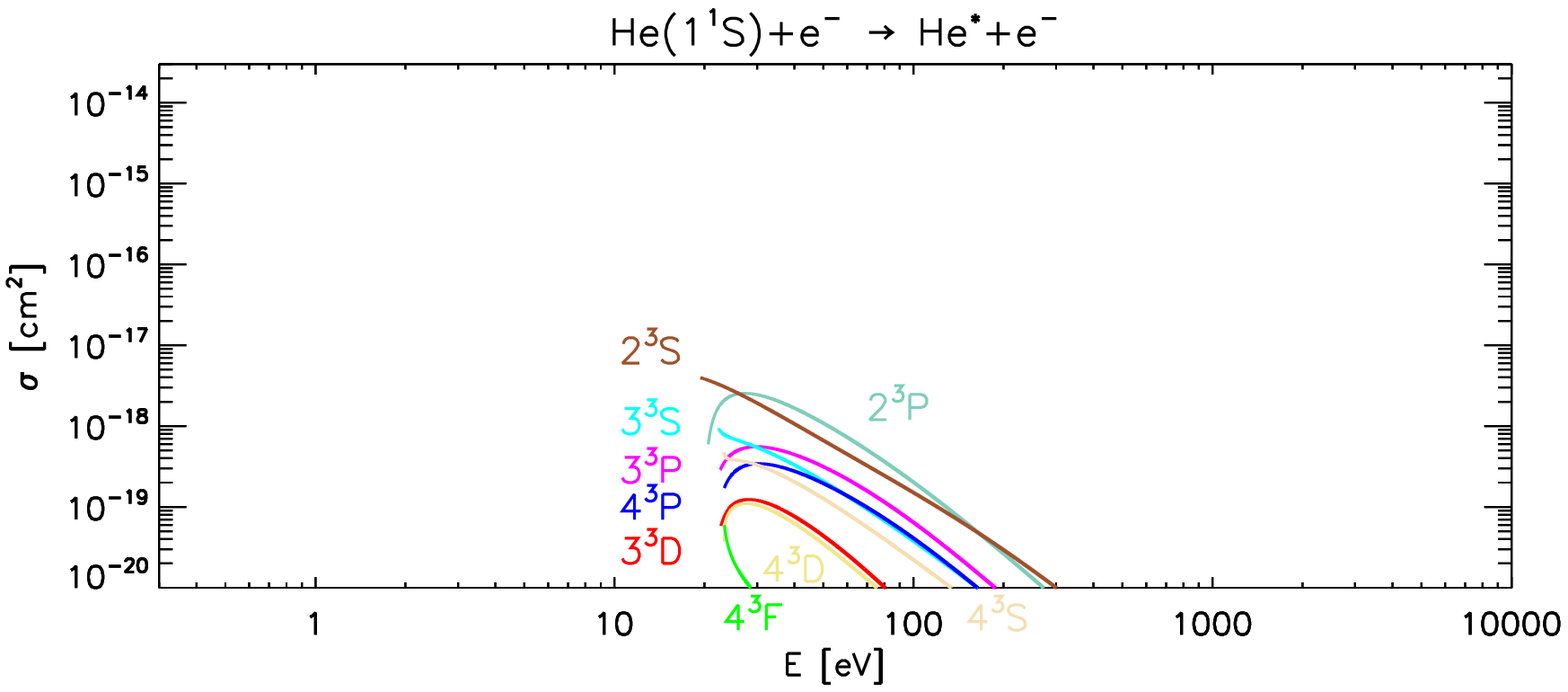} \\ 
        
    \caption{Momentum transfer, (de)excitation and ionization cross sections for channels involving ground levels H(1) and He(1$^1S$). The deexcitation data for H(2$s$) and H(2$p$) (second graph from the top) are shown in short dashed lines with the same color as the associated excitation data.
    The top graph includes the electron-electron cross sections from  Eq. \ref{sigmaee_eq} for $f_{\rm{ee}}$=0.05 and some values of $T_{\rm{e}}$ and $n_{\rm{e}}$. 
    Plots based on the energy grid of bin size equal to 0.01 eV.
    } 
    \label{xs_fig1}
\end{figure}

\begin{figure}[b]
\centering
\includegraphics[width=13cm]{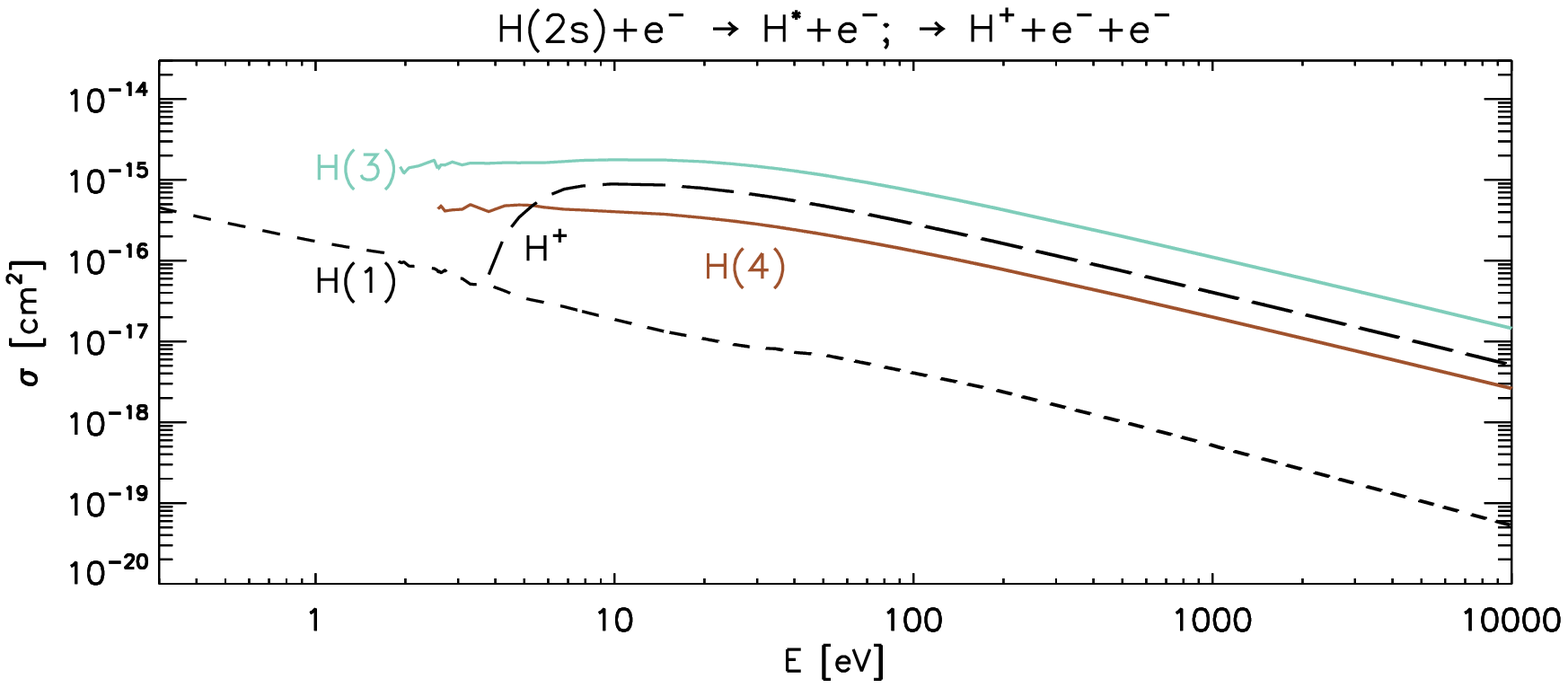} \\ 
\vspace{-0.2cm}
\includegraphics[width=13cm]{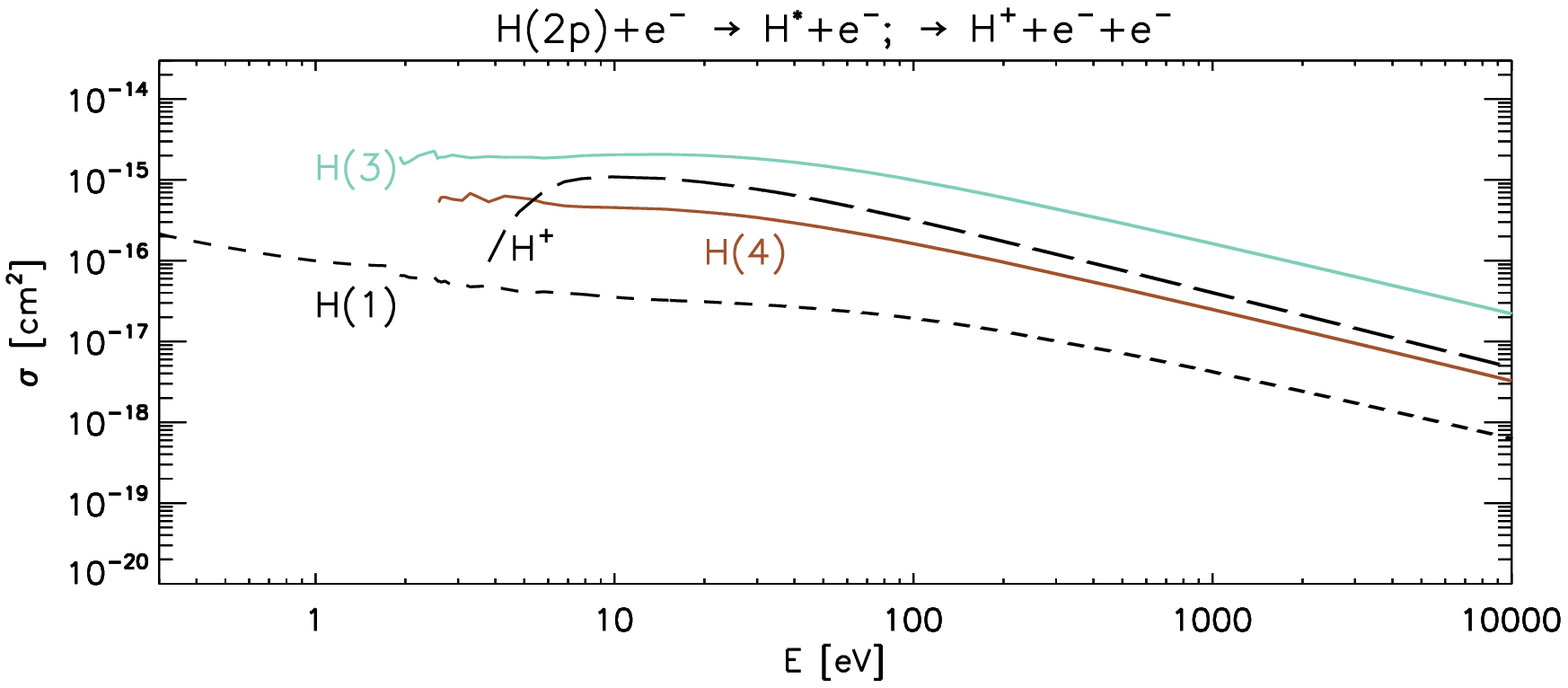} \\ 
        
    \caption{
    (De)excitation and ionization cross sections for channels involving  excited levels H(2$s$) and H(2$p$).
    The deexcitation data are shown in short dashed lines. Plots based on the energy grid of bin size equal to 0.01 eV.   
    } 
    \label{xs_fig2}
\end{figure}

The implemented cross sections were obtained by sampling the original data over uniform grids of bin sizes $\Delta_g E$=0.01, 0.1 and 1 eV. The MC model uses the piecewise description   $\sigma$($E$)=$\sigma$($E_i$) for $E${$\in$}[{$E_i$}{$-$}{$\Delta_g E$/2},$E_i$+{$\Delta_g E$/2}[, where $E_i$ is an element of the energy grid, 
and thus the bin size dictates the amount of detail that is resolved. 
We adopted the grid of bin size 1 eV as reference for most of the calculations, but 
did also a few tests with the finer grids. 
We found that the results obtained with the three grids deviated by typically less than a few percent.
The current implementation of
the MC scheme assumes a uniform energy grid, but it 
could easily be modified to use any format. 
Because at each simulated collision the MC scheme must look up the relevant cross sections at the specified energy, 
it seems reasonable to use
energy grids that are describable analytically and are quick to search. 
Grids whose bin sizes increase with energy following a power or logarithmic law
might represent good options, but we did not explore them yet. One such energy grid would ideally resolve in detail the lower energies where the cross sections exhibit some structure while relaxing the memory requirements in the model by taking larger bin sizes at high energies where the cross sections exhibit little structure.
The implemented cross sections are shown in Figs. \ref{xs_fig1}-\ref{xs_fig2} of this Appendix.
\\

The outcome of an ionization collision is two photoelectrons. 
To sample the energy $E''$ of the slowest of the two post-collision photoelectrons, we used probability distributions \citep{opaletal1971,shull1979}: 
\begin{equation*}
    P(E'' ; E')=\frac{1}{\bar{E} \arctan{((E'-{\rm{IP}}(X))/2\bar{E}})}
    \frac{1}{1+(E''/ \bar{E})^2},
\end{equation*}
which satisfy the normalization $\int_0^{(E'-{\rm{IP}}(X))/2}
P(E'' ; E') dE''=$1 and admit analytical integration. As usual $E'$ refers to the energy of the incident photoelectron. 
 The energy $E'''$ of the fastest of the two post-collision photoelectron is determined from energy conservation of the participating photoelectrons and the ionization potential of the heavy particle $X$,  $E'$+IP($X$)=$E''$+$E'''$.
 $\bar{E}$ [eV] is a constant specific to each heavy particle.

\subsection{Electron-electron collisions}

\citet{swartzetal1971} reported a fit to the loss function $L_{ee}$($E$) for the deposition of photoelectron energy in collisions with thermal electrons. 
We derived pseudo-discrete cross sections for electron-electron collisions
\citep{nagybanks1970} from the equalities $-${$dE$}/{$dx$}=$L_{ee}$($E$){$n_{\rm{e}}$}={$\sigma_{\rm{ee}}$}{$\Delta E_{\rm{ee}}$}{$n_{\rm{e}}$}. 
For the energy transferred by the photoelectrons to the thermal electrons, 
we take a constant fraction  of the incident energy,
{$\Delta E_{\rm{ee}}$}=$f_{\rm{ee}}${$E$} 
\citep{habinggoldsmith1971,shull1979}. 
We used $f_{\rm{ee}}$=0.05, but 
confirmed by running a few examples with $f_{\rm{ee}}$ in the range 0.01--0.2 the finding of past authors \citep{habinggoldsmith1971,shull1979,furlanettostoever2010}
that the calculations are not strongly sensitive to $f_{\rm{ee}}$.
The reason for this is possibly that the slowing down of the photoelectrons at small energies is dominated by the product of 
$\sigma_{\rm{ee}}$ and {$\Delta E_{\rm{ee}}$}, which remains independent of $f_{\rm{ee}}$, rather than by either $\sigma_{\rm{ee}}$ or {$\Delta E_{\rm{ee}}$}.
Summarizing, the cross sections for electron-electron collisions are calculated from: 
\begin{equation}
    \sigma_{\rm{ee}}(E; n_{\rm{e}}, T_{\rm{e}}, f_{\rm{ee}}) \; {[\rm{cm}^2]} =\frac{3.37\times10^{-12}}{f_{\rm{ee}} E^{1.94} n^{0.03}_{\rm{e}}} \left( \frac{E-8.618\times10^{-5} T_{\rm{e}}}{E-4.568\times10^{-5} T_{\rm{e}}} \right)^{2.36},
\label{sigmaee_eq}    
\end{equation}
where $E$ [eV] specifies the incident photoelectron energy and 
$T_e$ [K] is the temperature of the thermal electrons.\\

A remark is due on the dependence of $\sigma_{\rm{ee}}$ on the number density of the thermal electrons
$n_{\rm{e}}$. 
The numerical solution from the MC scheme to the slowing down of photoelectrons involves comparing 
at each simulated collision
the probabilities $p_j$ introduced in \S\ref{subsec:cascadesimulation}.
This is equivalent to comparing products of the number densities of the
relevant target particles and their cross sections. If all the cross sections are independent of the number densities, the solution will only depend on the ratio of the number densities and not on their absolute values. 
$\sigma_{\rm{ee}}$ is the only cross section in the implementation that depends on an absolute density. 
This dependence is however very weak and to most practical effects it is justified to think that the MC solutions remain dictated by the ratios of number densities.

\subsection{Electron-H atom collisions}

We considered elastic and inelastic collisions with the ground level H(1). 
We also considered a partial set of channels for collisions 
with the excited sublevels H(2$s$) and H(2$p$) that result in their (de)excitation and ionization.
 We adopted IP(H)=13.6057 eV, IP(H)(1$-$1/$n^2$) for the energy of bound levels with principal quantum number $n$, and $\bar{E}$(H)=8 eV in the description of collisional ionization.\\

For momentum transfer in elastic collisions with H(1), we adopted the cross sections reported by \citet{itikawa1974} below  10 eV and \citet{dalgarnoetal1999} from there to 1 keV, extrapolating at higher energies. 
For ionization and excitation into bound levels with $n${$\le$}4, we borrowed from ALADDIN\footnote{https://www-amdis.iaea.org/ALADDIN/} the cross sections reported by \citet{braystelbovics1995} below 1 keV and \citet{janevsmith1993} at higher energies. 
We kept separate cross sections for excitation into H(2$s$) and H(2$p$), but summed over the orbital quantum number the cross sections for excitation into levels with $n$=3-4. 
For excitation into levels with $n$=5-10, we adopted the cross sections by \citet{janevsmith1993} (summed over the orbital quantum number in their original format) for all energies. 
Further, we lumped into a super-level the levels with $n$=6--10 by 
adding up their cross sections.
Summarizing, the upper levels that are explicitly resolved on the list of collisional channels 
are those with $n$=1, 3, 4 and 5, the sublevels 2$s$ and 2$p$, and the high-energy pseudo-level.
\\

We adopted from ALADDIN the ionization cross sections for sublevels 
H(2$s$) and H(2$p$) reported by \citet{braystelbovics1995},  extrapolating them above 1 keV. 
We also collected the cross sections for their excitation into H(3) and H(4), which are comparable in magnitude to the ionization cross sections. 
Lacking other information, we adopted $\bar{E}$(H(2$s$))=$\bar{E}$(H(2$p$))=0.6$\times$IP(H(2)){$\approx$}2.04 eV \citep{kozmafransson1992}. 
This choice has no practical consequence as the production of photoelectrons from collisional ionization of the excited levels is typically negligible.

\subsection{Electron-He atom collisions}

We considered elastic and inelastic collisions with the ground level He(1$^1S$), assuming that excitation  occurs within the neutral atom and that the resulting singly He$^+$ is in its ground level.
We  omitted double ionization of He(1$^1S$) into He$^{++}$ because the cross sections are small \citep{genevriezetal2019}.
We also omitted inelastic collisions with He$^{+}$, whose influence is negligible \citep{dalgarnoetal1999};  indeed, their cross sections are smaller than those for neutral He and for moderate fractional ionizations the slowing down of the photoelectrons rapidly becomes dominated by electron-electron collisions, at least at small and moderate energies.
We adopted IP(He)=24.5874 eV from  NIST \citep{kramidaetal2021}, and 
used the NORAD database \citep{nahar2010,nahar2020} for the energies of the bound levels.
\\

We adopted the momentum transfer cross sections below 10 eV reported by \citet{itikawa1978} and \citet{dalgarnoetal1999} from 10 eV to 1 keV, extrapolating at higher energies.
For excitation, we implemented separately the cross sections recommended by \citet{ralchenkoetal2008} for each dipole-allowed, dipole-forbidden and spin-forbidden channel that connects the 
ground level He(1$^1S$) with any of the following excited levels:
 2$^1P$, 3$^1P$, 4$^1P$, 2$^1S$, 3$^1S$, 3$^1D$, 4$^1S$, 4$^1D$, 4$^1F$, 2$^3P$, 2$^3S$, 3$^3D$, 3$^3P$, 3$^3S$, 4$^3D$, 4$^3F$, 4$^3P$, 4$^3S$. For collisional channels that connect the ground level with an excited level of principal quantum number  $n${$\ge$}5, \citet{ralchenkoetal2008} propose cross sections that scale as $(4/n)^3$. We do not explicitly consider such high-energy levels, but take them into account by multiplying by {$4^3$}{$\sum_{n=5}^{\infty} n^{-3}$}{$\approx$1.56} the excitation cross section of each channel ending at $n$=4.
We borrowed the ionization cross sections for He(1$^1S$)
from \citet{ralchenkoetal2008}, and adopted  $\bar{E}$(He(1$^1S$))=15.8 eV from \citet{opaletal1971}. 
\\

\newpage

\clearpage
\section{\label{validation_appendix} Validation material}

This Appendix collects an extensive set of calculations for validation of the multi-score MC scheme. Each table/figure refers to the original reference and model setup. 
In the preparation of this material, 
we have in general retained the nomenclature used in the original reference even if it differs from the nomenclature used elsewhere in this work. When needed, we digitized the figures from the original references using  WebPlotDigitizer \citep{rohatgi2020}. 

\begin{table}[h]
\caption{Fractional energy depositions at $E_0$=2 keV calculated with the multi-score MC scheme for a gas of H atoms and thermal electrons. 
The calculations may be compared to those in Table 1 of \citet{xumccray1991}.
The table reports separately the depositions into H(2$s$) and H(2$p$), but merges the energy deposited into H($n${$\ge$6}). 
}             
\label{xumccray1991_table1}      
\centering                          
\begin{tabular}{c c c c c c c c c c}        
\hline
$\chi_e$ & 10$^{-10}$ & 10$^{-4}$ & 10$^{-3}$  & 10$^{-2}$ & 0.05 & 0.1 & 0.5 & 1 & 10  \\    
\hline
$\eta_h$      & 1.13($-$1) & 1.29($-$1) & 1.84($-$1) & 3.44($-$1) & 5.72($-$1) & 6.91($-$1) & 8.98($-$1) & 9.44($-$1) & 9.93($-$1) \\ 
$\eta_c$      & 3.66($-$1) & 3.65($-$1) & 3.53($-$1) & 2.94($-$1) & 1.93($-$1) & 1.40($-$1) & 4.61($-$2) & 2.55($-$2) & 2.96($-$3) \\  
$\eta_{2s}$   & 3.52($-$2) & 3.17($-$2) & 2.40($-$2) & 1.43($-$2) &	7.94($-$3) & 5.48($-$3) & 1.69($-$3) & 9.28($-$4) &	1.07($-$4) \\  
$\eta_{2p}$   & 3.44($-$1) & 3.36($-$1) & 3.10($-$1) & 2.47($-$1) &	1.62($-$1) & 1.17($-$1)	& 3.87($-$2) & 2.14($-$2) &	2.49($-$3) \\
$\eta_{3}$    & 7.69($-$2) & 7.55($-$2)	& 6.97($-$2) & 5.45($-$2) &	3.51($-$2) & 2.53($-$2)	& 8.29($-$3) & 4.58($-$3) &	5.33($-$4) \\
$\eta_{4}$    & 2.84($-$2) & 2.80($-$2) & 2.60($-$2) & 2.04($-$2) &	1.31($-$2) & 9.42($-$3) & 3.09($-$3) & 1.70($-$3) &	1.98($-$4) \\
$\eta_{5}$    & 1.75($-$2) & 1.72($-$2) & 1.57($-$2) & 1.14($-$2) & 6.83($-$3) & 4.82($-$3) & 1.54($-$3) & 8.45($-$4) &	9.79($-$5) \\  
$\eta_{6-10}$ & 1.85($-$2) & 1.84($-$2) & 1.78($-$2) & 1.47($-$2) &	9.52($-$3) & 6.85($-$3) & 2.24($-$3) & 1.23($-$3) &	1.44($-$4) \\  
 
\hline                        

\end{tabular}
\end{table}

\begin{figure}[h]
    \centering
    \includegraphics[width=8.9cm]{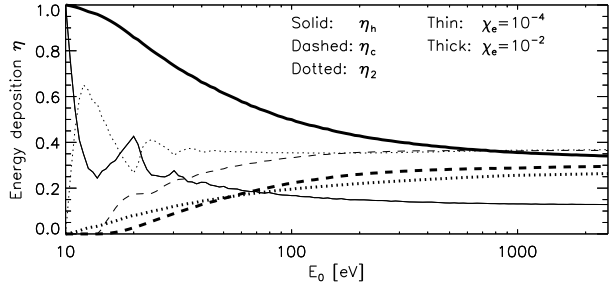}
    \caption{Fractional energy depositions calculated with the multi-score MC scheme for a gas of H atoms at two fractional ionizations. The calculations may be compared to those in Figs. 4-5 of \citet{xumccray1991}.
    } 
    \label{xumccray1991_figures45}
\end{figure}

\vspace{+0.5cm}

\begin{figure}[h]
    \centering
    \includegraphics[width=8.9cm]{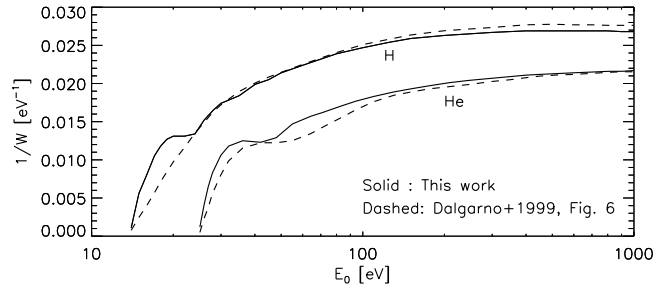}
    \caption{Number of ions (H$^+$ or He$^+$) per incident energy $E_0$ produced in a pure neutral gas of H or He atoms.
    \textit{\underline{Note.}} The quantities shown in the vertical axis are equivalent to $\Phi_{\rm{H}(1) \rightarrow \rm{H}^+}$/$E_0$
    and $\Phi_{\rm{He}(1^1S) \rightarrow \rm{He}^+}$/$E_0$ as defined in our work.
    } 
    \label{dalgarnoetal1999_fig6}
\end{figure}

\newpage


\begin{figure}[h]
    \centering
    \includegraphics[width=8.9cm]{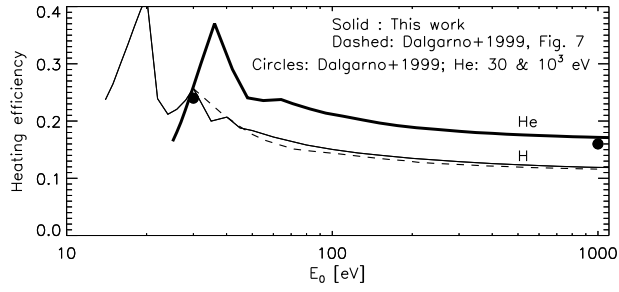}
    \caption{Heating efficiency for a pure neutral gas of H or He atoms.} 
    \label{dalgarnoetal1999_fig7}
\end{figure}

\vspace{+0.5cm}

\begin{figure}[h]
    \centering
    \includegraphics[width=8.9cm]{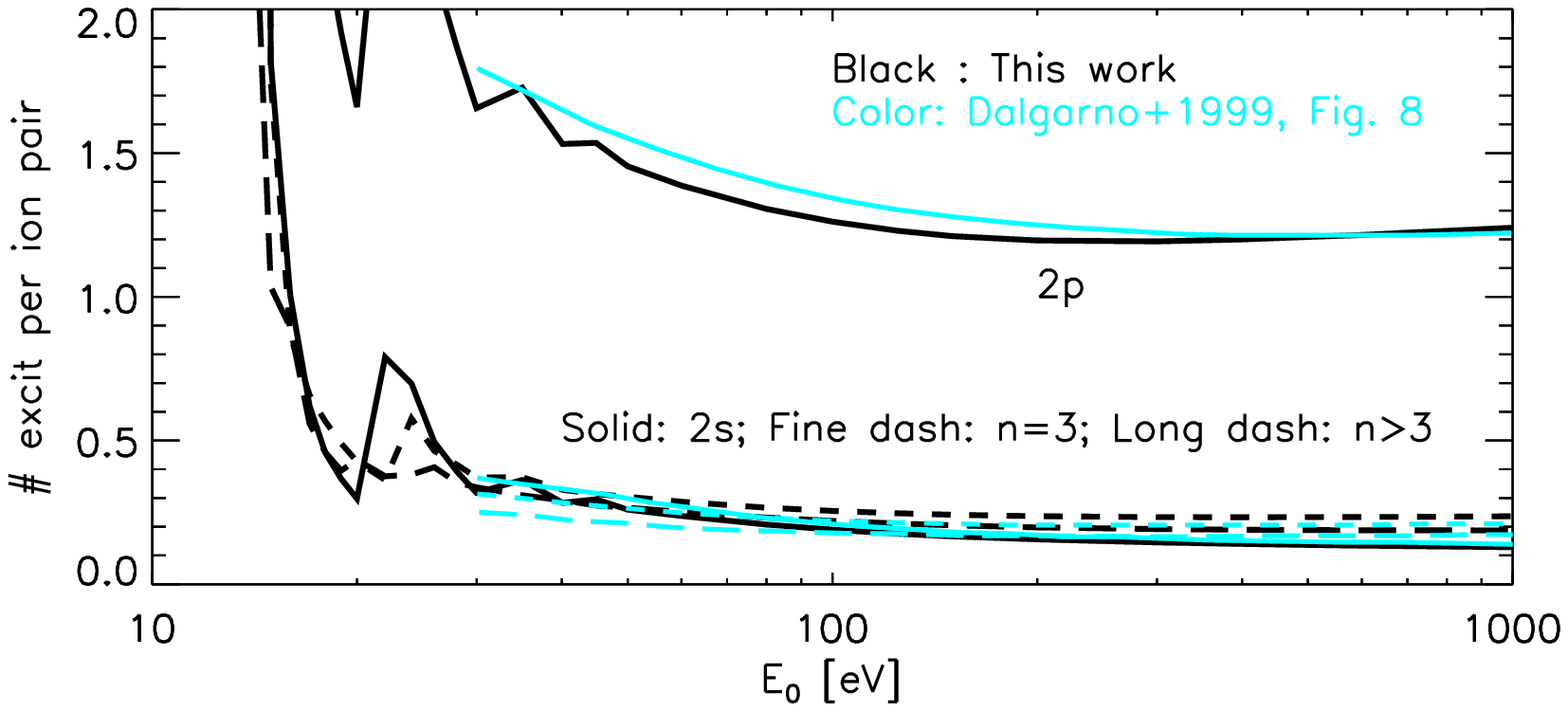}
    \caption{Number of excitations per ion pair for a pure neutral gas of H atoms.
    \textit{\underline{Note.}} The quantity shown in the vertical axis is equivalent to
    $\Phi_{\rm{H}(1) \rightarrow \rm{H}^*}$/$\Phi_{\rm{H}(1) \rightarrow \rm{H}^+}$ as defined in our work.    
    } 
    \label{dalgarnoetal1999_fig8}
\end{figure}

\vspace{+1.0cm}

\begin{figure}[h]
    \centering
    \includegraphics[width=8.9cm]{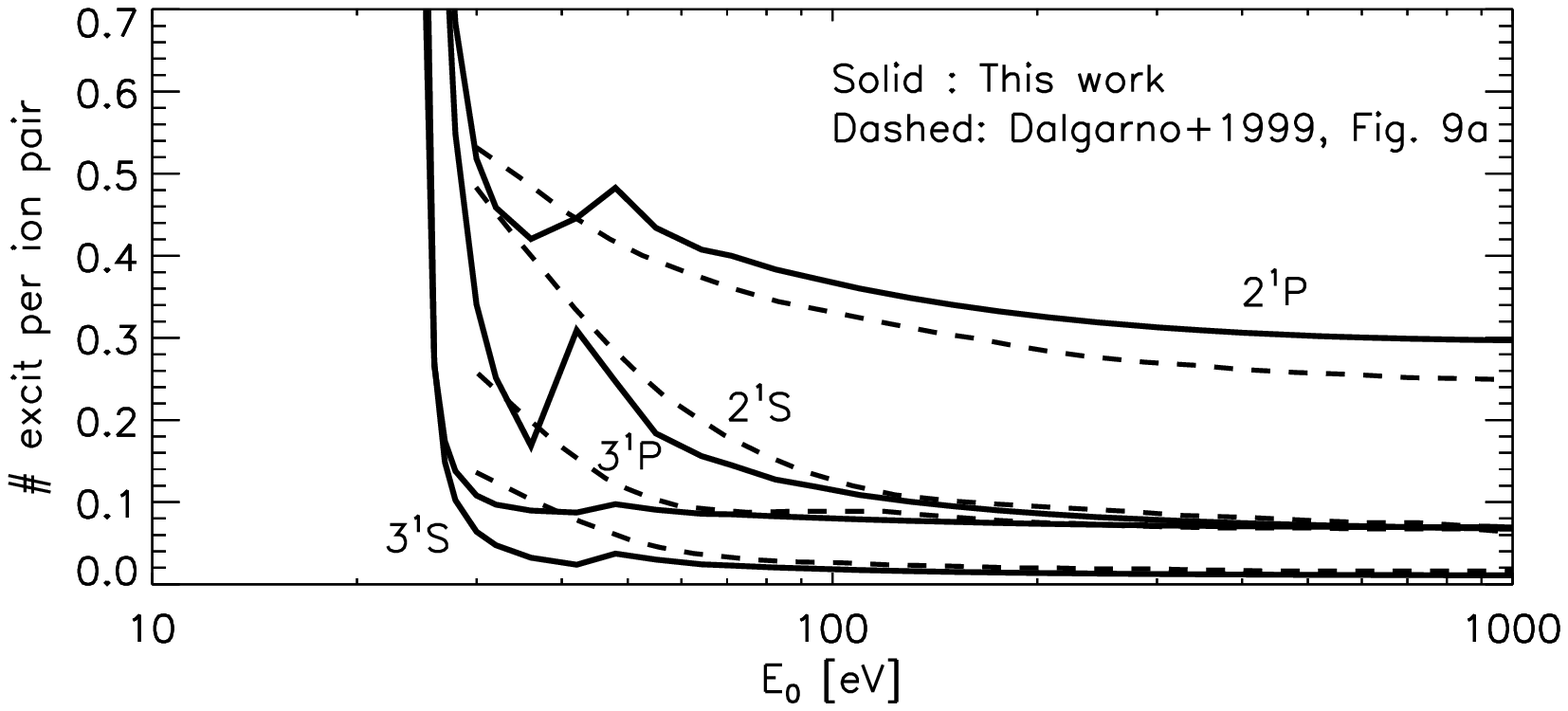} 
    \includegraphics[width=8.9cm]{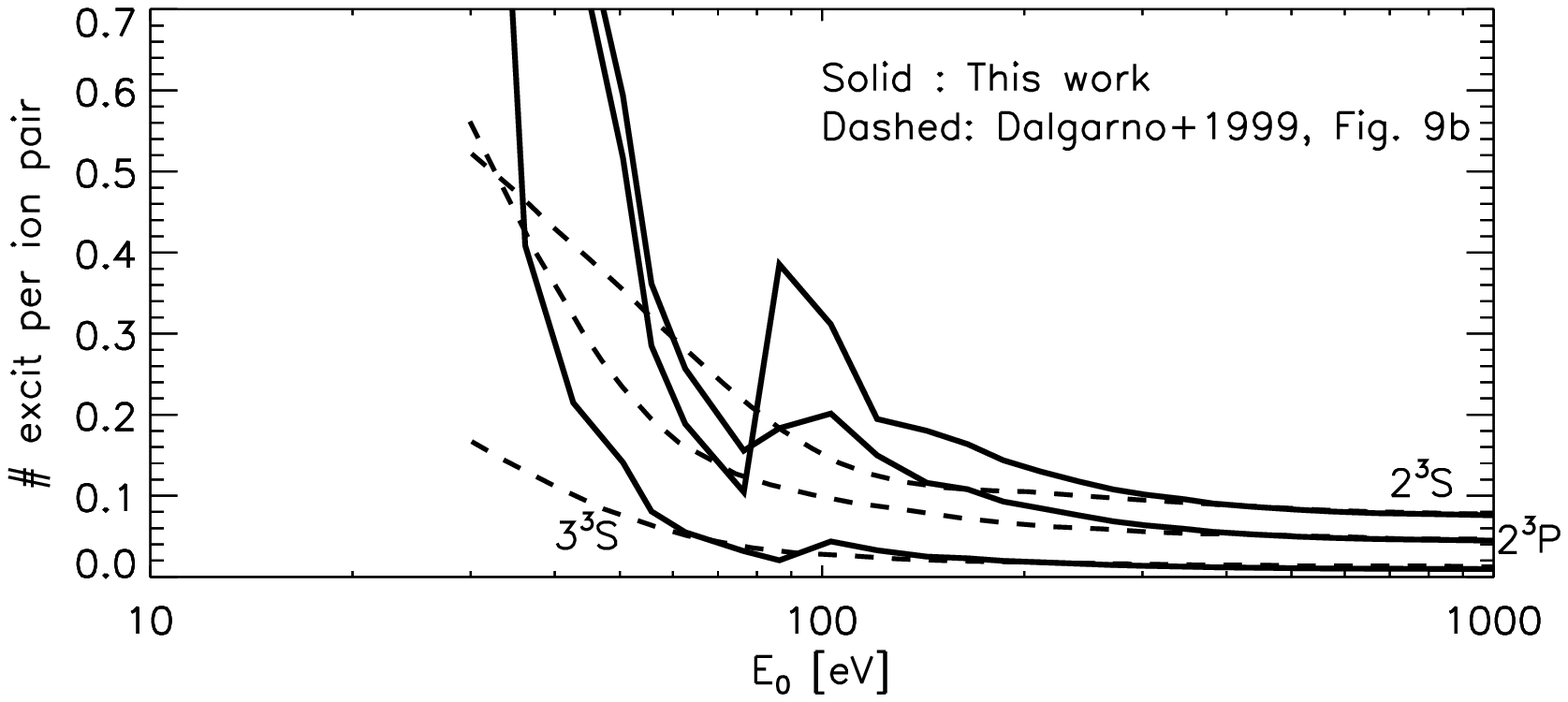} \\
    \caption{Number of excitations per ion pair for a pure neutral gas of He atoms.
    \textit{\underline{Note.}} The quantities shown in the vertical axis are equivalent to
    $\Phi_{\rm{He}(1^1S) \rightarrow He^*}$/$\Phi_{\rm{He}(1^1S) \rightarrow \rm{He}^+}$ as defined in our work.} 
    \label{dalgarnoetal1999_fig9}
\end{figure}

\vspace{+1.0cm}

\begin{figure}[h]
    \centering
    \includegraphics[width=8.9cm]{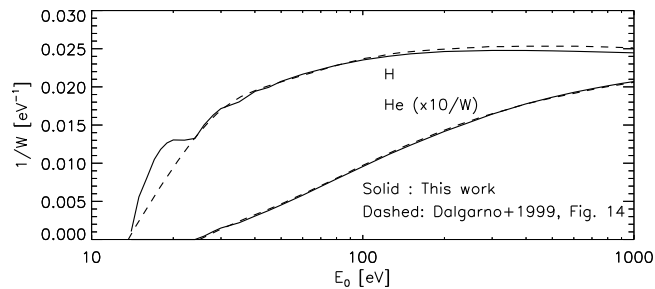}
    \caption{Number of ions (H$^+$ and He$^+$) per incident energy $E_0$ produced in a neutral H:0.1He mixture of atoms.
\textit{\underline{Note.}} The quantities shown in the vertical axis are equivalent to 
$\Phi_{\rm{H}(1) \rightarrow \rm{H}^+}$/$E_0$ and $\Phi_{\rm{He}(1^1S) \rightarrow \rm{He}^+}$/$E_0$  
as defined in our work.    
    } 
    \label{dalgarnoetal1999_fig14}
\end{figure}

\begin{figure}[h]
    \centering
    \includegraphics[width=8.9cm]{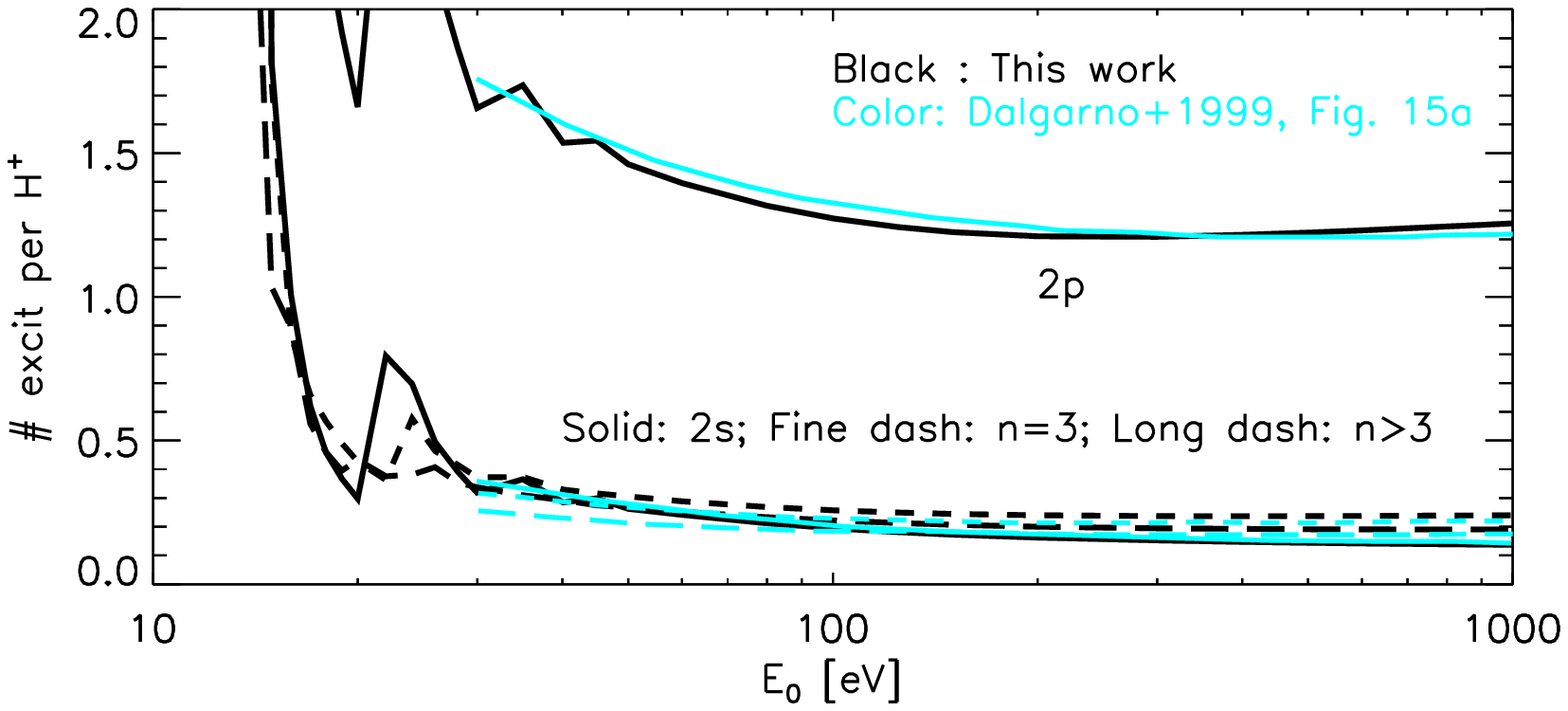}
    \caption{Number of excitations per H$^+$ for a neutral H:0.1He mixture of atoms.
\textit{\underline{Note.}} The quantity shown in the vertical axis is equivalent to
    $\Phi_{\rm{H}(1) \rightarrow \rm{H}^*}$/$\Phi_{\rm{H}(1) \rightarrow \rm{H}^+}$ as defined in our work.     
    } 
    \label{dalgarnoetal1999_fig15a}
\end{figure}

\vspace{+2.0cm}

\begin{figure}[h]
    \centering
    \includegraphics[width=8.9cm]{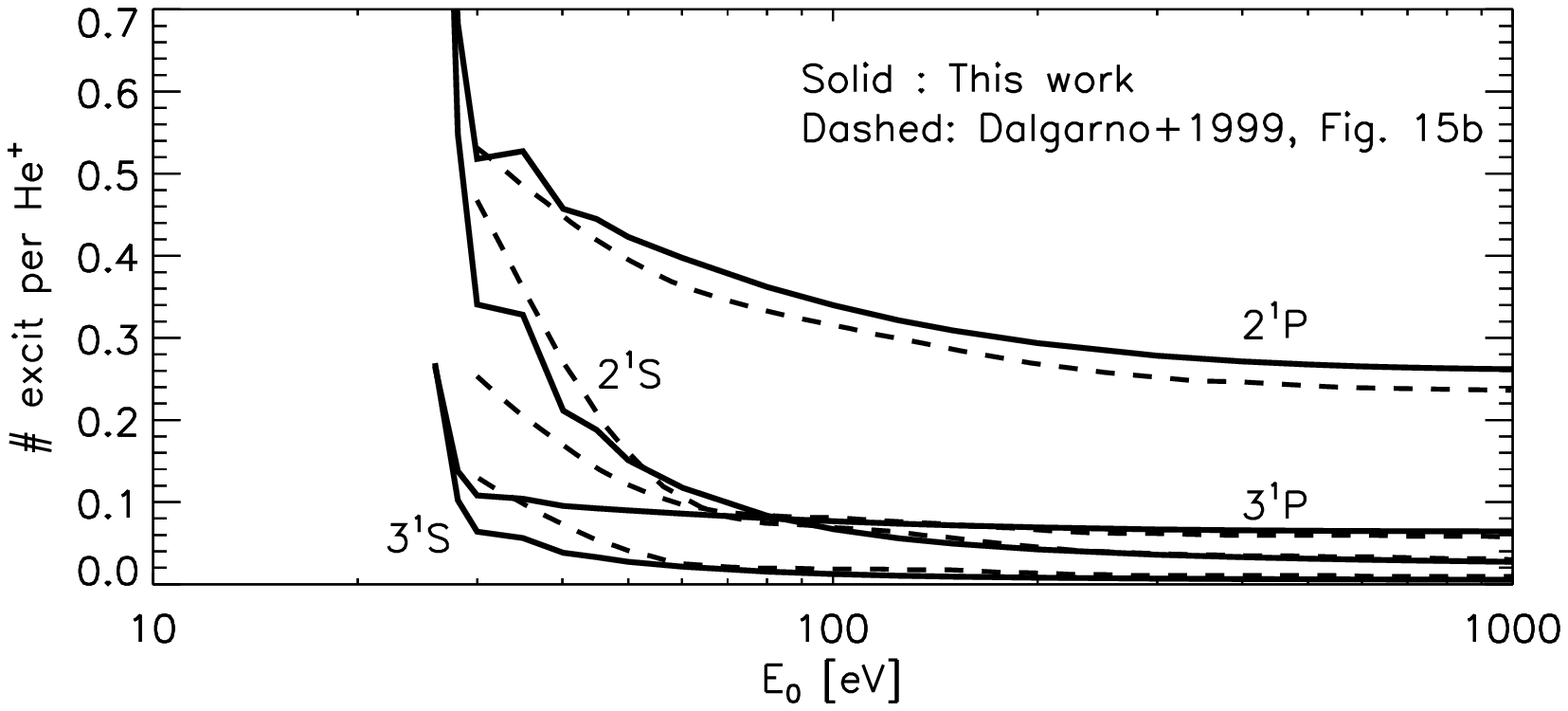} 
    \includegraphics[width=8.9cm]{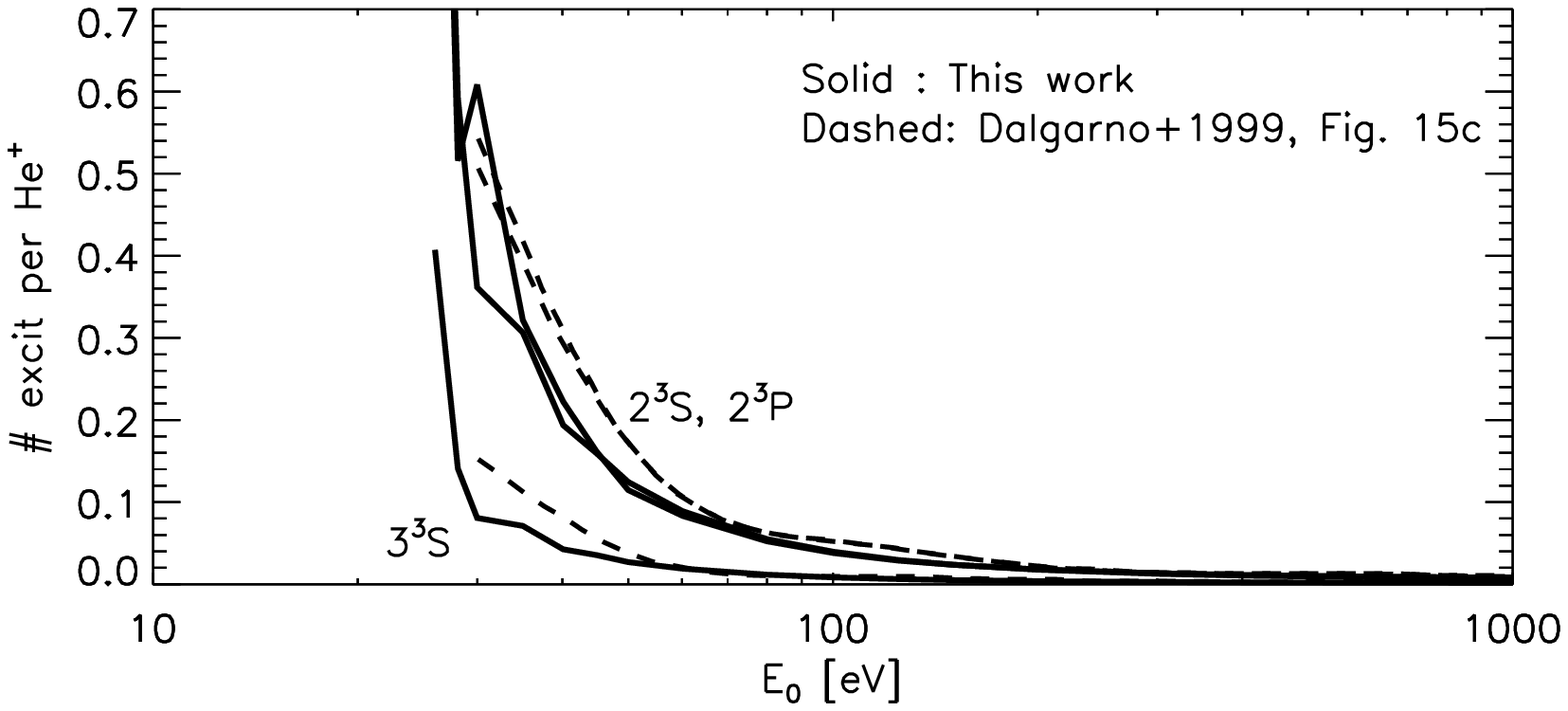} \\
    \caption{Number of excitations per He$^+$ for a neutral H:0.1He mixture of atoms.
    \textit{\underline{Note.}} The quantity shown in the vertical axis is equivalent to
    $\Phi_{\rm{He}(1^1S) \rightarrow \rm{He}^*}$/$\Phi_{\rm{He}(1^1S) \rightarrow \rm{He}^+}$ as defined in our work.    
    } 
    \label{dalgarnoetal1999_fig15bc}
\end{figure}

\clearpage

\begin{table}[h]
\caption{Top. H excitation yields per H$^+$ for a neutral H:0.1He mixture of atoms as obtained here and as reported in Table 1 of \citet{dalgarnoetal1999}. These quantities are equivalent to
    $\Phi_{\rm{H}(1) \rightarrow H^*}$/$\Phi_{\rm{H}(1) \rightarrow \rm{H}^+}$ as defined in our work.
Bottom. Heating efficiencies for a neutral 
H:0.1He mixture of atoms as obtained here and as reported in Table 3 of \citet{dalgarnoetal1999}
}             
\label{dalgarnoetal1999_table1}      
\centering                          
\begin{tabular}{c c c c c c c c}        
\hline
& \multicolumn{6}{c}{Energy} \\
\hline
& \multicolumn{6}{c}{Excitation yields} \\
\hline
Levels & 30 eV & 50 eV & 100 eV & 200 eV & 500 eV & 1000 eV & Ref.  \\    
\hline
H($2s$)      & 0.318 & 0.262 & 0.194 & 0.162 & 0.144 & 0.137 & This work \\
H($2s$)      & 0.349 & 0.272 & 0.198 & 0.162 & 0.142 & 0.134 & \citet{dalgarnoetal1999} \\ 
\hline
H($2p$)      & 1.657 & 1.461 & 1.273 & 1.211 & 1.223 & 1.255 & This work \\ 
H($2p$)      & 1.750 & 1.508 & 1.323 & 1.237 & 1.205 & 1.214 & \citet{dalgarnoetal1999} \\ 
\hline
H($n$=3)     & 0.372 & 0.306 & 0.257 & 0.240 & 0.237 & 0.240 & This work\\
H($n$=3)     & 0.309 & 0.258 & 0.219 & 0.204 & 0.204 & 0.209 & \citet{dalgarnoetal1999}\\ 
\hline
H($n${$\ge$}4) & 0.337 & 0.267 & 0.222 & 0.200 & 0.191 & 0.191 & This work\\
H($n${$\ge$}4) & 0.249 & 0.207 & 0.176 & 0.164 & 0.164 & 0.170 & \citet{dalgarnoetal1999}\\ 
\hline                        
& \multicolumn{6}{c}{Heating efficiency} \\
\hline                        
     & 0.257 & 0.183 & 0.150 & 0.134 & 0.123 & 0.118 & This work\\
     & 0.266 & 0.185 & 0.150 & 0.135 & 0.124 & 0.120 & \citet{dalgarnoetal1999} \\
\hline                        
\end{tabular}
\end{table}

\clearpage

\begin{figure}[h]
    \centering
    \includegraphics[width=8.9cm]{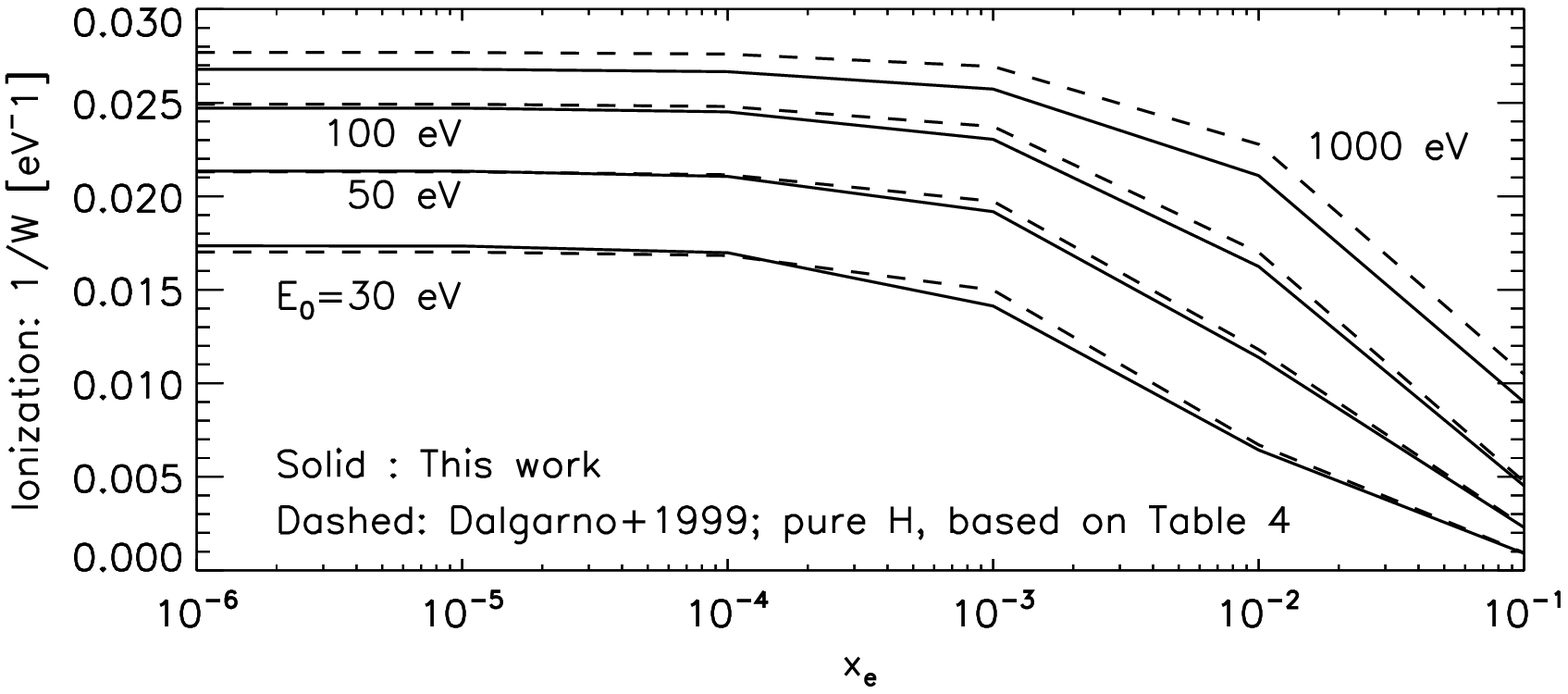}  \includegraphics[width=8.9cm]{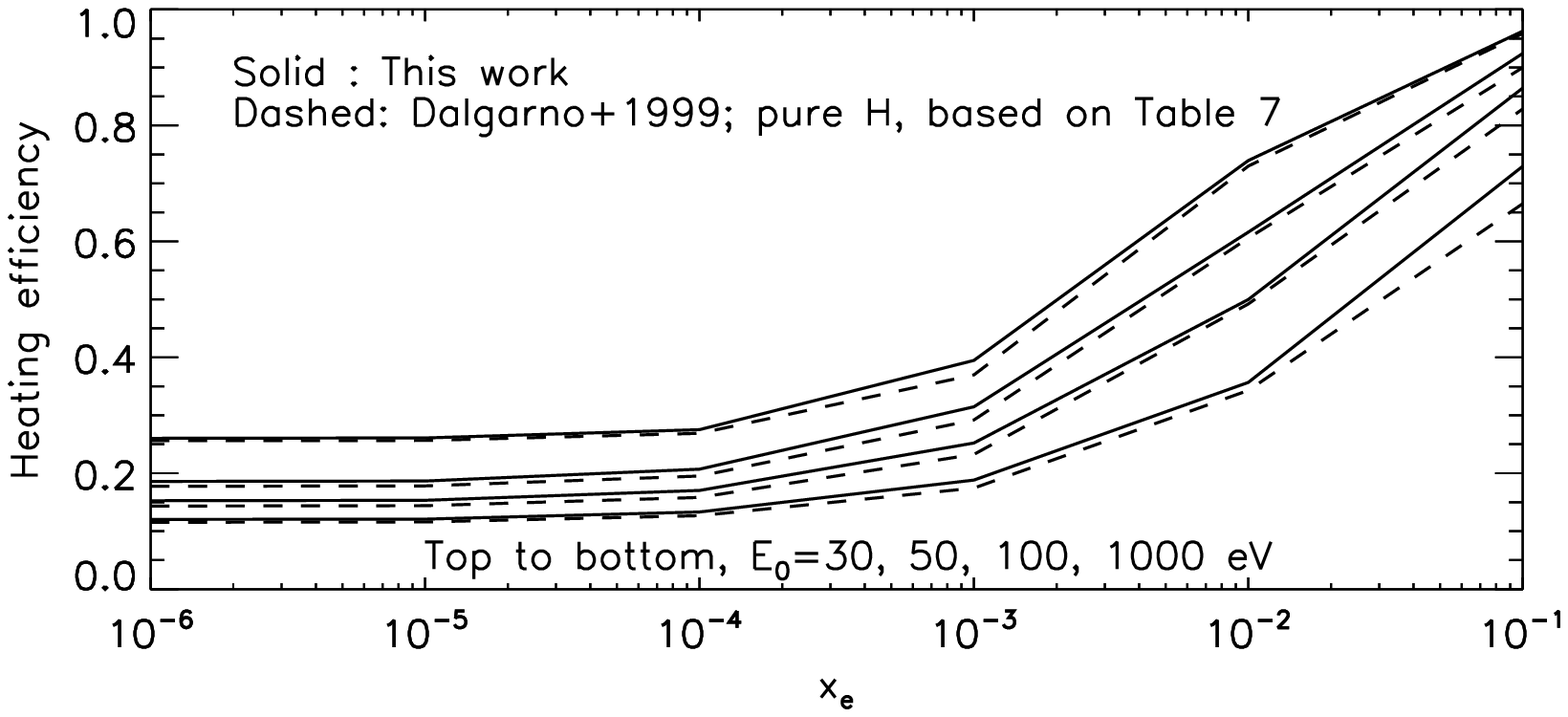} \\
    \includegraphics[width=8.9cm]{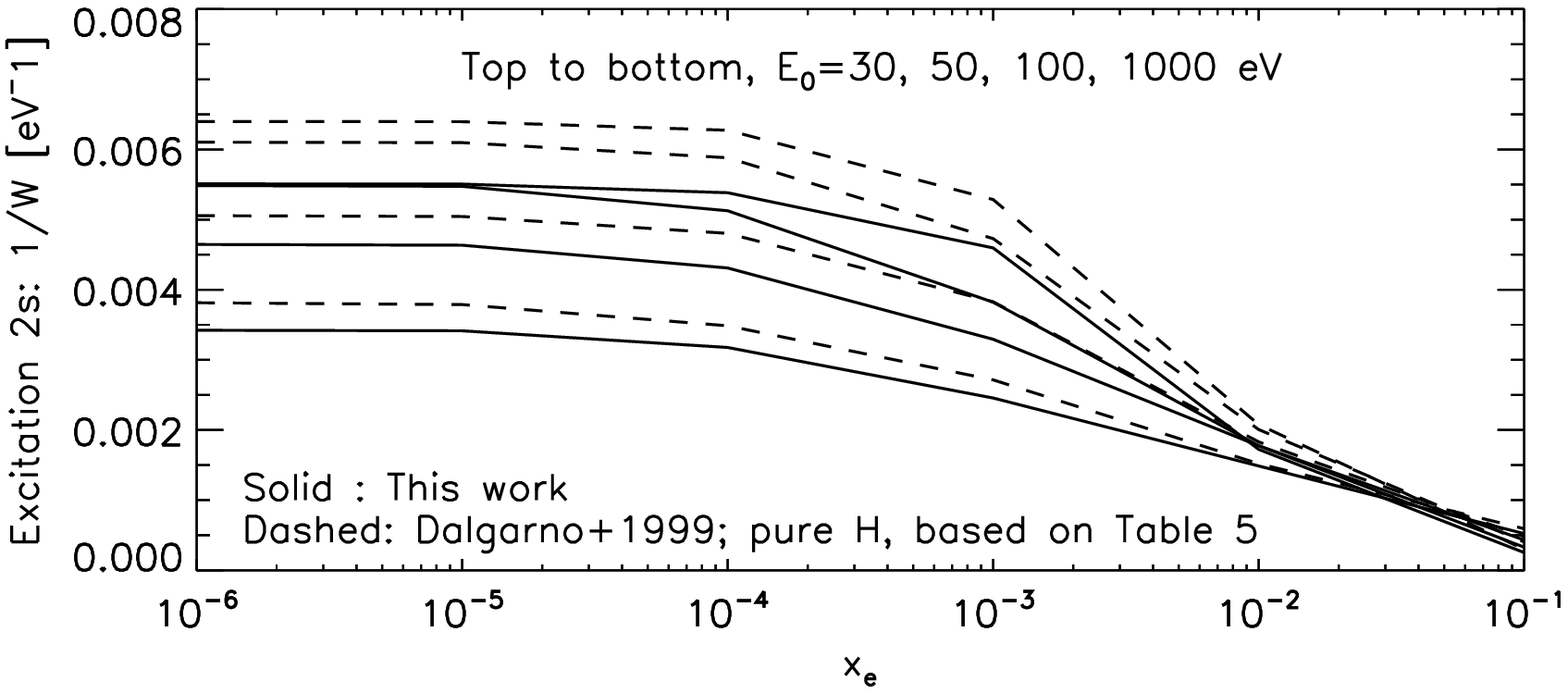}  \includegraphics[width=8.9cm]{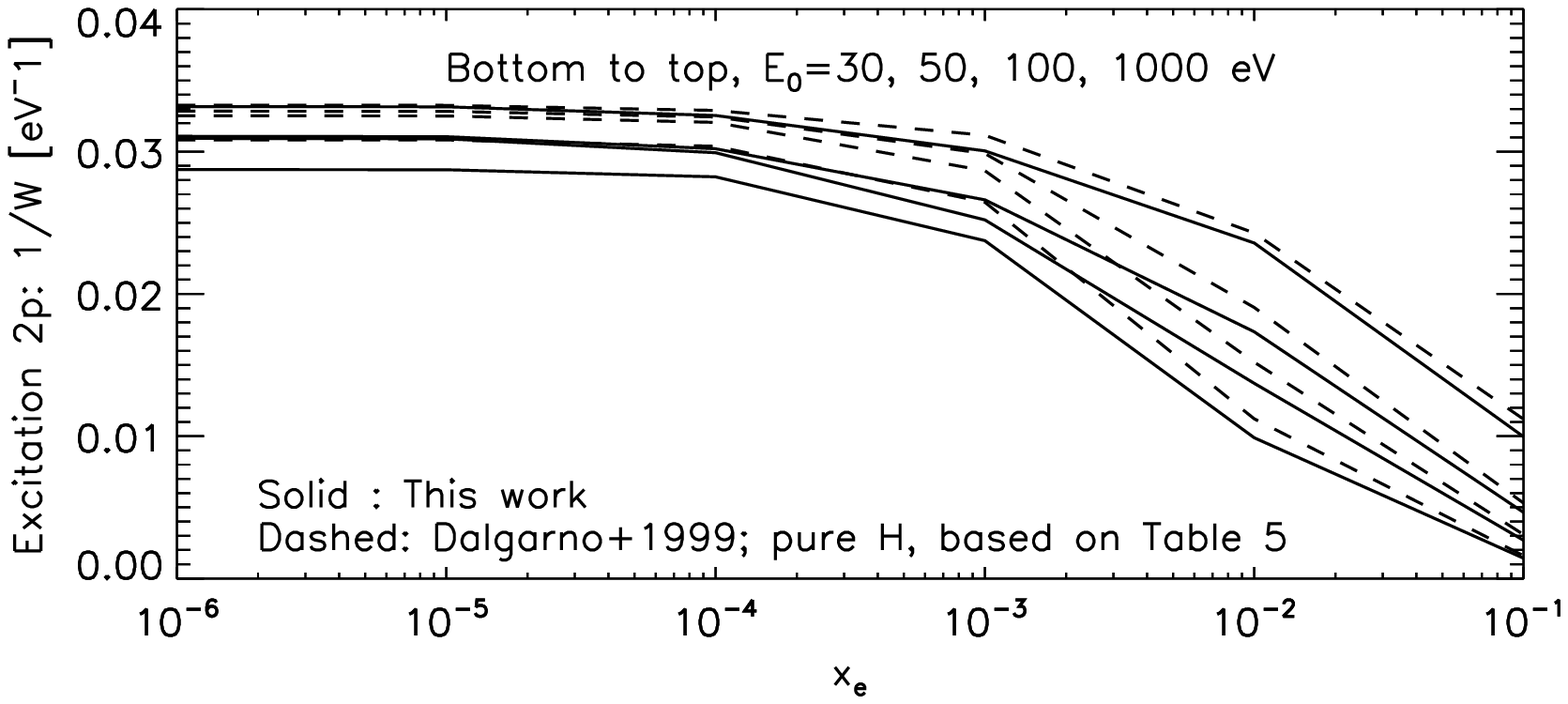} \\
    \caption{Ionization yield,  excitation yields for H(2$s$) and H(2$p$), and heating efficiency
    for a ionized gas of H atoms.
    \textit{\underline{Note.}} The yields are equivalent to
    $\Phi_{\rm{H}(1) \rightarrow \rm{H}^*}$/$E_0$ and 
    $\Phi_{\rm{H}(1) \rightarrow \rm{H}^+}$/$E_0$
    as defined in our work.        
    } 
    \label{dalgarnoetal1999_h-em}
\end{figure}

\begin{figure}[h]
    \centering
    \includegraphics[width=8.9cm]{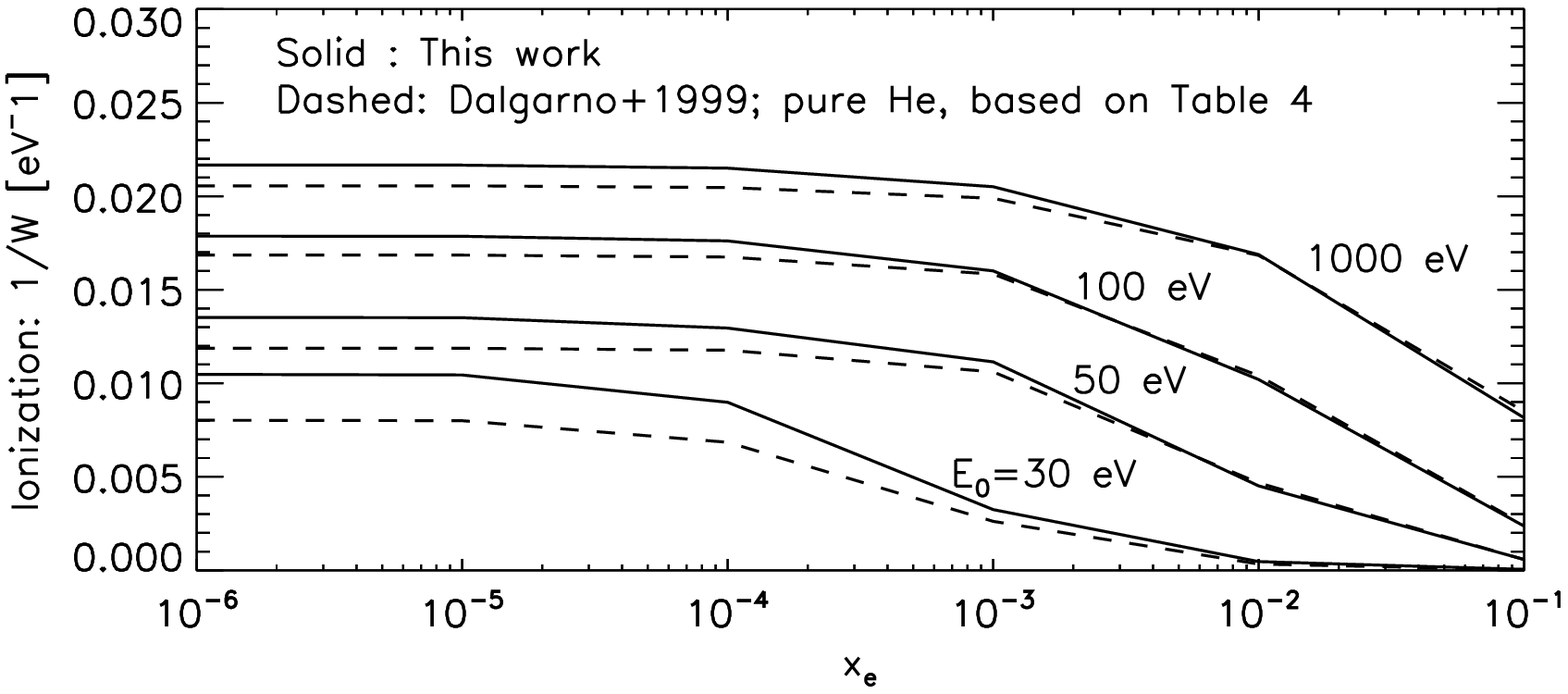}  \includegraphics[width=8.9cm]{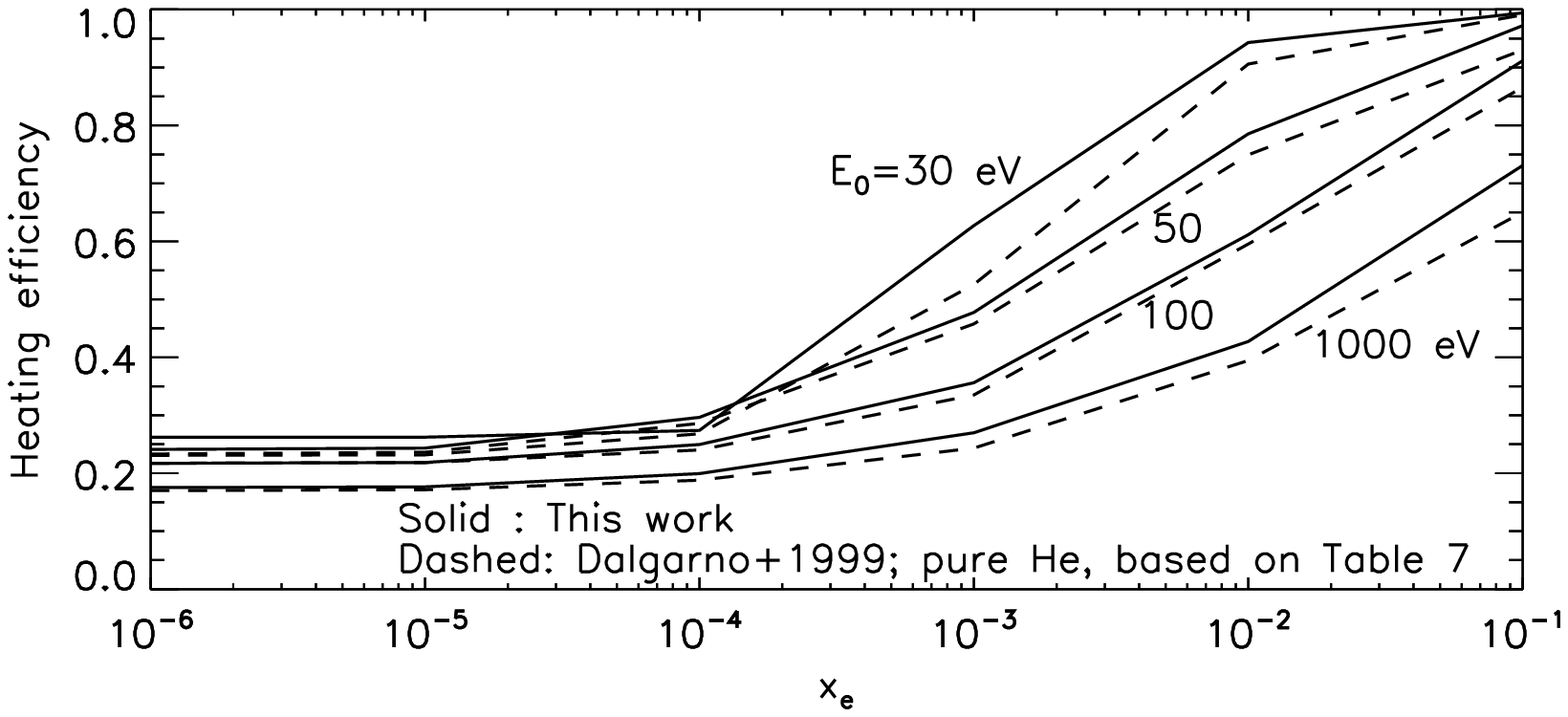} \\
    \includegraphics[width=8.9cm]{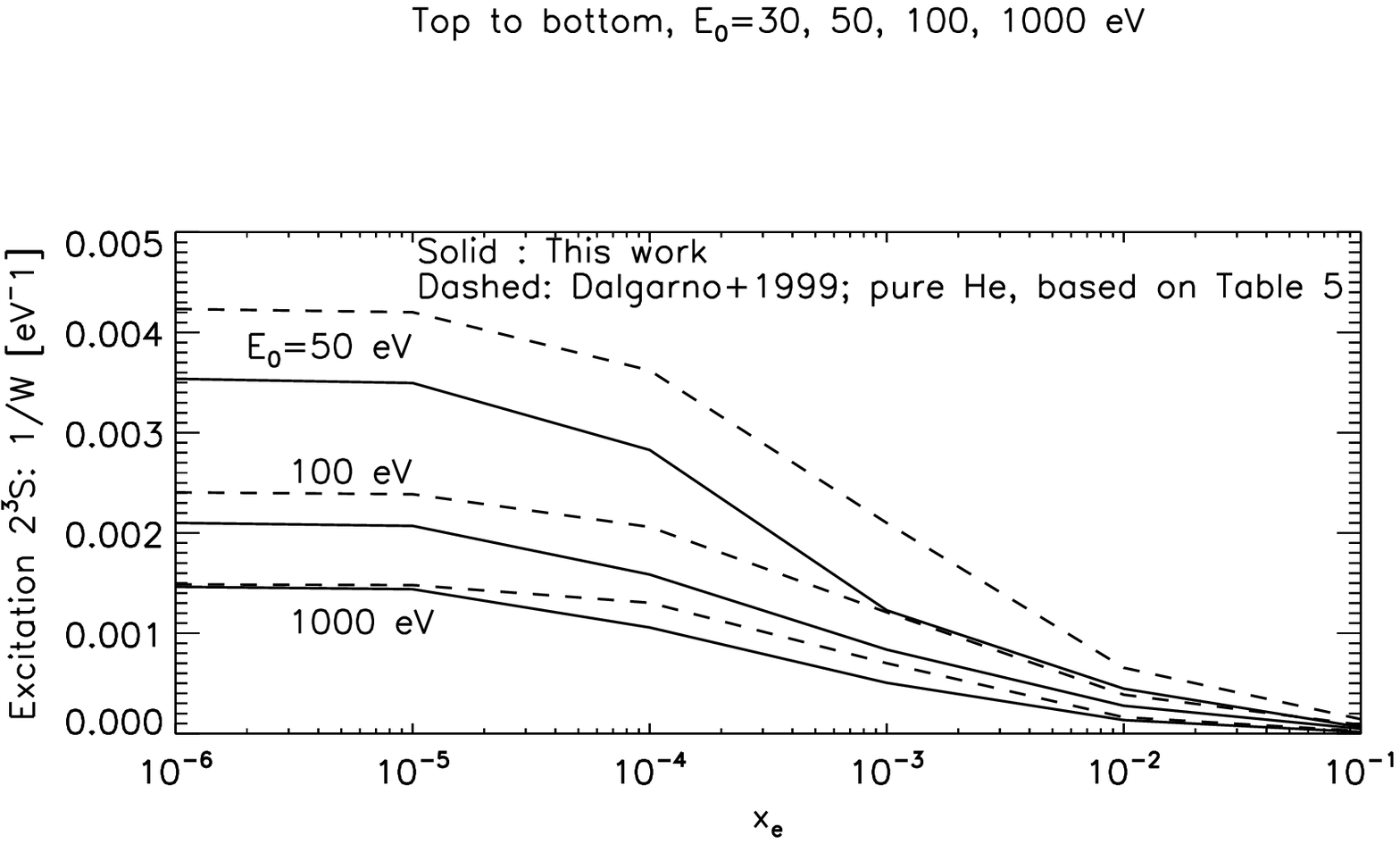}  \includegraphics[width=8.9cm]{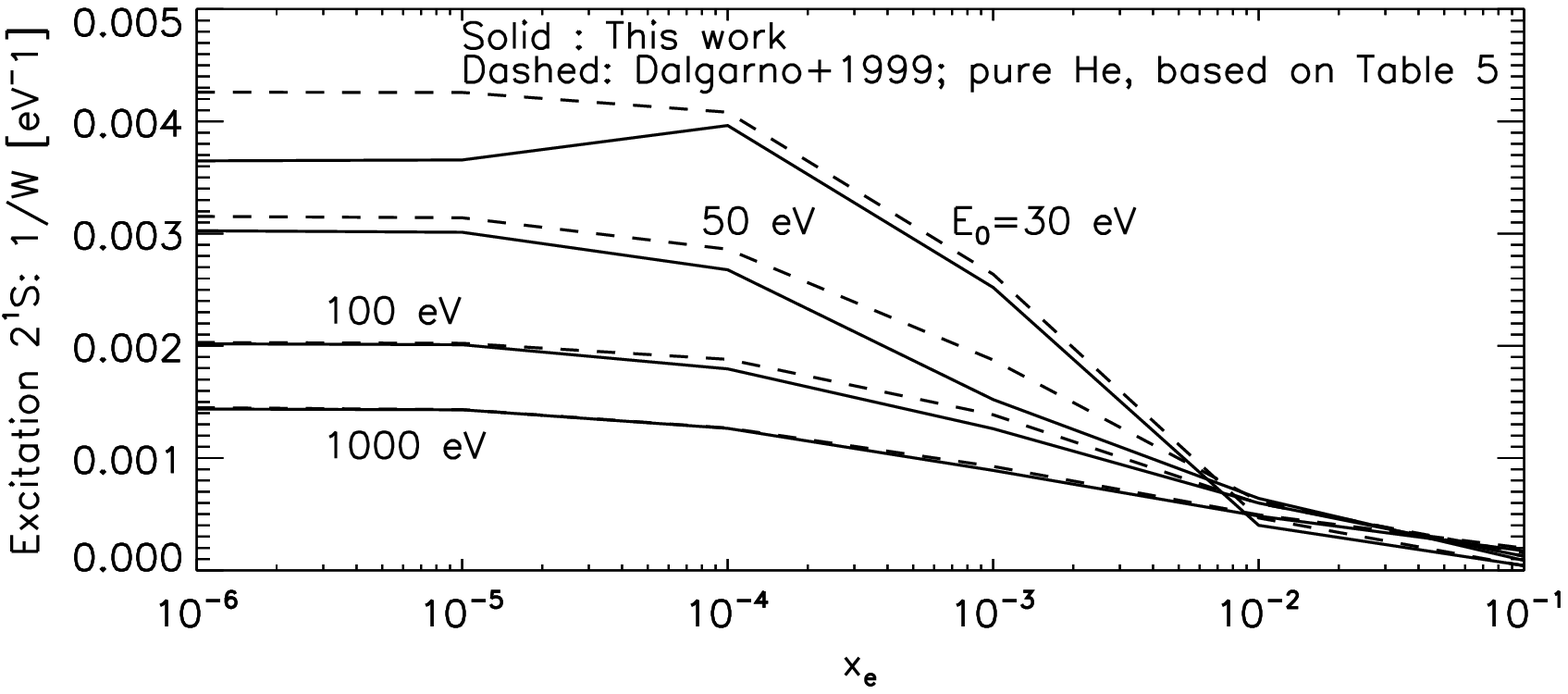} \\
    \caption{Ionization yield, excitation yields for He(2$^3S$) and He(2$^1S$), and heating efficiency
    for a ionized gas of He atoms.
    \textit{\underline{Note.}} The yields are equivalent to
    $\Phi_{\rm{He}(1) \rightarrow \rm{He}^*}$/$E_0$ and 
    $\Phi_{\rm{He}(1) \rightarrow \rm{He}^+}$/$E_0$
    as defined in our work.            
    } 
    \label{dalgarnoetal1999_he-em}
\end{figure}

\begin{figure}[h]
    \centering
    \includegraphics[width=8.9cm]{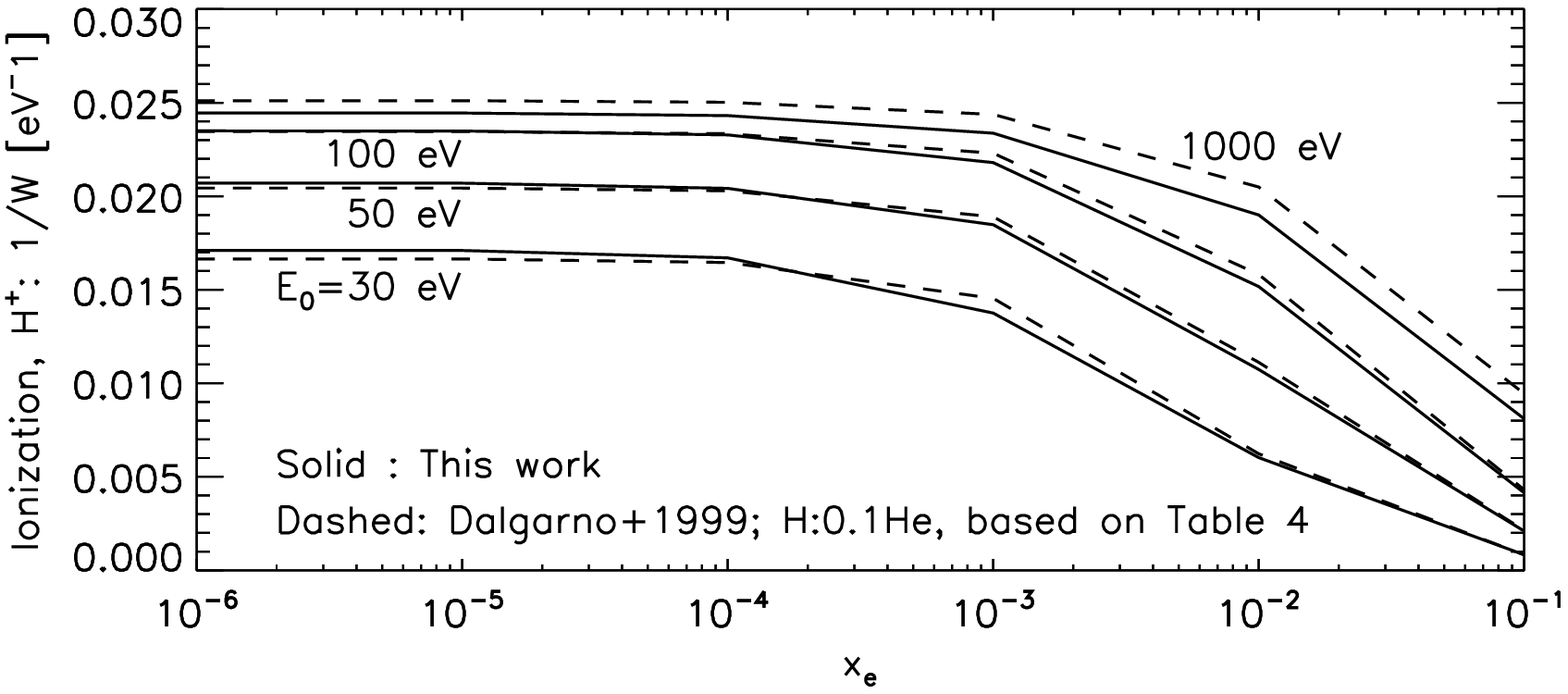} 
    \includegraphics[width=8.9cm]{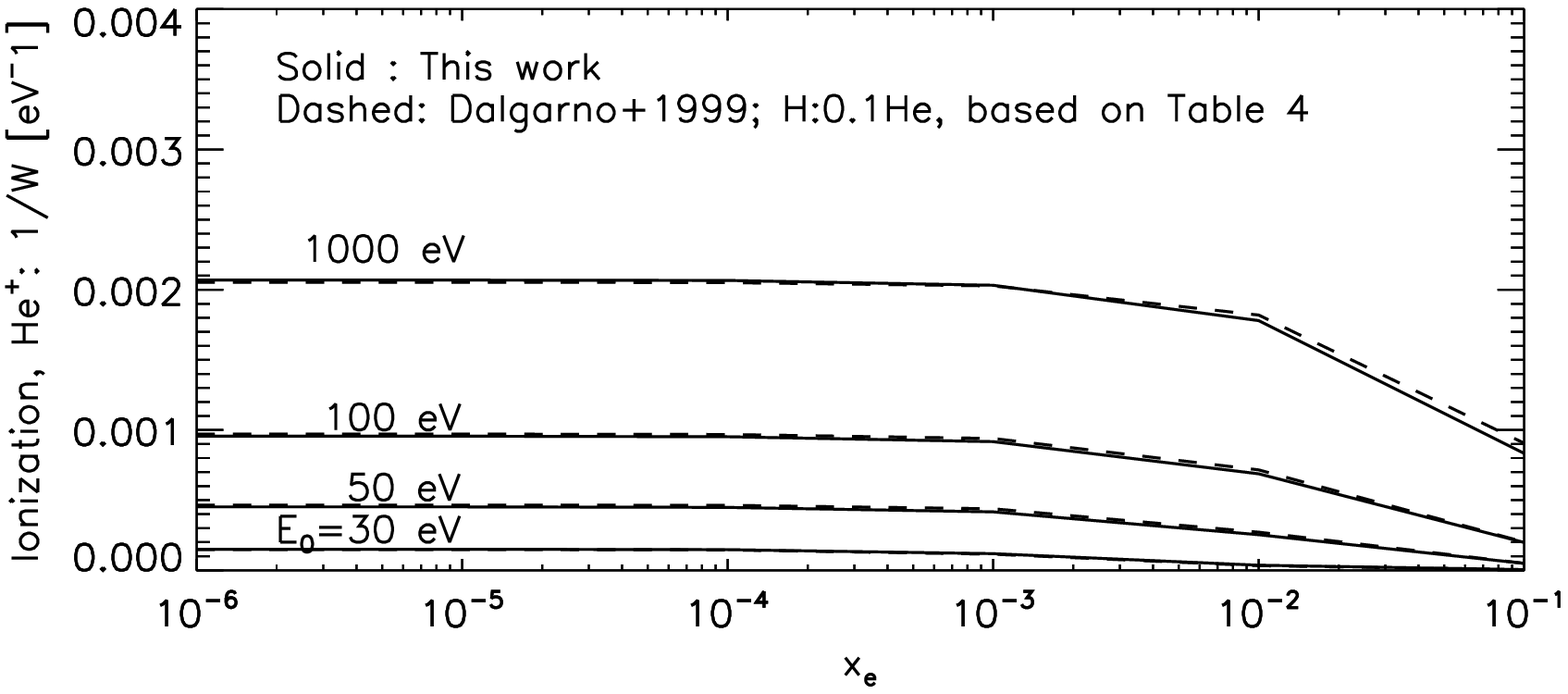} \\
    \includegraphics[width=8.9cm]{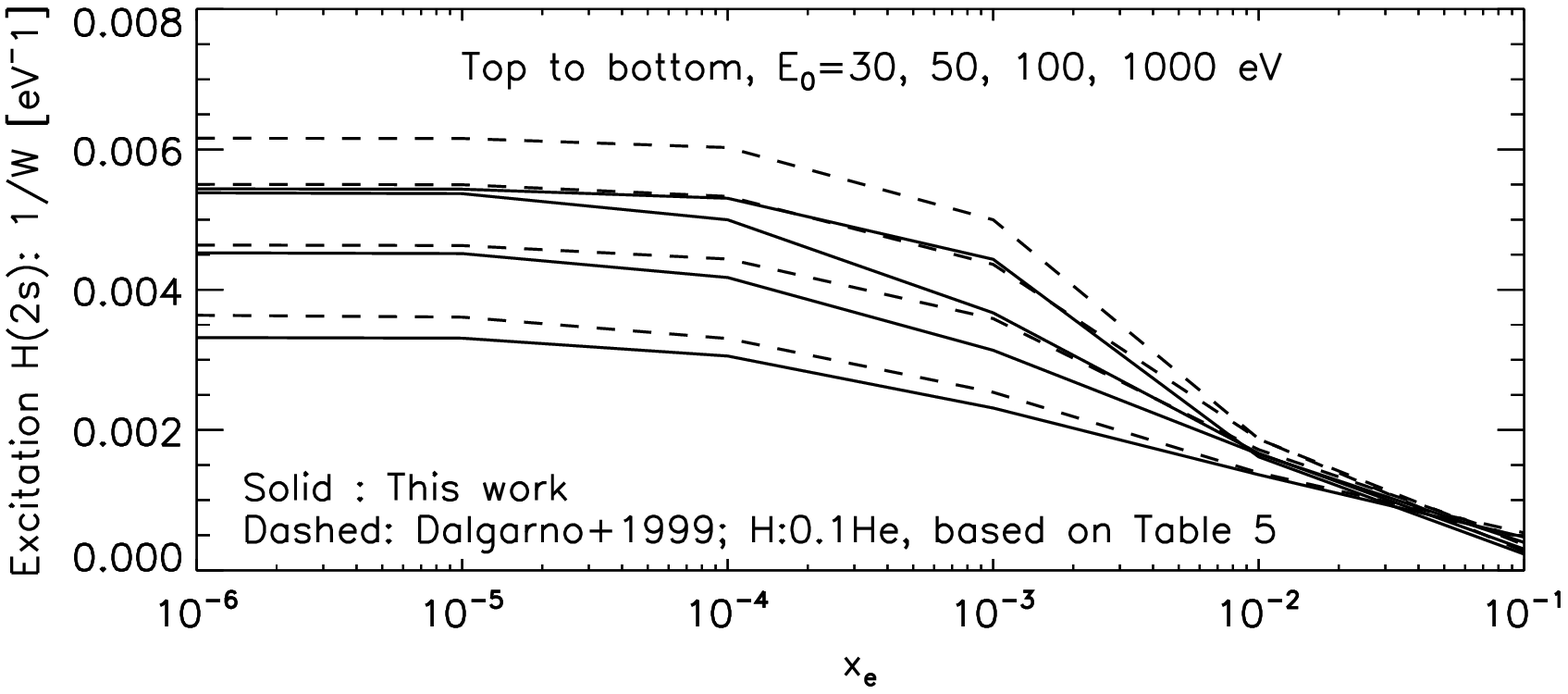}  \includegraphics[width=8.9cm]{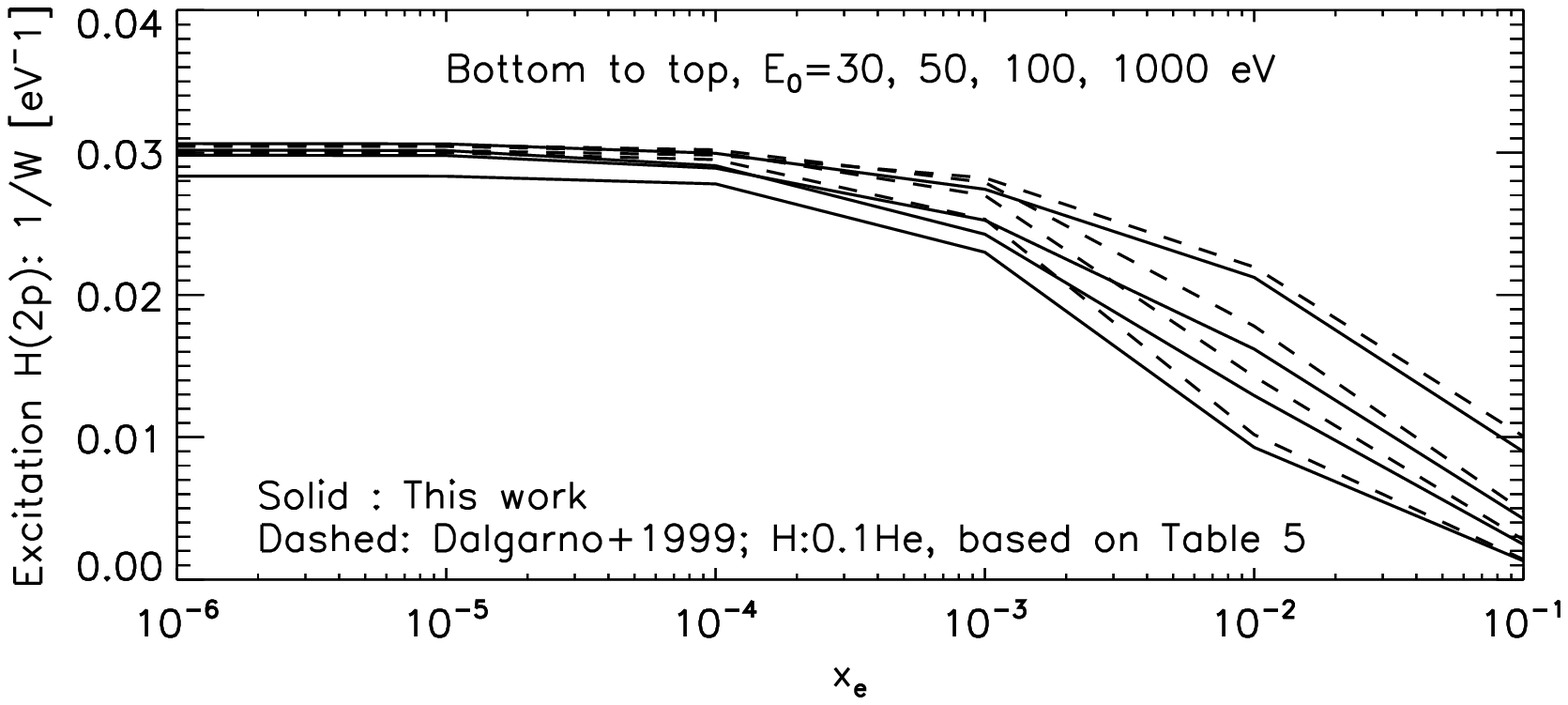} \\    
    \includegraphics[width=8.9cm]{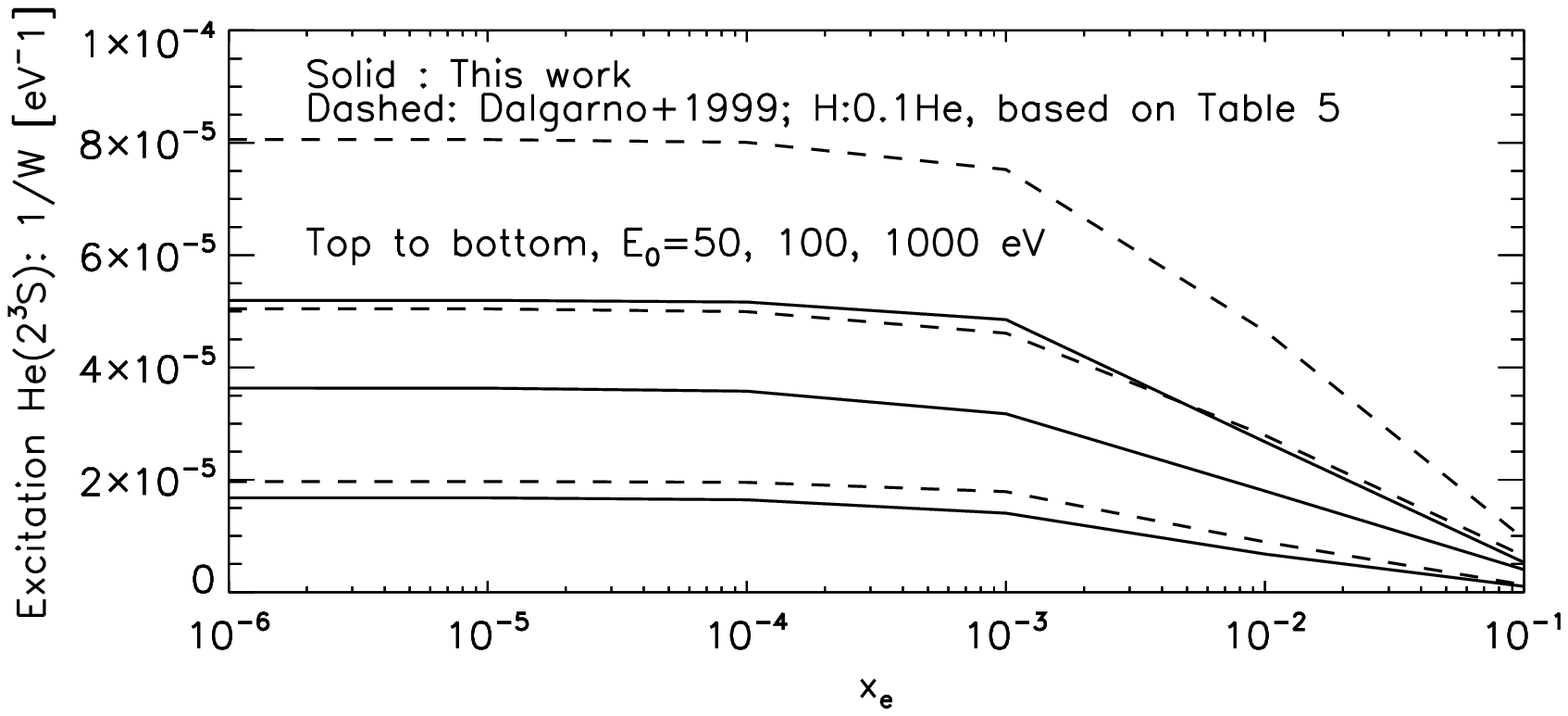}  \includegraphics[width=8.9cm]{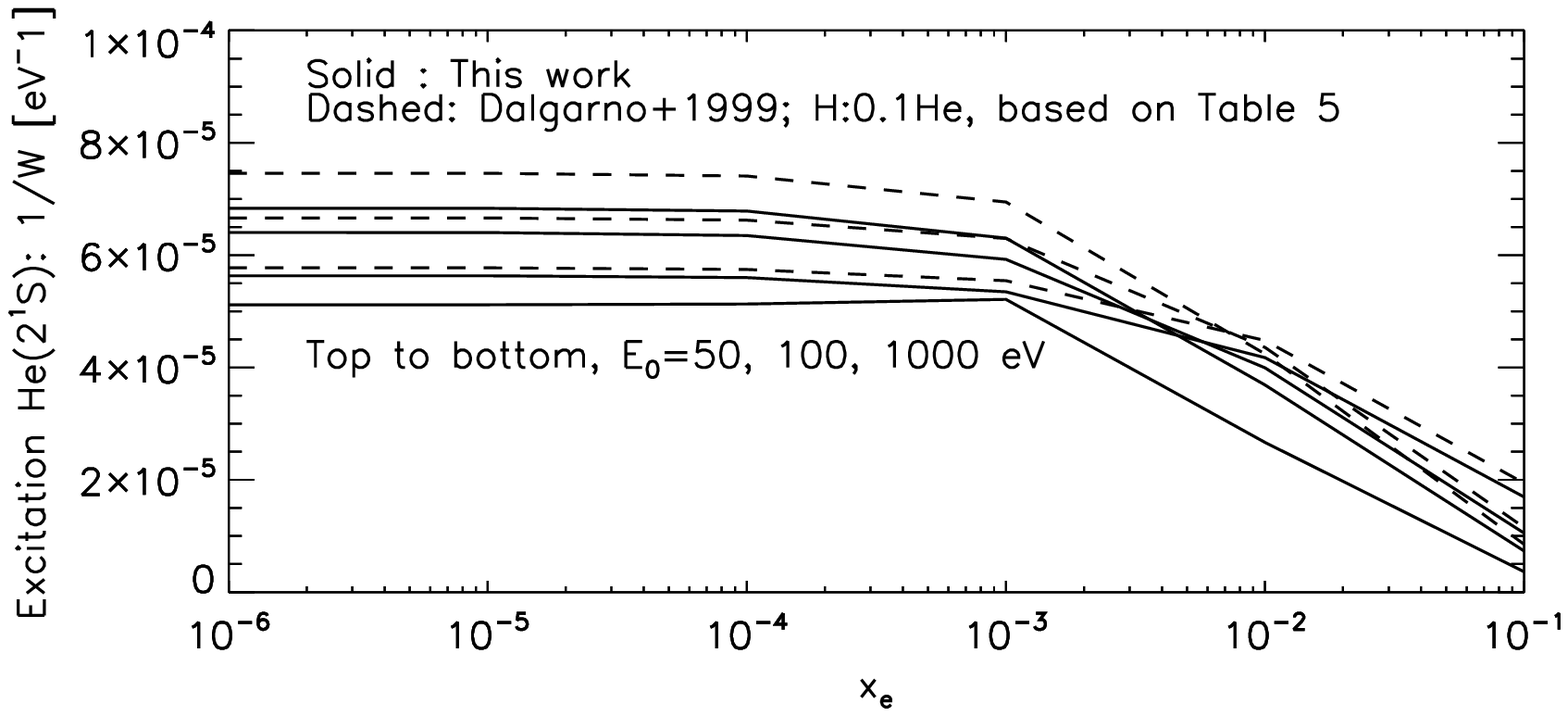} \\
\includegraphics[width=8.9cm]{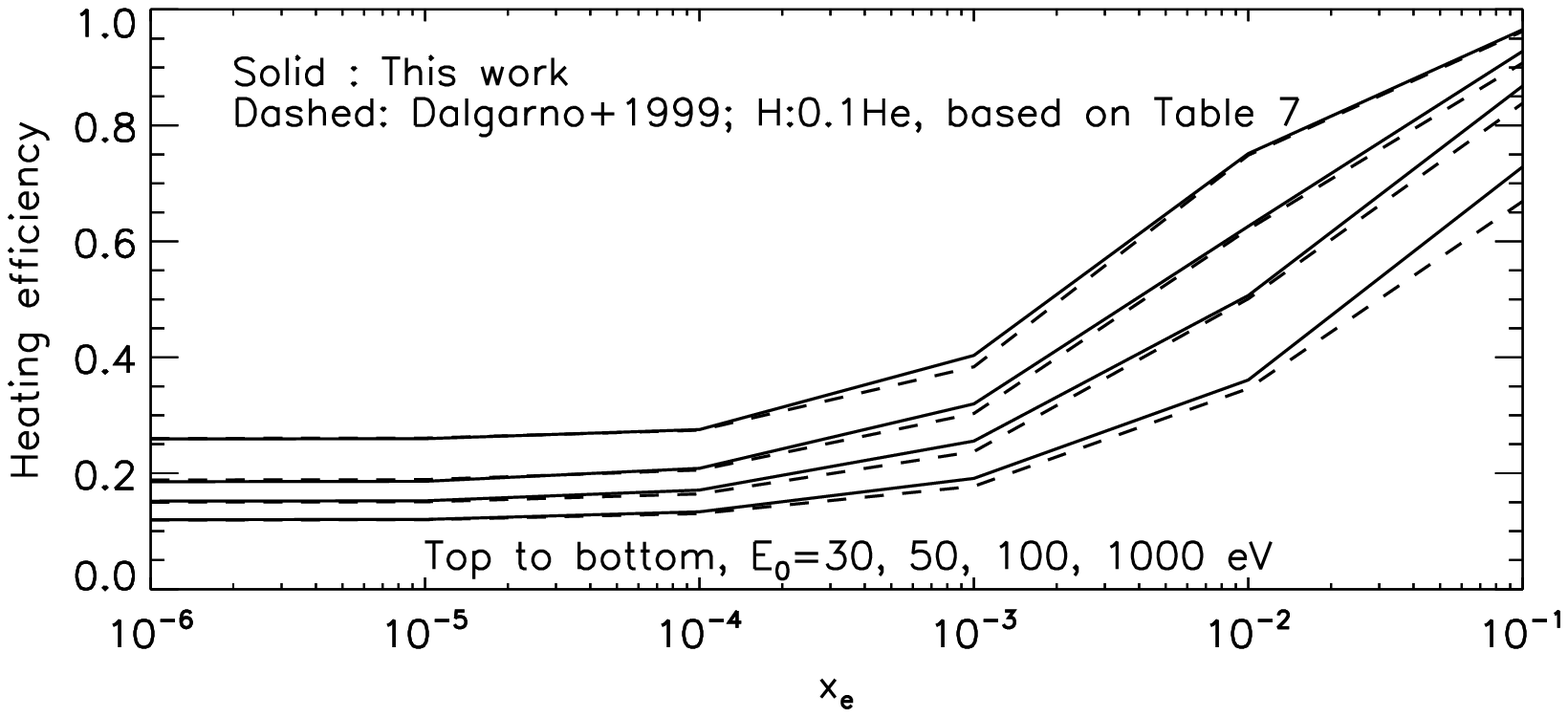} \\
        
    \caption{Ionization yields for H$^+$ and He$^+$,
    excitation yields for H(2$s$), H(2$p$), H(2$^3S$) and H(2$^1S$), and heating efficiency
    for a ionized H:0.1He mixture of atoms.
    \textit{\underline{Note.}} The yields are equivalent to
    $\Phi_{\rm{H}(1) \rightarrow \rm{H}^*}$/$E_0$, 
    $\Phi_{\rm{H}(1) \rightarrow \rm{H}^+}$/$E_0$,
    $\Phi_{\rm{He}(1^1S) \rightarrow \rm{He}^*}$/$E_0$ and 
    $\Phi_{\rm{He}(1^1S) \rightarrow \rm{He}^+}$/$E_0$
    as defined in our work.            
    } 
    \label{dalgarnoetal1999_h-he-em}
\end{figure}

\clearpage
\newpage
\bibliography{sample631}{}
\bibliographystyle{aasjournal}

\end{document}